\documentclass[aps,prfluids,reprint,superscriptaddress,showpacs,preprintnumbers,onecolumn,floatfix,longbibliography]{revtex4-1}
\usepackage{hyperref}
\usepackage{amsmath,latexsym}
\usepackage{graphicx}
\usepackage{siunitx}
\usepackage{textcomp}
\usepackage{wasysym}
\usepackage{amssymb}
\usepackage{MnSymbol}
\usepackage{oplotsymbl}
\usepackage{setspace}
\usepackage{svg}
\usepackage{mathtools}
\usepackage{wrapfig}
\usepackage{rotating}
\usepackage{epstopdf}
\usepackage{setspace}
\makeatletter
\let\@currsize\normalsize
\makeatother

\usepackage{etoolbox}
\usepackage{tikz}
\usepackage{multirow}
\usepackage{tablefootnote}
\usepackage{booktabs, makecell}

\usepackage{threeparttable}
\usepackage{setspace}
\usepackage{svg}
\usepackage{xparse}
\usepackage{chemformula}
\usepackage{vcell}
\usepackage{diagbox}

\definecolor{Muller}{RGB}{82,82,82}
\definecolor{Thiswork}{RGB}{26,111,223}

\definecolor{L25}{rgb}{0,0.25,1}
\definecolor{L50}{rgb}{0.5,1,0.5}
\definecolor{L100}{rgb}{1,0.75,0}
\definecolor{L200}{rgb}{1,0,0}
\definecolor{HPAM2}{rgb}{0.168627451,0.7568627451,0.6078431373}
\definecolor{HPAM1}{rgb}{0.7529411765,0.8705882353,0.1960784314}

\definecolor{cream}{RGB}{222,217,201}
\definecolor{PEO400}{RGB}{9,110,180}
\definecolor{XG25}{RGB}{178,12,12}
\definecolor{black}{RGB}{0,0,0}

\newrobustcmd*{\mysquare}[1]{\tikz{\filldraw[draw=#1,fill=#1] (0,0) rectangle (0.2cm,0.2cm);}}

\newrobustcmd*{\mycircle}[1]{\tikz{\filldraw[draw=#1,fill=#1] (0,0) circle [radius=0.1cm];}}

\newrobustcmd*{\mycirclesmall}[1]{\tikz{\filldraw[draw=#1,fill=#1] (0,0) circle [radius=0.07cm];}}

\newrobustcmd*{\mysquareExp}[1]{\tikz{\filldraw[draw=black,fill=#1] (0,0)
rectangle (0.2cm,0.2cm);}}

\newrobustcmd*{\mycircleExp}[1]{\tikz{\filldraw[draw=black,fill=#1] (0,0) circle [radius=0.1cm];}}

\newrobustcmd*{\mytriangleExp}[1]{\tikz{\filldraw[draw=black,fill=#1] (0,0) --
(0.2cm,0cm) -- (0.1cm,0.1732050808cm);}}

\newrobustcmd*{\mytriangle}[1]{\tikz{\filldraw[draw=#1,fill=#1] (0,0) --
(0.2cm,0cm) -- (0.1cm,0.1732050808cm);}}

\NewDocumentCommand{\myhollows}{ O{0pt} m }{
  \tikz[baseline=0, yshift=#1, inner sep=0pt, outer sep=0pt]{
    \draw[#2, line width=1pt] (0,0) circle [radius=0.15cm];
  }
}

\usepackage{todonotes}
\usepackage{comment}
\allowdisplaybreaks[1]
\begin{document}
\title{Tuning the shear and extensional rheology of semi-flexible polyelectrolyte solutions}
\date{\today}

\author{Hyeokgyun Moon}
\affiliation{School of Mechanical Engineering, Sungkyunkwan University, Suwon 16419, South Korea}
\author{Sami Yamani}
\affiliation{Hatsopoulos Microfluids Laboratory, Department of Mechanical Engineering, Massachusetts Institute of Technology, Cambridge, Massachusetts 02139, USA}
\author{Jinkee Lee}
\email[Corresponding author. ]{lee.jinkee@skku.edu}
\affiliation{School of Mechanical Engineering, Sungkyunkwan University, Suwon 16419, South Korea}
\affiliation{Institute of Quantum Biophysics, Sungkyunkwan University, Suwon 16419, South Korea}
\author{Gareth H. McKinley}
\email[Corresponding author. ]{gareth@mit.edu}
\affiliation{Hatsopoulos Microfluids Laboratory, Department of Mechanical Engineering, Massachusetts Institute of Technology, Cambridge, Massachusetts 02139, USA}

\begin{abstract}

\noindent Semi-flexible polyelectrolytes are a group of biopolymers with a wide range of applications from drag reducing agents in turbulent flows to thickening agents in food and cosmetics. In this study, we investigate the rheology of aqueous solutions of xanthan gum as a canonical semi-flexible polyelectrolyte in steady shear and transient extensional flows via torsional rheometry and dripping-onto-substrate (DoS), respectively. The high molecular weight of the xanthan gum and the numerous charged groups on the side branches attached to the backbone allow the shear and extensional rheology of the xanthan gum solutions to be tuned over a wide range by changing the ionic strength of the solvent. In steady shear flow, increasing the xanthan gum concentration increases both the zero shear viscosity and the extent of shear-thinning of the solution. Conversely, increasing the ionic strength of the solvent by addition of sodium chloride (NaCl) decreases both the zero shear viscosity and the level of shear-thinning. In transient extensional flow, increasing the xanthan gum concentration changes the dynamics of the capillary thinning from an inelastic power-law (IP) response to an elastocapillary (EC) balance, from which an extensional relaxation time can be measured based on the rate of filament thinning. Increasing the NaCl concentration decreases the extensional relaxation time and the transient extensional viscosity of the viscoelastic solution. Based on the dynamics of capillary thinning observed in the DoS experiments, we provide a relationship for the smallest extensional relaxation time that can be measured using DoS. We suggest that the change in the dynamics of capillary thinning from an IP response to an EC response can be used as an easy and robust experimental method for identifying the rheologically effective overlap concentration of a semi-flexible polyelectrolyte solution, \textit{i.e.}, the critical concentration at which polymer molecules start to interact with each other to produce a viscoelastic strain-stiffening response (often perceived as "stringiness") in transient extensional flows such as those involved in dripping, dispensing and filling operations.\\

\end{abstract}





\maketitle
\section{\label{sec_1}Introduction}

Polyelectrolytes are polymers with ionizable functional groups that form charged polymeric chains in a suitable solvent. Examples span a wide range of synthetic and natural polymers such as nucleic acids, proteins, polypeptides, and polysaccharides. The charges on the repeat unit of a polyelectrolyte play an important role in determining the macromolecular conformation, intermolecular interactions, and physico-chemical properties of the solutions, which are dependent on the solvent ionic strength, pH, temperature, and the presence of surfactants or other polymers~\cite{Lapasin2012, Alexandridis2022, Kesharwani2017}. These unique properties of polyelectrolytes make them suitable for various applications in food formulations~\cite{Elhamirad2021, ArjonaRoman2021}, cosmetics~\cite{JimenezKairuz2015}, drag reduction~\cite{Mckinley2019, Soares2013, Grisel2015}, pharmaceuticals~\cite{Zhang2022,Harada2020}, and oil recovery~\cite{Zhang2021} industries.

Xanthan gum is a polyelectrolyte obtained from the bacterium \textit{Xanthomonas compestris} through a fermentation process in a batch reactor containing the necessary substrates and nutrients~\cite{Lapasin2012,Chatterji2023}. It is a semi-flexible biopolymer with good solubility in aqueous solution because of its ability to form multiple hydrogen bonds with water molecules. Xanthan gum is commonly used as a rheological modifier in various products and considered as a suitable alternative to synthetic polymers due to its biodegradability and low price~\cite{Mckinley2019}. The rheological versatility of xanthan gum is a result of the high charge density of the anionic polymer side branches, the lack of crystallinity, the strong propensity for hydrogen bonding as well as its semi-flexible structure in aqueous solution.

\begin{figure*}[h]
\centering
  \includegraphics[width=0.99\textwidth]{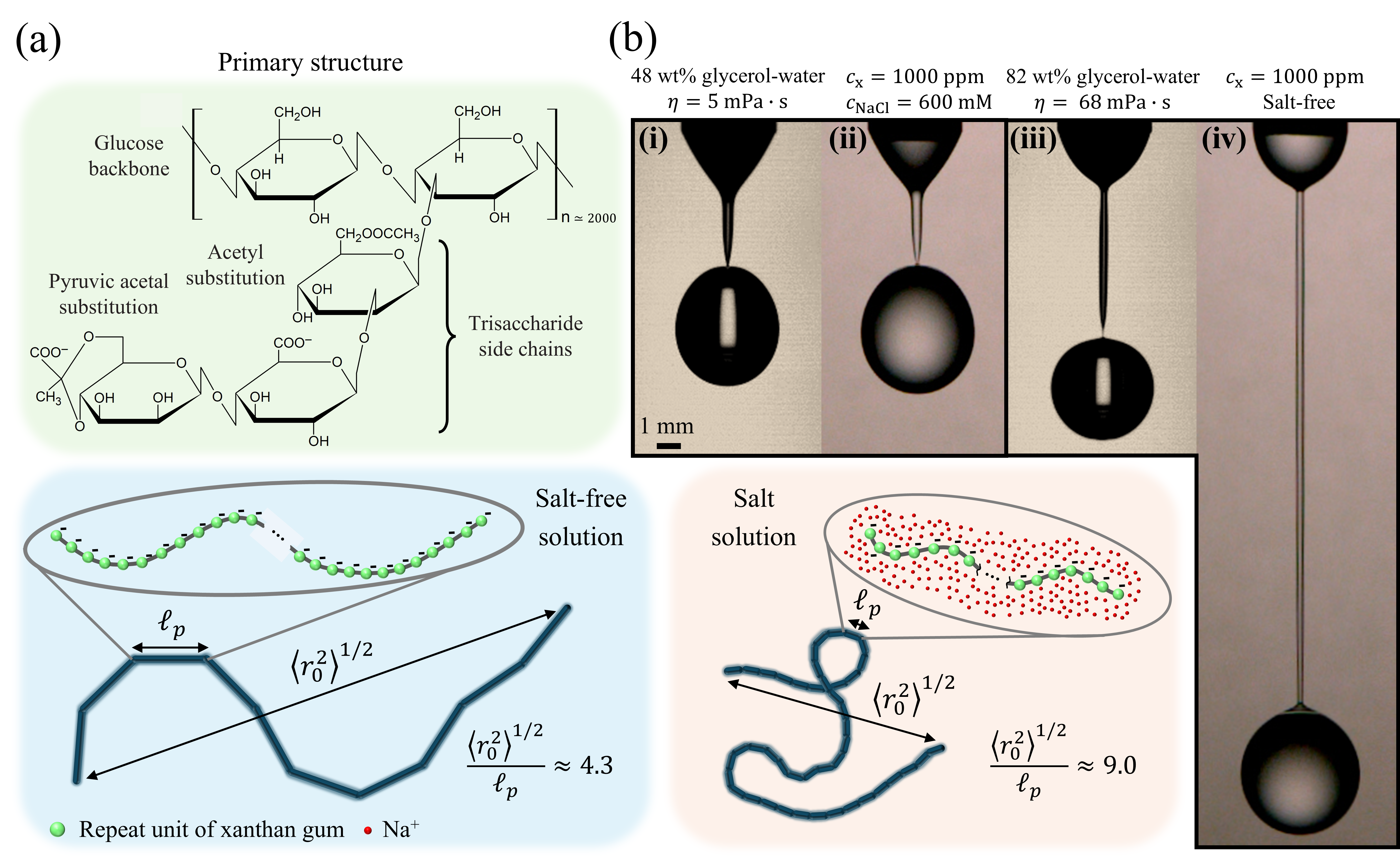}
  \caption{(a) Molecular structure of xanthan gum showing different conformations depending on the ionic strength. (b) Illustrative Liquid dripping experiments of glycerol-water mixture and xanthan gum solutions with matched viscosities at flow rate $Q=5$ ml/min: (\lowercase\expandafter{\romannumeral1}) glycerol-water mixture (48/52 w/w), (\lowercase\expandafter{\romannumeral2}) 1000 ppm xanthan gum solution with 600 mM NaCl, (\lowercase\expandafter{\romannumeral3}) glycerol-water mixture (82/18 w/w), and (\lowercase\expandafter{\romannumeral4}) 1000 ppm xanthan gum solution without salt.}
  \label{fgr_1}
\end{figure*}

The primary structure of xanthan gum consists of a 1,4-linked $\beta$-D-glucose backbone and charged trisaccharide side chains ($\beta$-D-mannose-(1–4)-$\alpha$-D-glucuronic acid-(1–2)-$\alpha$-D-mannose) linked in the C(3) position of every alternate glucose residue. The pyruvic acetal substitution is linked to the O(4) and O(6) of the terminal mannose unit and the acetyl substitution is linked to the O(6) of the mannose unit adjacent to the backbone, as shown in Fig.\,\ref{fgr_1}(a). The degree of pyruvate and acetyl substitution may vary depending on the fermentation process~\cite{Lapasin2012,Alexandridis2022,Liberatore2009}.
Due to the charges on these short side branches, the persistence length of the polymer chain (\textit{i.e.}, the distance over which correlations in the direction vector tangent to the chain are lost), and the polymer conformation change depending on the presence of salt. The persistence length $\ell_p$ characterizes quantitatively the local chain flexibility. The persistence length is commonly decomposed into two parts: (\textit{i}) the structural contribution to the persistence length ($\ell_{p,s}$) due to the steric hindrance of both the backbone and the bulky side chains and (\textit{ii}) the electrostatic persistence length ($\ell_{p,e}$) due to electrostatic repulsion between charged groups~\cite{Chauveteau1986}. In a salt-free solution disordered chains minimize contributions to the steric hindrance. However, negative charges along the side chain enhance the electrostatic repulsion, leading to an expanded state. In a salt-free solution, the contribution from $\ell_{p,s}$ is small and $\ell_{p,e}$ is large. In the presence of salt, the charge screening induced by salt ions reduces the electrostatic repulsion and enables formation of a locally rigid helix structure. This results in a dominant contribution from $\ell_{p,s}$ and negligibly small value of $\ell_{p,e}$. For xanthan gum, the total persistence length $\ell_{p}=\ell_{p,s}+\ell_{p,e}$ decreases from approximately 210 nm in a salt-free solution to 52 nm in a 1 mM sodium chloride (NaCl) solution~\cite{Chauveteau1986}. This decrease is due to the dramatic reduction of the electrostatic persistence length $\ell_{p,e}$ resulting from charge screening~\cite{Chauveteau1986}.

In a salt-free solution, long xanthan gum molecules have a worm-like chain conformation that is relatively stiff and this leads to extended structures with large value of the root-mean-square end-to-end distance $\left<r_{0}^{2} \right>^{1/2}$~\cite{Brant1990}. In contrast, in the presence of salt, xanthan gum chains adopt a more compact and flexible conformation. An important characteristic of xanthan gum as a polyelectrolyte is the tunability of this persistence length, which can be achieved by controlling the concentration of salt added to the solvent (see the \hyperref[ESI]{Supplemental Material} Fig.\,\hyperlink{supp:fig_1}{S1} for additional details). The change in the local flexibility and equilibrium size of the chain is manifested at the continuum level as a dramatic change in the rheological properties.

Since the discovery of xanthan gum in 1961~\cite{Jeanes1961}, there have been a number of studies investigating its rheological properties in dilute~\cite{Liberatore2009, Macosko1978}, semi-dilute~\cite{Liberatore2011, Macosko1978} and concentrated solutions~\cite{Liberatore2011, Macosko1978, missi2024thermo}. At sufficiently high concentrations xanthan gum solutions become elastoviscoplastic and a yield stress develops~\cite{missi2024thermo}. In the present work, we remain far below this concentration. The majority of these earlier studies focused on the impact of diverse physico-chemical parameters, such as polymer concentration~\cite{Macosko1978, missi2024thermo, Liberatore2009, Liberatore2011}, temperature~\cite{Gomez2000,Morris2019,missi2024thermo}, 
 presence of surfactants~\cite{Alexandridis2022}, molecular weight~\cite{Fujita1984scattering,Fujita1984, Holzwarth1978}, and solvent ionic strength~\cite{Boger2000,Chauveteau1986,Rinaudo1989,Bean2007,Bercea2014, Liberatore2009, Liberatore2011}, on the shear rheology of xanthan gum solutions. However, the extensional rheology of xanthan gum solutions has been largely overlooked, despite the significant importance of extensional flows in numerous industrial applications~\cite{Sharma2020pof,Rahatekar2012} and complex fluid dynamical phenomena such as friction reduction. An important characteristic of xanthan gum in drag reduction applications is that it is much less susceptible to flow-induced degradation~\cite{cussuol2023polymer}. The nonlinear extensional behavior of polymer solutions is commonly characterized by the dimensionless finite extensibility of the polymer molecule $L$, defined as the ratio of the polymer contour length $r_{\textrm{max}}$, (\textit{i.e.}, the length of the polymer in its fully extended state), to the root-mean-square end-to-end distance of the polymer chain at equilibrium. Changes in this ratio can dramatically modify the resulting extensional rheology of a solution~\cite{larson2015modeling}. 
 
 To illustrate this, we show in Fig.\,\ref{fgr_1}(b) several snapshots of the liquid dripping dynamics observed for Newtonian (glycerol-water mixture) and xanthan gum solutions exiting from a nozzle with outer radius of $R_0 = 2$ mm at a slow flow rate of $Q=5$ ml/min. Dripping of a glycerol-water mixture with viscosity $\eta=5$ mPa$\cdot$s is compared to dripping of a xanthan gum solution with xanthan gum concentration $c_{\textrm{x}}=1000$ ppm dissolved in salt water with NaCl concentration $c_{\textrm{NaCl}}=600$ mM, which is comparable to the salinity of sea water~\cite{McDougall2008}, as shown in Fig.\,\ref{fgr_1}(b-\lowercase\expandafter{\romannumeral1}) and Fig.\,\ref{fgr_1}(b-\lowercase\expandafter{\romannumeral2}), respectively. Similarly, dripping of a glycerol-water mixture with viscosity $\eta=68$ mPa$\cdot$s is compared to dripping of a xanthan gum solution with $c_{\textrm{x}}=1000$ ppm in deionized (DI) water as solvent, as shown in Fig.\,\ref{fgr_1}(b-\lowercase\expandafter{\romannumeral3}) and Fig.\,\ref{fgr_1}(b-\lowercase\expandafter{\romannumeral4}), respectively. Here, the viscosities of the Newtonian water-glycerol mixtures shown in Fig.\,\ref{fgr_1}(b-\lowercase\expandafter{\romannumeral1}) and (b-\lowercase\expandafter{\romannumeral3}) are tuned by varying the concentration of glycerol to have matching shear viscosities with the xanthan gum solutions shown in Fig.\,\ref{fgr_1}(b-\lowercase\expandafter{\romannumeral2}) and (b-\lowercase\expandafter{\romannumeral4}) at the shear rate $\dot{\gamma} \simeq 8Q/3\pi R_0^3=9\,\textrm{s}^{-1}$ that the flow experiences at the nozzle. The values of the Reynolds number at the nozzle $\textrm{Re}_{\textrm{drip}} = \rho U R_0/\eta(\dot{\gamma})$ are $\textrm{Re}_{\textrm{drip}} = 2.97$ for $\eta = 5$ mPa$\cdot$s and $\textrm{Re}_{\textrm{drip}} = 0.24$ for $\eta = 68$ mPa$\cdot$s, respectively, where $\rho$ is the fluid density, and $U=Q/\pi R_0^2$ is the nominal flow velocity at the nozzle. At comparable values of the viscosity, the xanthan gum solution with DI water as its solvent exhibits strong viscoelastic strain-stiffening and filament formation~\cite{McKinley2005, clasen2006dilute} (Fig.\,\ref{fgr_1}(b-\lowercase\expandafter{\romannumeral4})) compared to the Newtonian fluid (Fig.\,\ref{fgr_1}(b-\lowercase\expandafter{\romannumeral3})). Even though $c_{\textrm{x}}$ is the same for both solutions, the presence of salt decreases the shear viscosity but also the nonlinear elasticity dramatically (Fig.\,\ref{fgr_1}(b-\lowercase\expandafter{\romannumeral2})) and the thread pinches-off more rapidly. The stringiness of the fluid thread formed is markedly reduced which is important in many dispensing operations~\cite{clasen2012dispensing}. These dripping experiments illustrate visually the viability of tuning both shear and extensional rheology of semi-flexible polyelectrolyte solutions by the addition of salt. In this manuscript, we quantify these dramatic effects by measuring the steady shear and transient extensional rheology of xanthan gum solutions as a function of polymer concentration and ionic strength.

 One important parameter in characterizing the dynamics of a polymer solution is  the overlap concentration $c^\ast$, representing the concentration at which polymer chains begin to interact with each other in solution~\cite{Rubinstein1995}. The overlap concentration distinguishes the dilute regime from the semi-dilute regime for polymer solutions. It is essential to accurately evaluate the overlap concentration given its critical role in understanding the dominant physics governing the changes in rheology of polymer solutions. Extensive research has been conducted to estimate the overlap concentration for various types of polymers, including flexible polymers~\cite{DeGennes1979}, semi-flexible polymers~\cite{Chu1987}, rigid-rod polymers~\cite{Edwards1988}, and flexible polyelectrolytes~\cite{Rubinstein1995,Rubinstein2005}. Semi-flexible polyelectrolytes, however, have received limited attention. In addition, the high sensitivity of semi-flexible polyelectrolytes such as xanthan gum to environmental conditions, such as solvent ionic strength, further complicates the governing theory for determining the overlap concentration of semi-flexible polyelectrolytes~\cite{Boger2000}. The overlap concentration of xanthan gum has been evaluated experimentally~\cite{Bercea2014,Liberatore2009,Liberatore2011}, but these evaluations may be subject to experimental error depending on the experimental technique utilized and the data analysis procedure~\cite{coviello1986solution}. 

In this study, we investigate the influence of the ionic strength of the solvent on the microstructure of xanthan gum and its subsequent manifestation in the macroscopic shear and extensional rheological properties of xanthan gum solutions. As we have discussed above, the persistence length of xanthan gum chains in aqueous solution can be tuned by the addition of NaCl, resulting in a progressive conformation transition~\cite{camesano2001single} and a systematic change in rheology~\cite{Liberatore2009}. Using both torsional rheometry and dripping-onto-substrate (DoS) experiments, we characterize the rheological response of xanthan gum solutions in shear and extensional flows over a wide range of polymer and salt concentrations and compare them to the shear and extensional rheology of dilute solutions of polyethylene oxide (PEO), a neutral flexible polymer. We also provide estimates of the smallest extensional relaxation time that can be measured for a given DoS setup and suggest an experimental method based on DoS rheometry for calculating the rheologically-effective overlap concentration of semi-flexible polyelectrolytes in solvents of different ionic strength.

\section{\label{sec_2}Materials and Methods}
\subsection{Materials}
We formulate our semi-flexible polyelectrolyte solutions using commercial xanthan gum (Keltrol T 622, CP Kelco) with reported molecular weight $M_w\approx2\times10^6$ g/mol and polydispersity index $\sim$2~\cite{Liberatore2009}. Xanthan gum powder is dissolved in DI water (UltraPure Type 1, ChemWorld) with resistivity greater than 18.2 M$\Omega~$cm and NaCl-water solutions with different NaCl concentrations. Stock solutions of xanthan gum with $c_{\textrm{x}}=1000$ ppm in DI water and NaCl-water solutions with $1 \textrm{ mM}\leq c_{\textrm{NaCl}}\leq 600\textrm{ mM}$ are gently mixed at room temperature for 1 day using a roller mixer. The maximum salt concentration utilized in this study is the salt concentration of sea water, $c_{\textrm{NaCl}} = 600\textrm{ mM}$. Samples at different xanthan gum concentrations are prepared by successive dilution of the stock solution with the solvent. Samples are then gently mixed on a roller for 1 day. To avoid degradation of the xanthan gum, measurements are performed within four days of solution preparation. 
 The number of NaCl ions per unit xanthan gum chain, per repeat unit, and per unit persistence length at different xanthan gum and NaCl concentrations are shown in Tables\,\hyperlink{supp:tbl_1}{S\,\MakeUppercase{\romannumeral1}}-\hyperlink{supp:tbl_3}{S\,\MakeUppercase{\romannumeral3}} of the \hyperref[ESI]{Supplemental Material}, respectively. Importantly, even at a salt concentration of $c_\textrm{NaCl} = 1$ mM and xanthan gum concentration $c_\textrm{x} = 1000$ ppm, there are 2000 $\textrm{Na}^+$ ions available per chain, corresponding to one $\textrm{Na}^+$ ion per repeat unit as shown in Fig.\,\ref{fgr_1}(a) and Tables\,\hyperlink{supp:tbl_1}{S\,\MakeUppercase{\romannumeral1}} and \hyperlink{supp:tbl_2}{S\,\MakeUppercase{\romannumeral2}}.
 
 For comparison, PEO with $M_w\approx2\times10^6$ g/mol (Sigma-Aldrich) was used to prepare dilute solutions of a flexible and highly extensible polymer. PEO is added to DI water and mixed for 3 days to make an 0.5\% wt. stock solution. Aqueous dilute solutions of PEO at different concentrations $c_{\textrm{PEO}}$ are then prepared by successive dilution.

\subsection{Torsional shear rheometry}
The rate-dependent shear viscosity of the polymer solutions was measured using a double-wall concentric cylinder geometry with inner diameter $d_i=29.4$ mm, outer diameter $d_o=32.1$ mm and gap size of 2 mm on a torsional rheometer (ARES-G2, TA instruments) at $25~^{\circ}\textrm{C}$. The measurements were performed for shear rates in the range $10^{-2}~\textrm{s}^{-1} \leq \dot{\gamma} \leq 10^3~\textrm{s}^{-1}$. The measured viscosity is reported over a wide range of shear rates between the minimum shear rate corresponding to the low-torque limit of the rheometer and the critical shear rate corresponding to the onset of inertial secondary flows beyond a critical Taylor number~\cite{Macosko1994}. The smallest viscosity that can be measured based on the low torque limit of the rheometer is, $\eta_{\textrm{min}}= F_{\tau}T_{min}/\dot{\gamma}$, where $F_{\tau}$ = 6119.19 Pa/N$\cdot$m is the stress constant for our double-wall concentric cylinder geometry and $T_{\textrm{min}}$ = 1$\mu$N$\cdot$m is the low torque limit of the rheometer. The calibrated value of $T_{\textrm{min}}$ is typically 10 times larger than the instrument resolution specification due to additional resistive torque produced for example by surface tension effect~\cite{Caretta2015}. The largest viscosity that can be accurately measured before the secondary flows induced by flow inertia and curved streamlines increase the measured torque, leading to an overestimation of the viscosity, is $\eta \lesssim \kappa \dot{\gamma}$, where $\kappa$ is determined based on the geometry and fluid properties. For a Newtonian fluid in a single-wall concentric cylinder, $\kappa=\rho(r_o-r_i)^{5/2}/(1700r_i)^{1/2}$, where $r_o$ and $r_i$ are the radii of cup and bob, respectively. However, the coefficient, $\kappa$, will be different in other systems such as a double-wall concentric cylinder geometry~\cite{Caretta2015}. We use an empirically derived value of $\kappa=4\times10^{-6}$ kg/m to constrain the maximum shear rate and viscosity relationship. 

\subsection{Dripping-onto-substrate (DoS) rheometry}
\begin{figure*}[h]
\centering
  \includegraphics[width=0.99\textwidth]{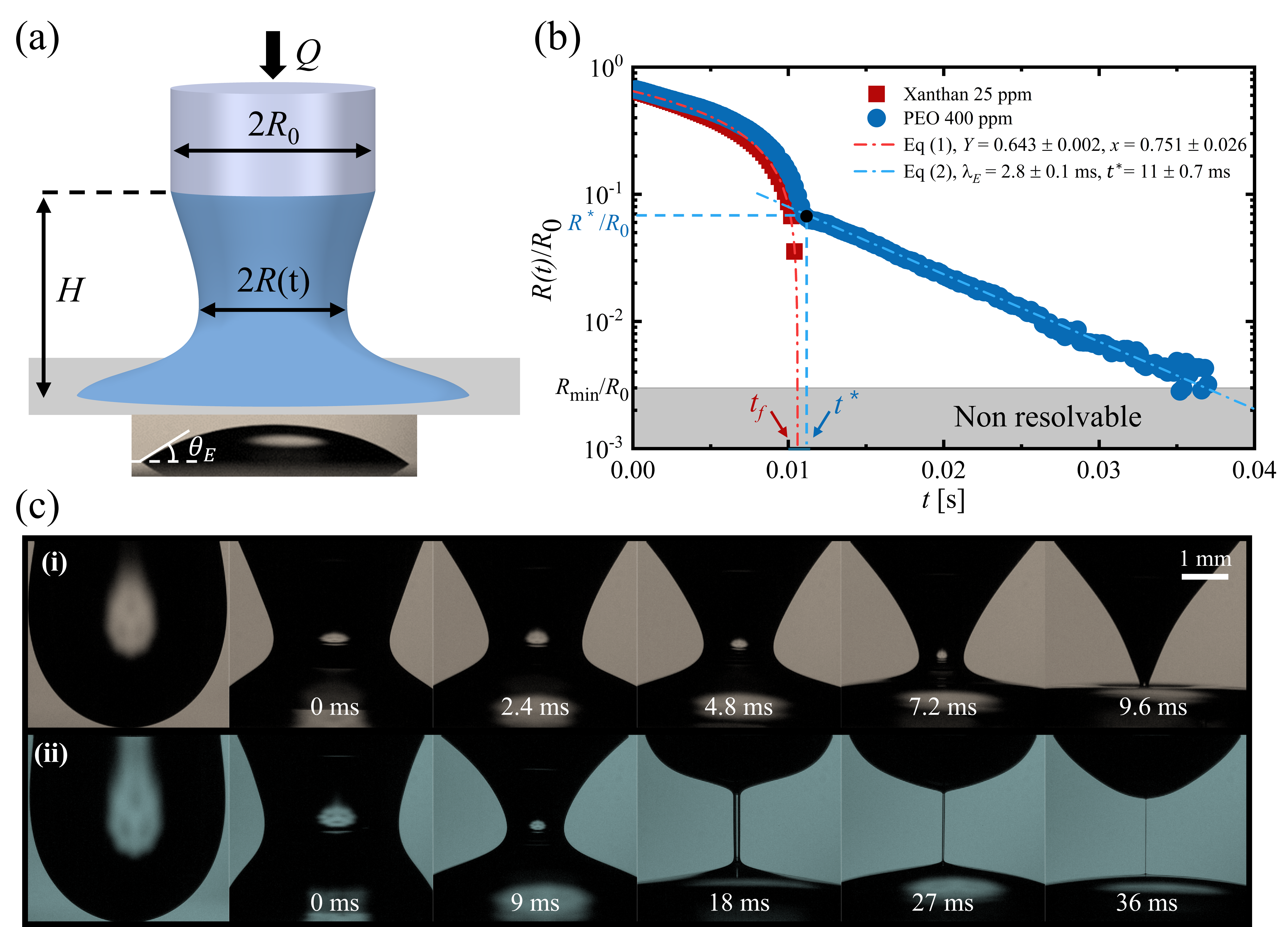}
  \caption{(a) Schematic diagram of the DoS rheometry setup. The inset shows the equilibrium contact angle of a drop of xanthan gum solution on a clean glass substrate; $\theta_E=35^{\circ}$. (b) Temporal evolution of the filament radius at the neck, $R(t)$, for a salt-free xanthan gum (\mysquare{XG25}) and a PEO solution (\mycircle{PEO400}) in the dilute regime ($c/c^\ast\approx0.5$). (c) A sequence of image from DoS experiments corresponding to (b): (\lowercase\expandafter{\romannumeral1}) salt-free xanthan gum solution with $c_{\textrm{x}}=25$ ppm and (\lowercase\expandafter{\romannumeral2}) PEO solution with $c_{\textrm{PEO}}=400$ ppm.}
  \label{fgr_2}
\end{figure*}

DoS rheometry~\cite{Sharma2015,Sharma2017} is used for characterizing the extensional rheology of complex fluids with low viscosity ($\eta<20$ mPa $\cdot$s) and can be especially useful for low elasticity fluids (extensional relaxation time, $\lambda_E<1$ ms)~\cite{Sharma2020pof}. It can thus extend the range of material properties accessible with other extensional rheometers such as the capillary breakup extensional rheometer (CaBER)~\cite{McKinley2005}. Visualization of capillary thinning and pinch-off dynamics at high spatio-temporal resolution, together with detailed image processing of the evolving liquid bridge, is used to characterize the evolution of the liquid filament profile $R(z,t)$. The experimental setup consists of a syringe dispenser, as shown in Fig.\,\ref{fgr_2}(a), and a high-speed digital imaging system. In the present experiments, a steel nozzle with outer diameter $D_o=2R_0=4$ mm and an inner diameter $D_i=3.5$ mm is used, and the nozzle height from the substrate is set at $H=2R_0$. A small volume of the solution ($V=5\,\mu$l) is dispensed at a very low flow rate ($Q = 0.05$ ml/min) to form a pendant drop. This drop then contacts a partially wetting glass substrate with equilibrium contact angle $\theta_E=35^{\circ}$ to form a liquid bridge, which then undergoes capillarity-driven thinning and breakup. The evolution of the filament profile is recorded using a high-speed camera (Phantom Miro 320S) at 5000 frames per second (fps). The video is analyzed using an in-house developed MATLAB$^{\textrm{\textregistered}}$ code to calculate the evolution of the filament profile $R(z,t)$ with time. Each experiment is repeated four times, and good reproducibility is observed.

Figure\,\ref{fgr_2}(b) shows the temporal evolution of the minimum filament radius at the neck, $R(t)=\textrm{min}\{R(z,t)\}$, for dilute solutions of xanthan gum and PEO with similar molecular weight $M_w \approx 2\times10^6$ g/mol, and concentrations such that $c/c^\ast \simeq 0.5$ for both solutions. These filament thinning profiles are achieved by analyzing a sequence of images, a subset of representative frames are shown in Fig.\,\ref{fgr_2}(c).
The filament thinning profile observed for the dilute xanthan gum solution follows a power-law of the form,
\begin{equation}
\label{eq1}
\frac{R(t)}{R_0}=Y\left ( \frac{t_f-t}{t_R} \right )^x,
\end{equation}
\noindent expected for an inelastic fluid with a rate-dependent viscosity~\cite{McKinley2003}. In Eq.\,\ref{eq1}, the front factor $Y$ depends on the initial configuration, the power-law exponent $x$ varies in the range $2/3 \leq x \leq 1$, $t_f$ is the pinch-off time, and $t_R=(\rho R_0^3/\sigma)^{1/2}$ is the Rayleigh time scale. The red dash-dotted line in Fig.\,\ref{fgr_2}(b) is a power-law fit of the form shown in Eq.\,\ref{eq1} in the inelastic power-law (IP) regime. A power-law exponent $x = 2/3$ corresponds to an inertiocapillary (IC) response for a low viscosity Newtonian fluid~\cite{Calabrese2022, Sharma2021,day1998self} with $\textrm{Oh} = \eta_0/(\rho\sigma R_0)^{1/2} \ll 1$, where the Ohnesorge number characterizes the ratio of viscous dissipation to fluid inertia in a surface tension driven flow. 

The thinning dynamics of the low viscosity PEO solution initially also show an IC response, however this is followed by a subsequent transition to an elastocapillary (EC) response~\cite{Sharma2019}, as shown in Fig.\,\ref{fgr_2}(b). The filament radius in the EC regime evolves exponentially in time with dynamics give by
\begin{equation}
\label{eq2}
\frac{R(t)}{R_o}=\left ( \frac{G_ER_0}{2\sigma} \right )^{1/3} \exp \left ( -\frac{t-t^*}{3\lambda_E} \right ),
\end{equation}
\noindent where $G_E$ is an apparent or effective elastic modulus, $\sigma$ is the surface tension, $\lambda_E$ is the extensional relaxation time, and $t^*$ is the critical time at which the IC balance transitions to an EC balance and the EC regime begins as shown in Fig.\,\ref{fgr_2}(b). The blue dash-dotted line in the Fig.\,\ref{fgr_2}(b) is an exponential fit of the form shown in Eq.\,\ref{eq2} for the EC regime. The resulting effective extensional relaxation time determined from four repeated experiments for this PEO solution is $\lambda_E = 2.8\pm0.1$ ms.

\subsection{DoS limitations}\label{sec_2.4}

An important experimental constraint in these experiments is the shortest extensional relaxation time, $\lambda_{E,\textrm{min}}$, that can be measured in an extensional rheometer that relies on the establishment of an EC balance, \textit{e.g.}, in a CaBER or DoS device. The filament thinning of a prototypical low viscosity ($\eta<20$ mPa $\cdot$s) and low elasticity ($\lambda_E \sim \textrm{O}(1)$ ms) liquid, such as the PEO solution shown in Fig.\,\ref{fgr_2}(b), starts with an IC regime at short times that is governed by the Rayleigh time scale of the pendant drop and subsequently transitions to an EC regime at $t^*$. The filament radius at time $t^*$ can be written $R^*=R(t^*)=K R_{\textrm{min}}$, where $K$ is a constant greater than unity and $R_{\textrm{min}}$ is the minimum radius that can be confidently measured based on the spatial resolution of the imaging system. Considering Eq.\,\ref{eq1} and the PEO data shown in Fig.\,\ref{fgr_2}(b), if the EC regime is not established, the IC regime would end with a pinch-off at a fictitious time $t_f$ such that $(R^*/R_0) = Y((t_f-t^*)/t_R)^{2/3}$. Thus, $(t_f-t^*) = t_R(K R_{\textrm{min}}/R_0 Y)^{3/2}$. At $t^*$, a significant decrease in the instantaneous extensional strain rate $\dot{\varepsilon}(t)=(-2/R(t))(dR(t)/dt)$ is observed as the rapid transition from an IC regime with  $\dot{\varepsilon}_{\textrm{IC}} =4/3(t_f-t^*)$ to an EC regime with  $\dot{\varepsilon}_{\textrm{EC}}=2/3\lambda_E$ occurs~\cite{McKinley2024}. Considering that $ \dot{\varepsilon}_{\textrm{IC}} \geq \dot{\varepsilon}_{\textrm{EC}}$, a constraint for the minimum measurable extensional relaxation time thus satisfies, 
\begin{equation}
\label{eq3}
    \lambda_E \geq \frac{t_R}{2}\left(\frac{K R_{\textrm{min}}}{R_0 Y}\right)^{3/2}.
\end{equation}
Equation\,\ref{eq3} is a \textit{necessary} condition for an exponential capillary thinning response to be observable but not yet a \textit{sufficient} condition for the minimum measurable extensional relaxation time, as it assumes infinite temporal resolution of the imaging system, which is not a realistic assumption. Successful and accurate measurement of an extensional relaxation time from filament thinning in the EC regime also requires measuring a suitable number of data points (denoted $N_f$) that lie within the regime of exponential filament thinning. Returning to Eq.\,\ref{eq2}, which governs filament thinning in the EC regime, we may reasonably require that for a robust measurement of the EC balance we require a number of frames $N_f$ spanning over at least $1/3$ of a decade in $R/R_0$ and thus, because of the factor of 3 in the exponent of Eq.\,\ref{eq2}, spanning at least one relaxation time, \textit{i.e.}, $\Delta t=(t_2-t_1)\simeq \lambda_E$. Using this constraint, we write $t_1 = p/f$ and $t_2 = (p+ N_f -1)/f$ as the start and end time for data acquisition in the EC regime, respectively. Here, $p$ is the frame number associated with time $t_1$ and $f$ is the imaging frequency in fps. We combine these values for $t_2$ and $t_1$ with the expression $R(t_2)/R(t_1) = \exp(-(t_2-t_1)/3\lambda_E)$, and substitute $R(t_1)$ and $R(t_2)$ with $R^* = K R_{\textrm{min}}$ and $R_{\textrm{min}}$, respectively. Rearranging, we obtain the constraint
\begin{equation}
\label{eq4}
    \lambda_E \geq \frac{N_f - 1}{3f\ln K}.
\end{equation}
Thus, an extensional relaxation time that satisfies both Eq.\,\ref{eq3} and Eq.\,\ref{eq4} is measurable with an extensional rheometer that operates based on an EC balance in a filament thinning device.

In our DoS setup, $R_{\textrm{min}}=6~\mu$m and assuming an aqueous solution we obtain a Rayleigh time $t_R=10.56~\textrm{ms}$. Assuming acquisition of 5 data points, $N_f=5$, over $1/3$ of a decade in $R/R_0$, $R^* = K R_{\textrm{min}}\simeq 30~\mu\textrm{m}$, and $K\simeq5$. Consistent with the previously reported values~\cite{McKinley2005,Sharma2019,Bonn2018}, the numerical prefactor $Y \simeq 0.7$ in our setup. Based on Eq.\,\ref{eq3}, a viscoelastic polymer solution with $\lambda_E \geq 0.017~\textrm{ms}$ thus has a long enough EC regime for extensional relaxation time measurement. However, as noted above this assumes an infinite temporal resolution. Given that in our DoS setup the camera has a frame speed $f = 5000$ fps and considering Eq.\,\ref{eq4}, the measurable extensional relaxation times using our setup are constrained to $\lambda_E \geq 0.17~\textrm{ms}$. Measuring extensional relaxation times in the range $0.017~\textrm{ms}\leq \lambda_E < 0.17~\textrm{ms}$ would require increasing the frame rate at which filament thinning is recorded.

\section{\label{sec_3}Results and Discussion}
\subsection{Shear rheology of xanthan gum solution}\label{sec_3.1}
\begin{figure*}[h]
\centering
  \includegraphics[width=0.99\textwidth]{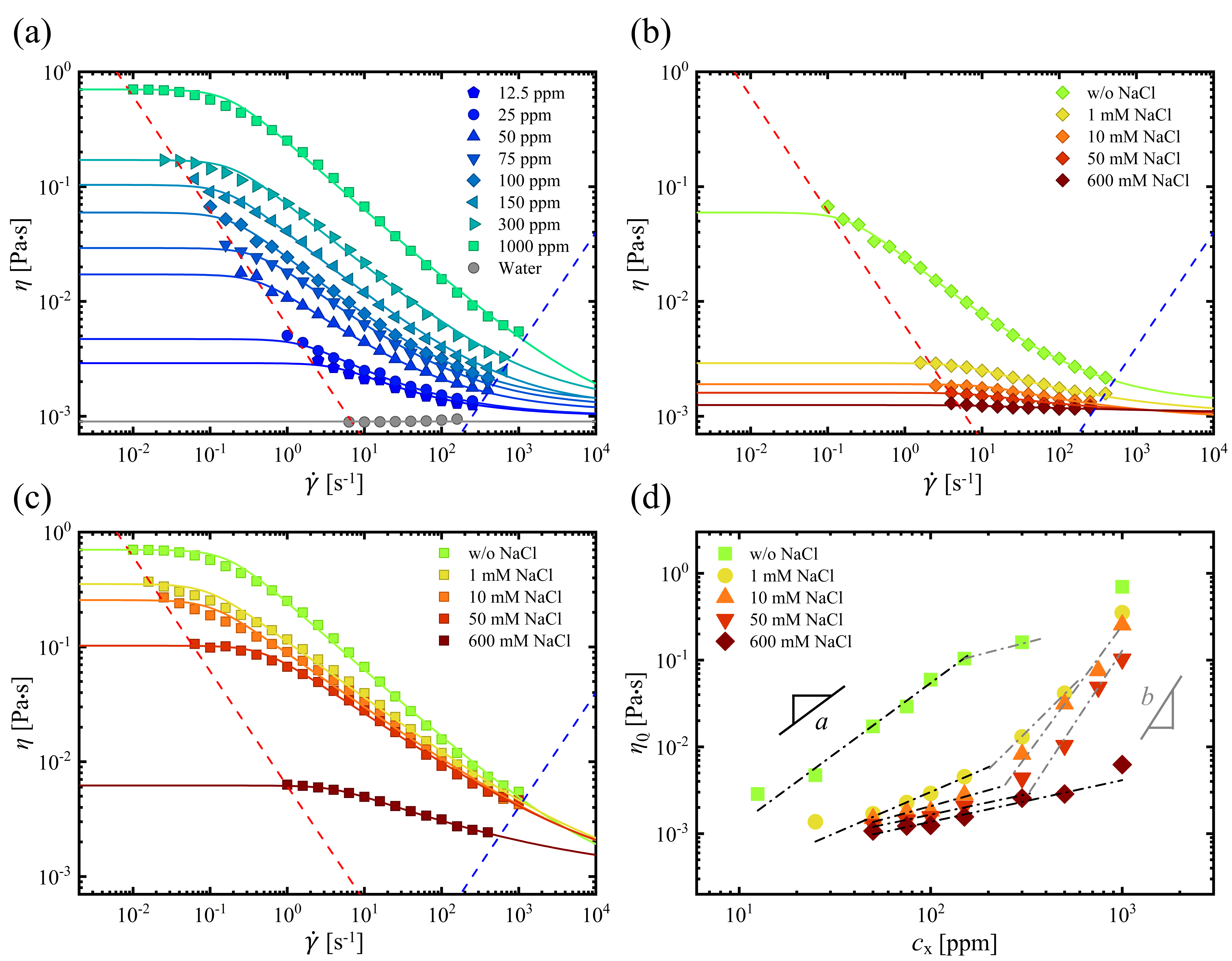}
  \caption{(a) Shear viscosity of salt-free xanthan gum solutions over a range of different xanthan gum concentrations. Shear viscosity of (b) 100 ppm xanthan gum solution and (c) 1000 ppm xanthan gum solution at different NaCl concentrations. Solid lines are fits based on the Carreau model and the red and blue dashed lines represent the minimum torque and onset of secondary flow limits, respectively. (d) Evolution in the zero shear viscosity of xanthan gum solutions with $c_{\textrm{x}}$ at different NaCl concentration. Slopes $a$ and $b$ mark the exponent of the power-law fits to the data below and above the critical overlap concentration $c^\ast_S$, respectively. The values of $a$ and $b$ are summarized in Table\,\ref{tbl_2}.}
  \label{fgr_3}
\end{figure*}

\begin{table*}
\small
  \caption{Rheological parameters of salt-free xanthan gum solutions in steady shear and transient extensional flows}
  \label{tbl_1}
  \renewcommand{\arraystretch}{1.3}
  \begin{tabular*}{\textwidth}{@{\extracolsep{\fill}}cccc|cccccc}
    \hline
    \hline
     & \multicolumn{3}{c|}{Shear flow}& \multicolumn{5}{c}{Extensional flow}\\
    \hline
    $c_{\textrm{x}}$~[ppm] & $\eta_0$~[Pa$\cdot$s] & $\lambda_S$~[s] & $n$ & $Y$ & $x$ & $\lambda_E$~[ms] & $t_f$~[ms] & Oh\\
    \hline
    ~~~~~~0.0 & 0.001 & - & - & 0.727 ± 0.008 & 0.653 ± 0.020 & - &  7.93 ± 0.54 & 0.002 \\
    ~~~~12.5 & 0.003 & 0.218 & 0.512 & 0.704 ± 0.008 & 0.689 ± 0.003 & - &  9.13 ± 0.22 & 0.008\\
    ~~~~25.0 & 0.005 & 0.613 & 0.522 & 0.643 ± 0.002 & 0.751 ± 0.026 & - &  10.2 ± 0.24 & 0.013\\
    ~~~~50.0 & 0.017 & 2.322 & 0.493 & 0.520 ± 0.009 & 0.952 ± 0.026 & - &  13.1 ± 0.38 & 0.045\\
    ~~~~75.0 & 0.029 & 2.404 & 0.467 & - & - & 1.62 ± 0.11 &  16.5 ± 0.41 & 0.076\\
    ~~~~~100 & 0.060 & 5.437 & 0.449 & - & - & 2.62 ± 0.23 &  18.9 ± 0.39 & 0.158\\
    ~~~~~150 & 0.104 & 5.894 & 0.448 & - & - & 4.32 ± 0.24 &  24.0 ± 1.20 & 0.274\\
    ~~~~~300 & 0.169 & 5.055 & 0.447 & - & - & 9.40 ± 0.58 &  36.3 ± 1.86 & 0.446\\
    ~~~1000 & 0.701 & 5.901 & 0.405 & - & - & 23.7 ± 1.82 &  91.3 ± 1.19 & 1.849\\
     \hline
     \hline
  \end{tabular*}
\end{table*}

The measured shear rheology $\eta(\dot{\gamma})$ for a wide range of xanthan gum solutions is summarized in Fig.\,\ref{fgr_3}. The molecular size of the xanthan gum macromolecules in ultrapure DI water is extended due to the electrostatic repulsion coming from the charged side branch of each repeat unit. This semi-rigid and extended structure is responsible for the high zero shear viscosity $\eta_0$ and strong shear-thinning observed in the xanthan gum solutions, even at low concentrations. By contrast, dilute solutions of flexible polymers such as PEO, have a small zero shear viscosity and only exhibit weak shear-thinning, as shown in Fig.\,\hyperlink{supp:fig_2}{S2}(a). Upon the application of shear in a xanthan gum solution, the semi-flexible polymer chains align in the direction of the flow resulting in a decrease in flow resistance, which leads to a reduction in viscosity. Increasing the xanthan gum concentration results in an even higher zero shear viscosity and stronger shear-thinning that extends at higher concentrations, as shown in Fig.\,\ref{fgr_3}(a).

The addition of salt screens intramolecular electrostatic repulsion, leading to a reduction in the persistence length and the equilibrium size of the polymer chains reduces. This more compact coil size increases the overlap concentration $c^\ast$. As a result, the shear viscosity of the xanthan gum solutions decreases significantly upon increasing the NaCl concentration, and the extent of shear-thinning weakens accordingly, as shown in Fig.\,\ref{fgr_3}(b, c) and Fig.\,\hyperlink{supp:fig_2}{S2}(b)-(e). The ionic strength required to achieve this reduction in viscosity depends on the number of polyelectrolyte chains per unit volume in the solution (see Tables\,\hyperlink{supp:tbl_1}{S\,\MakeUppercase{\romannumeral1}}-\hyperlink{supp:tbl_3}{S\,\MakeUppercase{\romannumeral3}} in the \hyperref[ESI]{Supplemental Material}). At low xanthan gum concentration, $c_{\textrm{x}}=100$ ppm, a notable decrease in zero shear viscosity, \textit{i.e.}, more than an order of magnitude, is observed even at low NaCl concentration, $c_{\textrm{NaCl}}=1$ mM (see Fig.\,\ref{fgr_3}(b)). However, at high xanthan gum concentration, $c_{\textrm{x}}=1000$ ppm, a corresponding decrease in zero shear viscosity is only observed at higher NaCl concentration, above $c_{\textrm{NaCl}}=50$ mM.

The shear-thinning behavior of these xanthan gum solutions can be characterized empirically using the Carreau model~\cite{Macosko1994}, 
\begin{equation}
\label{eq5}
\eta=\eta_\infty+\left(\eta_0{-\eta}_\infty\right)[1+(\lambda_S\Dot{\gamma})^2]^{(n-1)/2},
\end{equation}
where $\eta_\infty$ is the viscosity of the solution in the limit of infinite shear rate, $\eta_0$ is the zero shear viscosity, $\lambda_s$ is the characteristic shear relaxation time, and $n$ is the power index. The rheological properties of these xanthan gum solutions are summarized in Tables\,\ref{tbl_1}, \hyperlink{supp:tbl_4}{S\,\MakeUppercase{\romannumeral4}}, and \hyperlink{supp:tbl_5}{S\,\MakeUppercase{\romannumeral5}}, respectively. The viscosity of all polymer solutions measured at high shear rate converges towards the viscosity of the solvent, water. The zero shear viscosity increases rapidly and the strength of the shear-thinning increases as the xanthan gum concentration increases and as NaCl concentration decreases.

In Fig.\,\ref{fgr_3}(d), we summarize the evolution in the zero shear viscosity of xanthan gum solutions with different NaCl concentrations as a function of xanthan gum concentration. Unlike flexible polymers~\cite{Rubinstein1995, norisuye1993semiflexible}, there is currently no unified theoretical approach for evaluating the overlap concentration of semi-flexible polyelectrolytes. Specifically, our observation of zero shear viscosity increasing faster than linearly with concentration 
indicates that there must be some intermolecular interactions even at concentrations as low as 10 - 100 ppm. This is a reflection of highly expanded state of charged semi-flexible polyelectrolytes. Therefore, the overlap concentration is commonly estimated experimentally by measuring the zero shear viscosity for a wide range of polymer concentrations and identifying the concentration at which the power-law scaling of the zero shear viscosity with concentration changes~\cite{Sharma2020,Liberatore2009}. We define this concentration, commonly reported as the overlap concentration~\cite{Liberatore2009},  as a critical concentration determined from shear rheology, $c^\ast_{S}$. For $c_{\textrm{x}} < c^\ast_{S}$, $\eta_0 \sim c_{\textrm{x}}^a$, while for $c_{\textrm{x}} > c^\ast_{S}$, $\eta_0 \sim c_{\textrm{x}}^b$. This critical concentration is a strong function of ionic strength as shown in Fig.\,\ref{fgr_3}(d). The values of power-law scaling, $a$ and $b$, and the critical concentration obtained from steady shear flow experiments for different NaCl concentrations, $c^\ast_{S}$, are reported in Table\,\ref{tbl_2}. For a salt-free xanthan gum solution at low concentrations, $a=1.63 \pm 0.10$, while this power-law scaling changes to $b=0.57 \pm 0.03$ as xanthan gum concentration increases beyond $c^\ast_{S}$, as shown in Fig.\,\hyperlink{supp:fig_2}{S2}(f) and in close agreement to the work by Liberatore~\cite{Liberatore2009}. As salt concentration increases, the polymer chains contract due to charge screening, leading to an increase in $c^\ast_{S}$ and decrease in the exponent $a$. As the polymer concentration increases beyond the critical concentration, $c_{\textrm{x}} > c^\ast_{S}$, intermolecular interactions increase~\cite{Rubinstein1995} such that even at high salt concentrations a rapid increase in viscosity is observed leading to a new higher value for $b$, that increases as salt concentration increases.

\begin{table}
\small
  \caption{Scaling parameters for the evolution in the overlap concentrations of xanthan gum solution at different NaCl concentrations.}
  \label{tbl_2}
  \renewcommand{\arraystretch}{1.3}
  \begin{tabular*}{0.48\textwidth}{@{\extracolsep{\fill}}ccccc}
    \hline
    \hline
    \\[-1em]
    $c_{\textrm{NaCl}}$~[mM] & $a$ & $b$ & $c^\ast_{S}$~[ppm] & $c^\ast_{E}$~[ppm] \\
    \\[-1em]
    \hline
    \\[-1em]
    ~~~~0 & 1.63 ± 0.10 & 0.57 ± 0.03 & 172 ± 10.5 & ~~60 ±  10.7\\
    ~~~~1 & 0.95 ± 0.07 & 2.19 ±  0.02 & 203 ± 45.1 & ~~87 ± 17.3\\
    ~~10 & 0.61 ± 0.15 & 3.00 ±  0.06 & 230 ± 87.0 & 232 ± 20.9 \\
    ~~50 & 0.47 ± 0.08 & 3.37 ±  0.04 & 351 ± 56.5 & 615 ± 44.0\\
    600 & 0.48 ± 0.05 & - & - & -\\
    \hline
    \hline
  \end{tabular*}
\end{table}

\subsection{Extensional rheology of xanthan gum solutions}
\begin{figure*}[htb!]
\centering
  \includegraphics[width=0.99\textwidth]{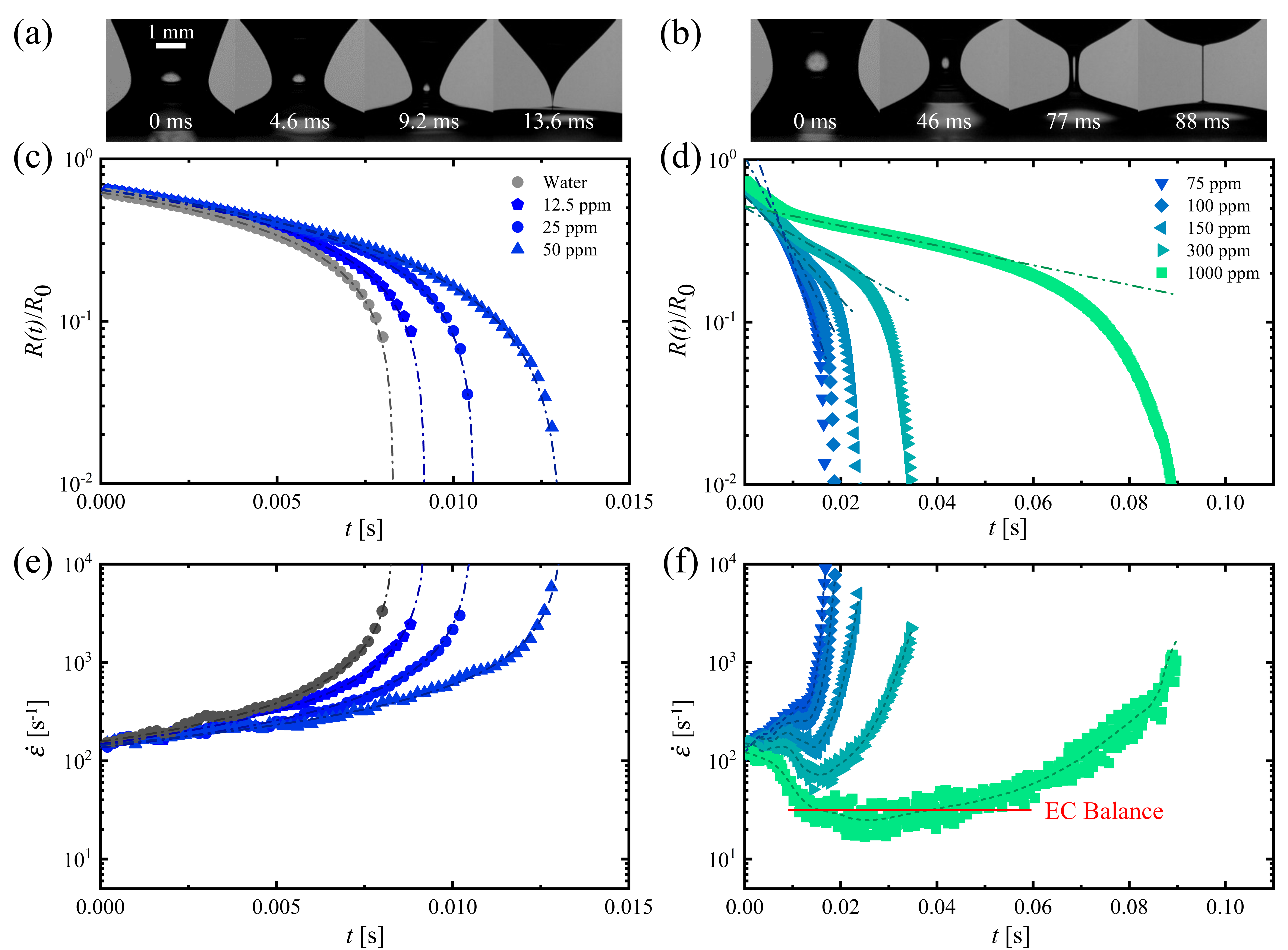}
  \caption{Concentration dependent capillary thinning dynamics of salt-free xanthan gum solutions. (a, b) Image sequence of filament thinning: (a) $c_{\textrm{x}}=50$ ppm and (b) $c_{\textrm{x}}=1000$ ppm. (c) Temporal evolution of filament radius at the neck for low concentration xanthan gum solutions that are well decribed by IP filament thinning profiles. The dash-dotted lines are power-law fits of the form shown in Eq.\,\ref{eq1}. (d) Temporal evolution of the filament radius at its neck for higher concentration xanthan gum solutions with an initial IP regime followed by a subsequent EC regime. The dash-dotted lines are exponential fits of the form shown in Eq.\,\ref{eq2}. The values of the corresponding rheological parameters are summarized in Table~\ref{tbl_1}. (e) Temporal evolution of the midpoint strain rate for xanthan gum solutions characterized by IP filament thinning profiles. The dash-dotted lines are fits of the form $\dot{\varepsilon} = 2x/(t_R-t)$ where $x$ is the power-law exponent given by Eq.\,\ref{eq1}. (f) Temporal evolution of strain rate for at higher concentration xanthan gum solutions showing an initial IP thinning followed by a subsequent EC balance with exponential like filament thinning profile. The dashed lines represent data smoothed with the the locally weighted scatter plot smoothing (LOWESS) method~\cite{fox2018r}.}
  \label{fgr_4}
\end{figure*}
Solutions of high molecular weight flexible polymers such as PEO show significant strain-stiffening in extensional flow, even in the dilute regime~\cite{Pastor2008,Shell2020}. At equilibrium, a flexible polymer chain adopts a random coil configuration due to its small monomer size and short persistence length. In a transient extensional flow such as that generated in a filament thinning experiment, the flexible polymer chains undergo a coil-stretch transition~\cite{de1974coil}, resulting in significant molecular unraveling. This then leads to establishment of a balance between elastic stresses and capillary pressure in the thinning filament~\cite{mckinley2005visco,entov1997effect}. This results in a long-lived EC thinning regime of a thin, approximately cylindrical, fluid ligament that follows an exponential profile of the form given in Eq.~\ref{eq2}, and shown in Fig.\,\ref{fgr_2}(c-\lowercase\expandafter{\romannumeral2}) and \hyperlink{supp:fig_3}{S3}(a). This long-lived EC balance is due to the large, but finite, extensibility of the high molecular weight flexible polymer chains. 

Xanthan gum molecules have a much smaller finite extensibility compared to PEO due to the high monomer mass of the xanthan gum repeat unit and the large persistence length of the chain. This results in markedly different nonlinear viscoelastic properties as compared to PEO~\cite{walters1990influence}. Using our DoS experiment, we find that the transient extensional rheology of dilute solutions of xanthan gum can be compactly and effectively characterized with a rate-dependent and inelastic generalized Newtonian fluid model~\cite{Sharma2017,McKinley2003}. As the concentration of xanthan gum increases, the shear viscosity increases and the pinch-off time of the liquid filament is systematically delayed until eventually a concentration-dependent, elastic response is observed.

Figure\,\ref{fgr_4} (a) and (b) compare the pinch-off dynamics for salt-free xanthan gum solutions with $c_{\textrm{x}}=50$ ppm and $c_{\textrm{x}}=1000$ ppm, respectively. In Fig.\,\ref{fgr_4}(c), we show temporal evolution of the filament radius at the neck location for DI water as well as for salt-free dilute solutions of xanthan gum with $c_{\textrm{x}}\leq 50$ ppm. This evolution follows a power-law fit of the form shown in Eq.\,\ref{eq1}. Consistent with the theory of low viscosity Newtonian fluids~\cite{Calabrese2022, Sharma2021}, the pinch-off dynamics of DI water follows an IC regime shown in Eq.\,\ref{eq1} with a power-law exponent $x=0.653 \pm 0.02$ (consistent with the theoretical expectation that $x=2/3$~\cite{day1998self}), as shown with dash-dotted gray line in Fig.\,\ref{fgr_4}(c). Addition of xanthan gum to DI water and increasing the concentration to $c_{\textrm{x}}= 50$ ppm, we find decreases the coefficient $Y$ and increases the exponent $x$ in the power-law fits. At $c_{\textrm{x}}= 50$ ppm, the power-law exponent $x$ approaches a value of unity as summarized in Table\,\ref{tbl_1}. 
The conical neck shape of the fluid filament during pinch-off at $c_{\textrm{x}}= 50$ ppm, shown in Fig.\,\ref{fgr_4}(a), and the low value of Ohnesorge number $\textrm{Oh} \ll 1$, suggest that the pinch-off dynamics of dilute xanthan gum solutions are all inelastic but also still experience significant inertia effect. we refer to this as an IP regime. The values of $Y$, $x$, and Oh for different concentrations of xanthan gum in salt-free solutions are summarized in Table.\,\ref{tbl_1}. 

At a critical concentration in the range $50~\textrm{ppm} < c_{\textrm{x}} < 75~\textrm{ppm}$, the purely IP response of the low viscosity fluids transitions to a more complex dynamics showing an initial IP response at short time followed by a subsequent EC response. In the EC regime observed at late time, the liquid filament exhibits a cylindrical neck shape as shown in Fig.\,\ref{fgr_4}(b). In Fig.\,\ref{fgr_4}(d), we show the temporal evolution of the filament radius at this neck location for salt-free xanthan gum solutions with $c_{\textrm{x}} \geq 75~\textrm{ppm}$. The latter stages of the filament thinning within the EC regime follow an exponential profile characterized by Eq.\,\ref{eq2}, as shown by the dash-dotted lines in Fig.\,\ref{fgr_4}(d). Near the critical concentration $c^\ast_{E}$ at $c_{\textrm{x}}=75$ ppm, only a very short-lived EC response is observed and accurate determination of an extensional relaxation time is limited by the experimental constraints discussed in section~\ref{sec_2.4}. As the xanthan gum concentration increases, the solutions become more viscous and also more elastic resulting in a long-lived EC response, which increases the extensional relaxation time, $\lambda_E$ and the pinch-off time, $t_f$. The values of $\lambda_E$ and $t_f$ for different concentrations of xanthan gum in salt-free solutions are summarized in Table\,\ref{tbl_1}.

Figures\,\ref{fgr_4}(e) and (f) show the temporal evolution of the local strain rate in the neck $\dot{\varepsilon}=(-2/R(t))(dR(t)/dt)$ for low concentration xanthan gum solutions in the IP regime and also for the higher concentration solutions that exhibit an EC regime, respectively. In good agreement with the experimental measurements, the dash-dotted lines in Fig.\,\ref{fgr_4}(e) are predictions of the functional form $\dot{\varepsilon}=2x/(t_R-t)$, obtained using Eq.\,\ref{eq1} and the values of $x$ given in Table\,\ref{tbl_1}. In the IP regime, the strain rate increases monotonically with time, indicating the acceleration in the local rate of filament thinning in the neck until pinch-off occurs at a critical time. In contrast, the temporal evolution of strain rate in the more concentrated solutions with an EC regime is non-monotonic, as shown in Fig.\,\ref{fgr_4}(f). The dashed lines represent data smoothed with the the locally weighted scatter plot smoothing (LOWESS) method~\cite{fox2018r}. At early times, \textit{i.e.}, before the EC balance is established, the strain rate increases for the low xanthan gum concentrations. Upon increase in the concentration of xanthan gum and at early times, the strain rate decreases. This short time evolution is followed by a local region at intermediate times when the EC balance is established and the strain rate is almost constant, $\dot{\varepsilon}=2/3\lambda_E$ denoted by the red solid line, corresponding to a constant local Weissenberg number $\textrm{Wi} = \dot{\varepsilon}\lambda_E=2/3$, as shown in Fig.\,\hyperlink{supp:fig_4}{S4}(a). This constant strain rate corresponds to the time period during which the thinning filament establishes an EC balance and the radial profile decays exponentially. At later times, the finite extensibility of the xanthan gum molecules results in a progressive deviation from exponential thinning and a subsequent increase in the strain rate until filament breakup occurs. Complementary to Fig.\,\ref{fgr_4} and  Fig.\,\hyperlink{supp:fig_4}{S4}(a), Fig.\,\hyperlink{supp:fig_4}{S4}(b) shows the evolution in the apparent extensional viscosity $\eta^+_{\textit{E}} = \sigma/(\dot{\varepsilon}(t)R(t))$, representing the ratio of the elastic tensile stress difference (assuming it is balanced by the capillary pressure) and the instantaneous strain rate, as a function of the time-evolving Hencky strain $\varepsilon = 2 \ln{(R_0/R(t))}$. The extensional viscosity of the xanthan gum solutions increases with the xanthan gum concentration, and in the EC regime, exhibits an exponential strain-stiffening response due to the constant strain rate.

\begin{figure}[h]
\centering
  \includegraphics[width=0.49\textwidth]{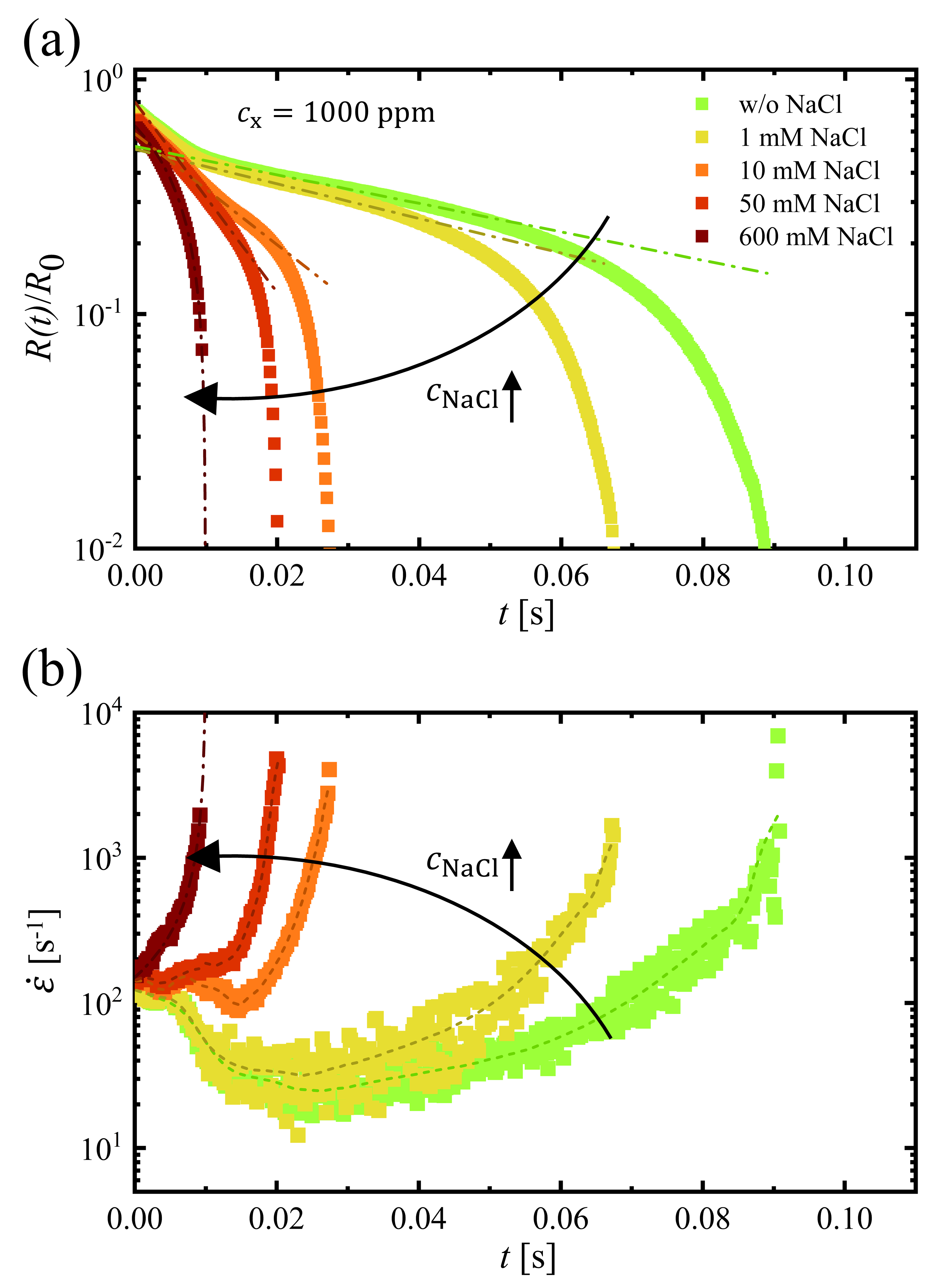}
  \caption{Impact of ionic strength on capillary thinning dynamics. Temporal evolution of (a) filament radius, and (b) strain rate for xanthan gum solutions with $c_{\textrm{x}}=1000$ ppm at different NaCl concentrations. The dash-dotted lines are exponential fits for $c_{\textrm{NaCl}}$ = 0 to 50 mM, and power-law fits for $c_{\textrm{NaCl}}=600$ mM, based on the parameters reported in Table\,\ref{tbl_s4}. The dashed lines shown in (b) denote data smoothed with the LOWESS method~\cite{fox2018r}.}
  \label{fgr_5}
\end{figure}

The addition of salt to the xanthan gum solutions dramatically decreases the extensional relaxation time. For xanthan gum solutions with $c_{\textrm{x}} = 1000$ ppm, this decrease in extensional relaxation time results in a transition from a long-lived EC regime in the absence of salt to an IP regime at the highest salt concentration, as shown in Fig.\,\ref{fgr_5} (a) and (b). At 600 mM of salt all evidence of viscoelasticity is lost and  the filament radius decreases as a power-law in time (Fig.\,\ref{fgr_5}(a)) and the strain rate increases correspondingly. This is a signature of an inelastic fluid response in which the extensional viscosity only depends weakly on the strain rate~\cite{mckinley2005visco}. The extensional relaxation time of xanthan gum solution at $c_{\textrm{x}}=1000$ ppm decreases from 23.7 ms in a salt-free solution to 3.60 ms in a solution with $c_{\textrm{NaCl}}=50$ mM. 

As the ionic strength increases, the pinch-off dynamics become increasingly rapid and the transient extensional viscosity decreases, as shown in Fig.\,\hyperlink{supp:fig_3}{S3}(b)-(f) and Fig.\,\hyperlink{supp:fig_5}{S5}(a). In Fig.\,\hyperlink{supp:fig_5}{S5}(b)-(d), we also show a similar decrease in the elasticity of xanthan gum solutions with $c_{\textrm{x}}=100~\textrm{ppm}$ as the ionic strength of the solvent increases. The values of the corresponding rheological parameters are summarized in Tables\,\hyperlink{supp:tbl_4}{S\,\MakeUppercase{\romannumeral4}} and \hyperlink{supp:tbl_5}{S\,\MakeUppercase{\romannumeral5}}. The addition of NaCl to xanthan gum solutions reduces the persistence length and increases the flexibility of the polymer chains, while decreasing coil-coil interactions. However, despite the increase in the flexibility of the chains, the extensional relaxation time and hence the nonlinear viscoelasticity of the xanthan gum solutions decreases. 

\begin{figure}
\centering
  \includegraphics[width=0.49\textwidth]{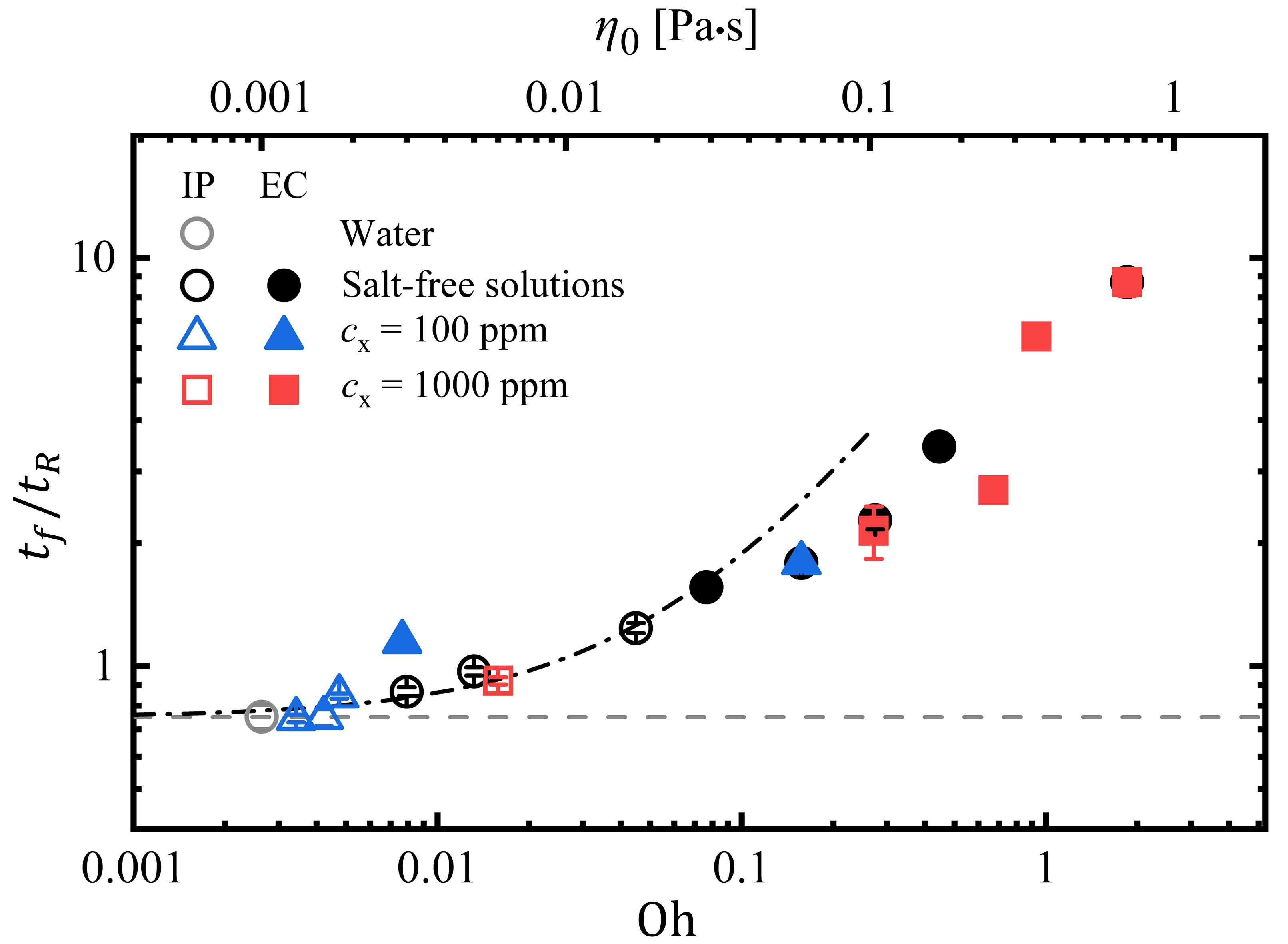}
  \caption{Evolution of the pinch-off time, normalized with the Rayleigh time scale, with Ohnesorge number (bottom axis) or with zero shear viscosity (top axis) for DI water (gray circle), salt-free xanthan gum solutions at different xanthan gum concentrations (black circles), xanthan gum solutions with $c_{\textrm{x}}=100$ ppm at different salt concentrations (blue triangles), and xanthan gum solutions with $c_{\textrm{x}}=1000$ ppm at different salt concentrations (red squares). The gray dashed line represents the minimum value of the pinch-off time of DI water. The black dash-dotted line is a linear fit to the data obtained for xanthan gum solutions showing IP dynamics (hollow symbols) given by $t_f/t_R=0.75+11.4\textrm{ Oh}$, where $t_R$ and $\textrm{Oh}$ are the the Rayleigh time scale and the Ohnesorge number, respectively. Viscoelastic solutions characterized by an EC response (filled symbols) deviate progressively from  this linear fit.}
  \label{fgr_7}
\end{figure}

It can be noted from Fig.\,\ref{fgr_3} that the shear viscosity also decreases significantly as salt concentration increases. For small values of the zero shear viscosity (and hence low Ohnesorge numbers), \textit{i.e.}, lower xanthan gum and higher salt concentrations, the pinch-off dynamics is predominantly IP in character and we find that the dimensionless pinch-off time increases approximately linearly with the increase in zero shear viscosity and Ohnesorge number, as shown by the hollow symbols in Fig.\,\ref{fgr_7}. As the pinch-off time and zero shear viscosity increases, \textit{i.e.}, at higher xanthan gum and lower salt concentrations, the pinch-off dynamics cross over to an EC balance resulting in a different relationship with zero shear viscosity and Ohnesorge number, as shown by the filled symbols in Fig.\,\ref{fgr_7}.  Thus, the increase in the finite extensibility of the molecules resulting from addition of NaCl does not appear to be the dominant factor that determines the elastic behavior of this semi-flexible polyelectrolyte, instead it is the loss of intermolecular entanglements between the semi-flexible chains in semi-dilute solutions that dominates the observed rheology.

\subsection{\label{sec_3.3}Polymer chain finite extensibility}

The dimensionless finite extensibility parameter, $L$, of a polymer molecule is the ratio of the contour length, $r_{\textrm{max}}$, to the root-mean-square end-to-end distance of the chain at equilibrium,
where the contour length $r_{\textrm{max}}$ is the product of monomer length and number of monomers and the root-mean-square end-to-end distance of a semi-flexible polymer chain is connected to the persistence length and the contour length by the relationship,
\begin{equation}
\label{eq6}
\left<r_{0}^{2} \right>^{1/2}=\sqrt{2r_{\textrm{max}}\ell_p-2\ell_p^2(1-e^{-r_{\textrm{max}}/\ell_p})},
\end{equation}
based on the worm-like chain model~\cite{Larson1999, norisuye1993semiflexible}. The finite extensibility, the contour length, and the root-mean-square end-to-end distance for the two polymers studied in this work are summarized in Tables\,\ref{tbl_3} and \hyperlink{supp:tbl_6}{S\,\MakeUppercase{\romannumeral6}}.

\begin{table}
\small
  \caption{Conformation parameters and finite extensibility of flexible and semi-flexible polymers.}
  \label{tbl_3}
  \renewcommand{\arraystretch}{1.3}
  \begin{tabular*}{0.48\textwidth}{@{\extracolsep{\fill}}crrrrrr}
    \hline
    \hline
    \\[-1em]
~~Polymer &\begin{tabular}[c]{@{}c@{}}$c_{\textrm{NaCl}}$\\{[}mM]\end{tabular} & \begin{tabular}[c]{@{}c@{}}$M_w$\\{[}g/mol]\end{tabular} & \begin{tabular}[c]{@{}c@{}}$r_{\textrm{max}}$\\{[}nm]\end{tabular}& \begin{tabular}[c]{@{}c@{}}$\ell_p$\\{[}nm]\end{tabular} & \begin{tabular}[c]{@{}c@{}}$\left<r_{0}^{2} \right>^{1/2}$\\{[}nm]\end{tabular} & \multicolumn{1}{c}{$L$}\\
    \hline
    \\[-1em]
~~PEO         & ~~0 & $2\times 10^6$ & 13182 & 0.48 & 112 & 117 \\
    \hline
            & ~~0 &  &  & ~210 & 901 & 2.38\\ 
            & ~~1 &  &  & ~~~52 & 466 & 4.59 \\
~~Xanthan gum & 10& $2\times 10^6$ & ~~2141 & ~~~50 & 457 & 4.68\\
            & ~~50 &  &  & ~~~47.5 & 446 & 4.80 \\
            & ~~600 &  &  & ~~~41 & 415 & 5.16 \\
    \hline
    \hline
  \end{tabular*}
\end{table}

High molecular weight PEO chains have large contour lengths and small root mean square end-to-end lengths, resulting in a large finite extensibility. This large finite extensibility ultimately leads to a long-lived EC regime in transient extensional flows, such as those generated in DoS or CaBER instruments. Xanthan gum, on the other hand, has a large monomer repeat unit size resulting in a relatively small contour length and a much smaller finite extensibility than flexible polymer chains of  the same molecular weight. The addition of salt to a xanthan gum solution decreases the persistence length. This decrease in persistence length, however, increases the finite extensibility by only a factor of 2 at the highest NaCl concentration, and thus the  finite extensibility of the xanthan gum molecules remains significantly lower than the finite extensibility of the studied PEO solutions, as shown in Table\,\ref{tbl_3}. The small finite extensibility of these xanthan gum solutions, even at high ionic strengths, makes it impossible for the coil-stretch transition of a dilute suspension of non-interacting chains to be the main reason for establishing an EC balance in a transient extensional flow, as observed in Fig.\,\ref{fgr_5}(a) for example. We therefore hypothesize that an EC balance for semi-flexible polyelectrolytes can only be established when transient intermolecular interactions commence and a transient polymer network is formed. Similar arguments have also recently been made to explain the development of a yield stress in xanthan gum solutions of even higher concentrations~\cite{missi2024thermo}. The simplest transient network theories~\cite{green1946new, tanaka1992viscoelastic, tripathi2006rheology} result in constitutive equations that closely resemble the Oldroyd-B equation and will again lead to an EC balance and exponential thinning of a fluid filament. Thus, the critical concentration determined from extensional rheology at which the observed pinch-off dynamics transition from an IP regime to an EC regime (denoted $c_{E}^{\ast}$), can be considered the rheologically-effective overlap concentration.

\subsection{The overlap concentration for semi-flexible polymer solutions}
\begin{figure}[h]
\centering
  \includegraphics[width=0.49\textwidth]{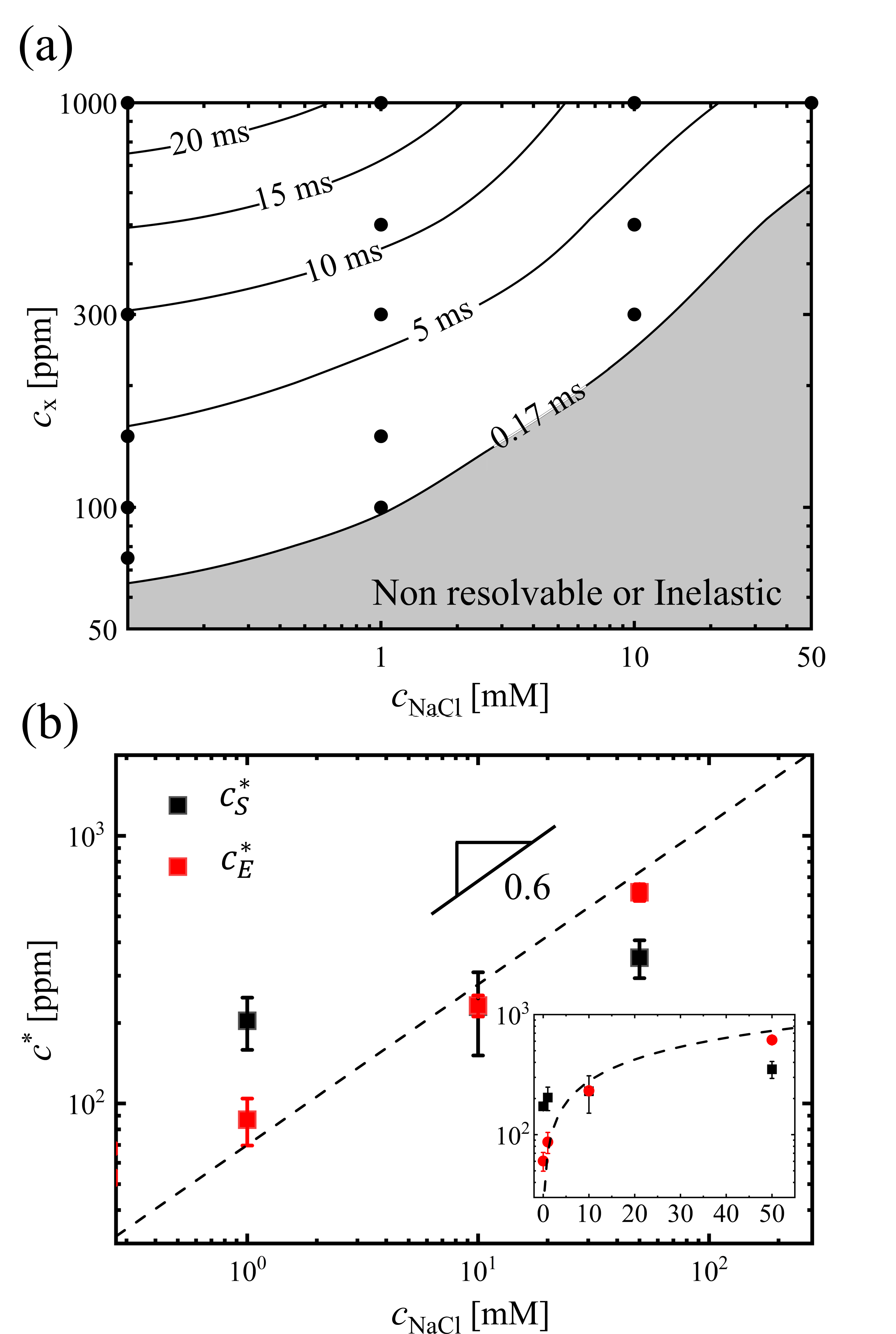}
  \caption{(a) Contour plot of the measured extensional relaxation time $\lambda_E$ for xanthan gum solutions at different xanthan gum and NaCl concentrations, based on DoS measurement data extracted from the EC regime (\mycirclesmall{black}). The solid lines represent data smoothed with the Savitzky-Golay filter~\cite{schafer2011savitzky}. (b) Critical overlap concentrations of xanthan gum solutions determined from both in shear flow ($c^\ast_S$) and extensional flow observation ($c^\ast_E$). The inset shows the same data in semi-log scale. The dashed line represents the theoretically predicted scaling of overlap concentration~\cite{Rubinstein1995}.}
  \label{fgr_6}
\end{figure}

\noindent In Fig.\,\ref{fgr_6}(a), we summarize our measurements of the extensional relaxation time for the xanthan gum solutions with different xanthan gum and NaCl concentrations in the form of a contour plot $\lambda_E(c_{\textrm{NaCl}},~ c_{\textrm{x}})$. The longest extensional relaxation time $\lambda_E=23.7$ ms is measured for the highest concentration $c_{\textrm{x}}=1000$ ppm solution in a salt-free solvent. Under such conditions, the xanthan gum coils are highly expanded and intermolecular interactions dominate. This leads to a high viscosity and extensive shear-thinning (Fig.\,\ref{fgr_3}(a)) as well as an EC response during filament thinning. At low xanthan gum concentrations and high ionic strengths, intermolecular interactions decrease, preventing the formation of a transient polymer network, and subsequently reducing the shear viscosity, the extent of shear-thinning, and the extensional relaxation time. Below a (salt-dependent) critical concentration $c_E^\ast$, filament thinning measurements reveal only IP dynamics and an extensional relaxation time cannot be measured, as represented by the grey area in Fig.\,\ref{fgr_6}(a). 

The critical concentration in shear flow, $c_S^\ast$ (\textit{cf.} Sec.\,\ref{sec_3.1}) and the critical concentration in extensional flow, $c_E^\ast$ (\textit{cf.} Sec.\,\ref{sec_3.3}) are compared in Table\,\ref{tbl_2}. The values of critical concentrations are calculated by fitting a polynomial to the data and determining the 95\% confidence intervals. While both $c_S^\ast$ and $c_E^\ast$ show monotonic increases as the salt concentration in the solvent increases, a quantitative difference between the two critical concentrations is observed. In Fig.\,\ref{fgr_6}(b), we compare the critical concentrations measured in shear and extensional flows with the scaling of overlap concentration predicted by the theory of flexible polyelectrolytes~\cite{Rubinstein1995}.
In the scaling model which is based on Debye-H$\ddot{\textrm{u}}$ckel theory~\cite{landau2013statistical}, the overlap concentration increases monotonically with the salt concentration and scales as $c^\ast \sim (c_\textrm{NaCl})^{3/5}$.
Even though this theory is developed for flexible polyeletrolytes, the increase in the flexibility of the xanthan gum chains upon the increase in salt concentration of the solvent and the decreasing values of $\ell_p / \left<r_{0}^{2} \right>^{1/2}$ shown in Table\,\ref{tbl_3} makes this theory a suitable first point of reference. The critical concentrations determined in extensional flows show good agreement with the scaling in overlap concentration predicted by the theory. DoS measurements are thus a simple and rapid rheological screening technique that can help identify the overlap concentration of semi-flexible polyelectrolytes. The inherent errors associated with extrapolation to determine zero shear viscosity and the perennial challenges associated with the low torque limits of modern torsional rheometers can result in uncertainty in determining the overlap concentration based on steady shear flow experiments. 
In contrast, the observation of pinch-off dynamics and the transition from an IP regime to an EC regime in DoS experiments is quite visually distinctive and robust. For example, in salt-free solutions, transition from an IP thinning response to an EC regime occurs at a concentration $ 50 \textrm{ ppm} \leq c_{\textrm{x}} \leq 75 \textrm{ ppm}$. The overlap concentration is thus assessed by interpolation to be $c_E^\ast = 60 \pm 10.7$ ppm.
By conducting additional experiments at intermediate concentration, we can refine this value for $c_E^\ast$, within the physical measurement limits of the DoS experimental setup $\lambda_E \geq 0.17$ ms discussed in section~\ref{sec_2.4}.

\section{Conclusions}

Xanthan gum is a semi-flexible  polyelectrolyte with a large persistence length compared to flexible polymers such as PEO. We illustrate the effect of solvent ionic strength on the nonlinear shear and extensional rheology of xanthan gum solutions through addition of NaCl and suggest a robust method for determining the rheologically-effective overlap concentration for solutions of semi-flexible polyelectrolytes.

In steady shear flows, increasing the xanthan gum concentration and decreasing the ionic strength of the solvent both act to increase the zero shear viscosity and strengthen shear-thinning. The zero shear viscosity $\eta_0(c_\textrm{X}, c_\textrm{NaCl})$ was evaluated for multiple concentrations of xanthan gum and over a range of ionic strengths. The critical polymer concentration at which the power-law governing the dependence of zero shear viscosity on xanthan gum concentration changes is commonly reported as the overlap concentration ($c_S^\ast$). Our measurements allow us to determine this concentration for different ionic strengths of the solvent. However, there exists an inherent error associated with estimating the zero shear viscosity of dilute and semi-dilute solutions of systems such as xanthan gum. This is due to the low torque limit of torsional rheometers and the slow approach to an asymptotic plateau value value of the viscosity for the highly expanded semi-flexible chains. This systematic error makes evaluating the rheologically-effective overlap concentration using this experimental method challenging and non-robust.

As an alternate approach, we determine the smallest extensional relaxation time that can be measured using DoS rheometry, by identifying through high-speed video imaging the transition to exponential filament thinning which indicates establishment of an EC balance. Increasing the xanthan gum concentration and decreasing the ionic strength of the solvent are both found to increase the extensional relaxation time of the solutions. At a fixed ionic strength, increasing the xanthan gum concentration beyond a critical concentration, denoted $c_E^\ast$, results in the development of a transient physical network and changes the filament thinning profile from an IP profile to a more complex response with an initial IP regime followed by a subsequent EC profile.  

At the microscale, increasing the ionic strength of the solvent screens the charges along the backbone and makes xanthan gum molecules more compact and flexible by decreasing their persistence length. Hence, the addition of salt increases the finite extensibility of xanthan gum solutions. This increase is, however, small and xanthan gum molecules in aqueous solution with salt concentrations of the level of sea water ($c_\textrm{NaCl} = 600$ mM) are still significantly more rigid and inextensible compared to flexible polymers such as PEO. We argue that the only way for solutions of weakly-extensible polymer such as xanthan gum to show an elastic behavior in extensional flows is to have a concentration above a critical overlap concentration such that the intermolecular interactions between neighboring xanthan gum molecules create a transient physical network that conveys elasticity to the semi-dilute polymer solution. Thus, the critical concentration at which a solution of a semi-flexible polyelectrolyte will begin to show an EC response in DoS experiments serves as a rheologically effective measure of the overlap concentration. This method for measuring the overlap concentration is experimentally simple to implement and is less vulnerable to experimental errors resulting from extrapolations inherent in determining the zero shear viscosity in steady shear experiments. Comparison of the overlap concentrations evaluated using DoS experiments with the only existing theory for the scaling of the overlap concentration for polyelectrolytes~\cite{Rubinstein1995} shows promising agreement. 

Looking further ahead, xanthan gum is cheap and biodegradable making it a suitable drag reducing alternative to conventional synthetic polymers made from hydrocarbon feedstocks~\cite{Soares2013,cussuol2023polymer}. Despite the importance of semi-flexible polyelectrolytes in several industries as well as recent developments in modeling them using Brownian dynamics simulations~\cite{pincus2020viscometric, pincus2023dilute}, there exists a need for closed-form constitutive equations to model dilute and semi-dilute semi-flexible polyelectrolytes at the continuum level. We envision using our experimental results obtained in canonical shear and extensional flows to develop and verify ensemble-averaged constitutive models based on finitely extensible bead-spring chains or transient network theories that as simple enough to be used, in conjunction with the Cauchy momentum equation (and conservation of mass), to simulate flows of semi-flexible polyelectrolyte solutions in complicated inhomogeneous flows with mixed kinematic character under both laminar and turbulent conditions.

\section*{Data availability}
The data that support the findings of this study are available within the article and the \hyperref[ESI]{Supplemental Material}.

\section*{Conflicts of interest}
There are no conflicts to declare.

\section*{Acknowledgements}
This research was supported by the Korean Institute for Advancement of Technology (KIAT) grant funded by the Korea Government (MOTIE) (P0017305, Human Resource Development Program for Industrial Innovation (Global)), by Basic Science Research Program through the National Research Foundation of Korea (NRF) grant funded by the Ministry of Science, ICT and Future Planning (NRF2020R1A2C3010568) and Korea Government(MIST)(No.2020R1A5A1019649), the National Science Foundation (NSF) Grant No. CBET-2027870 to MIT, and MathWorks Fellowship.

\newpage

\section*{\label{ESI}Supplemental Material}

\renewcommand{\thefigure}{S\arabic{figure}}
\renewcommand{\thetable}{S\,\Roman{table}}
\renewcommand{\theequation}{S\arabic{equation}}
\setcounter{figure}{0}
\setcounter{table}{0}

\hypertarget{supp:fig_1}{
\begin{figure}[htbp]
\centering
  \includegraphics[width=0.6\textwidth]{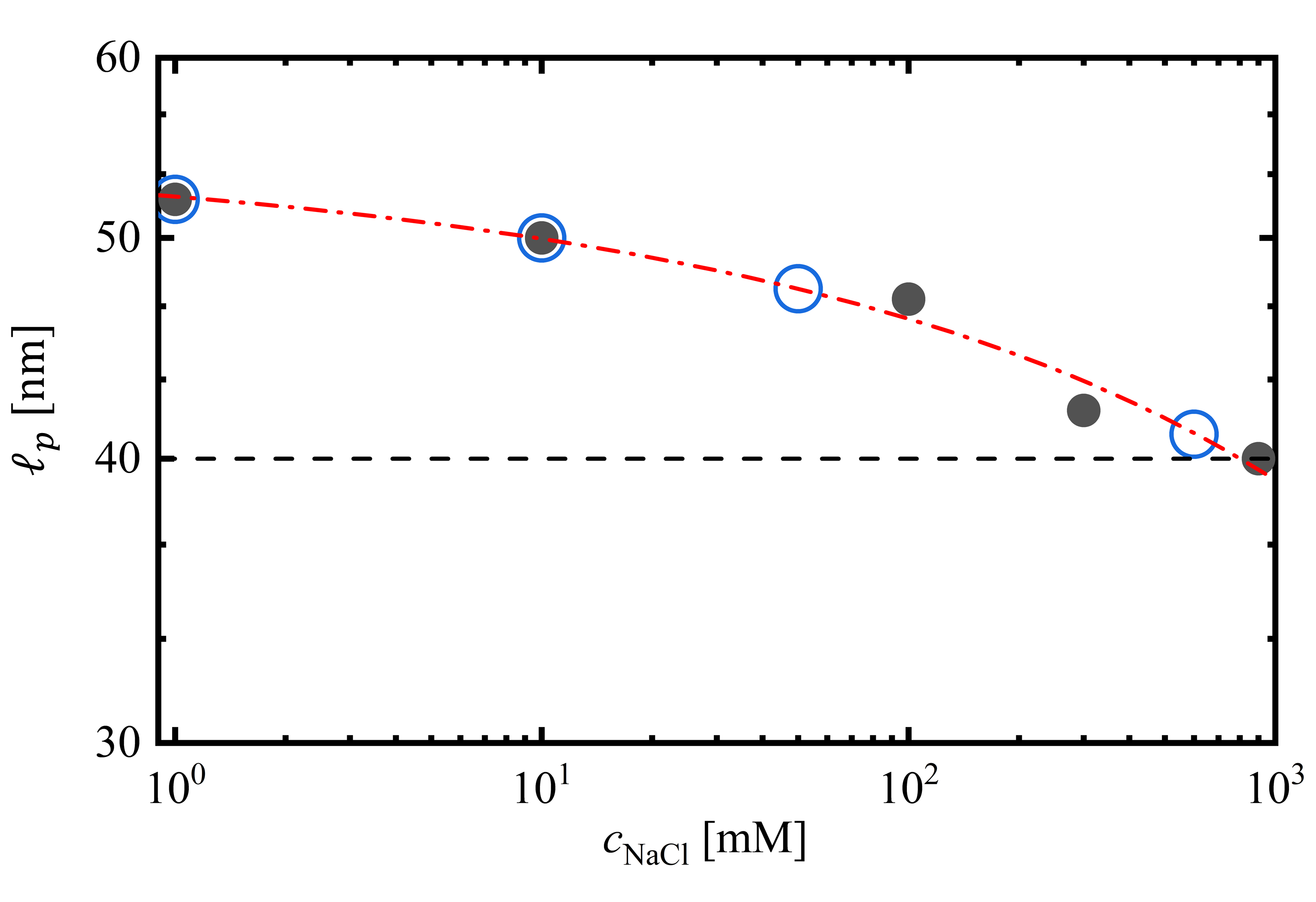}
\caption{Persistence length of xanthan gum chain at different salt concentrations. The persistence lengths for 0, 1, 10, 100, 300, and 900 mM NaCl concentrations are reported in~\cite{Chauveteau1986} (\,\mycircle{Muller}\,). The persistence lengths for 50 and 600 mM NaCl concentrations are interpolated based on the best power-law fit to the measurements reported in~\cite{Chauveteau1986} (\myhollows[2.5pt]{Thiswork}). The red dash-dotted line is the best power-law fit $y=A+Bx^C$, where $A = 55 \pm 3.8$, $B = -2.8 \pm 0.52$, and $C=0.25 \pm 0.03$. The black dotted line is the constant structural persistence length $\ell_{p,s} = 40$ nm of the xanthan gum chain in high ionic strengths.}
  \label{fgr_S1}
\end{figure}
}

\hypertarget{supp:fig_2}{
\begin{figure}[htbp]
\centering
  \includegraphics[width=0.99\textwidth]{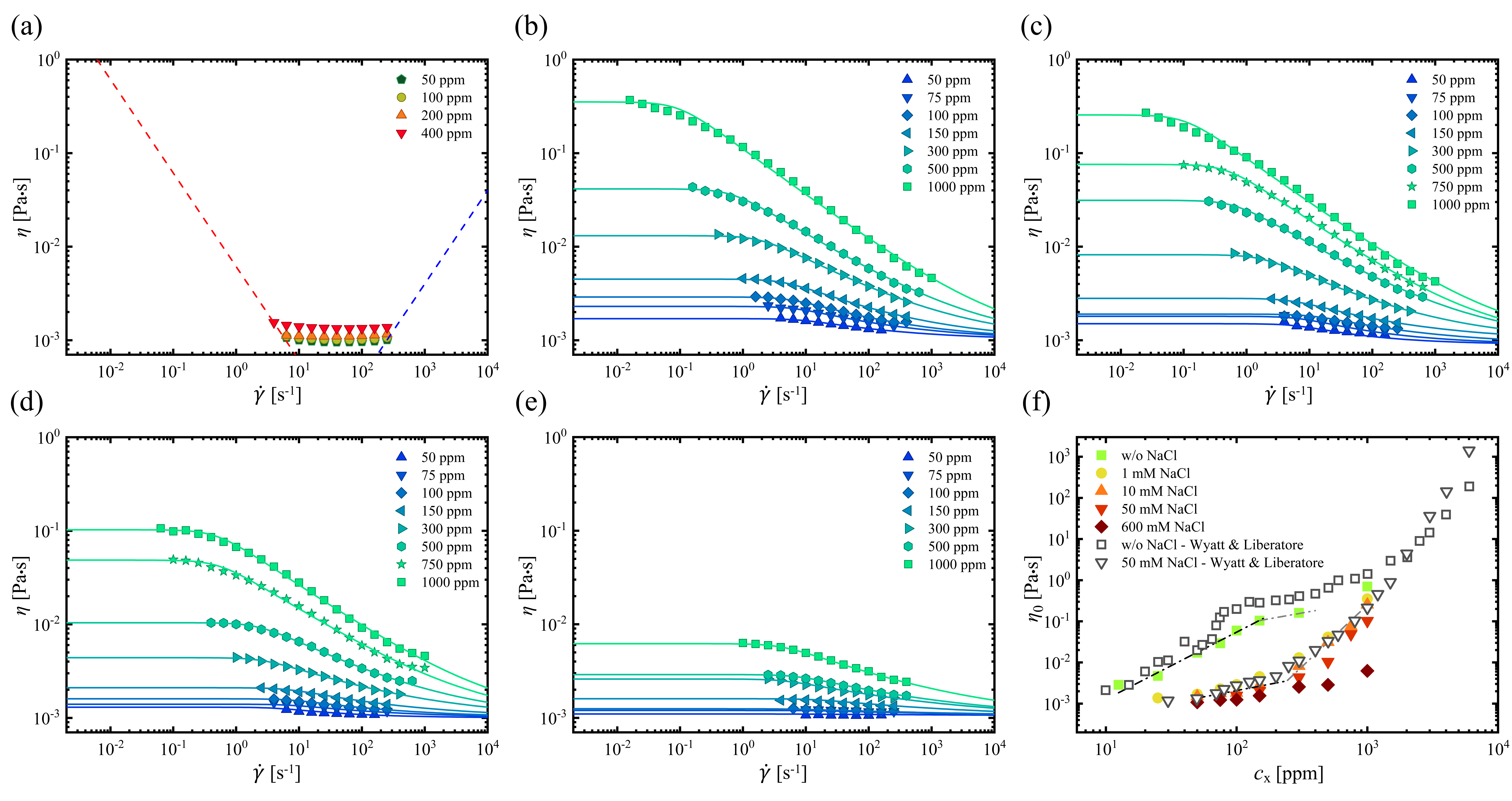}
  \caption{(a) Shear viscosity of PEO with $M_w \approx 2\times 10^6$ g/mol and concentrations below overlap concentration $c^*=858$ ppm. The overlap concentration of PEO is calculated based on $c^*=0.77/[\eta]$, where $[\eta]$ is the intrinsic viscosity calculated using the Mark-Houwink-Sakurada equation, $[\eta]=0.00072\times M^{0.65}_w$ for PEO.\cite{McKinley2006}. The red and blue dashed lines show the minimum torque and onset of secondary flow limits for the double-wall concentric cylinder geometry on the ARES-G2 rheometer. Shear viscosity of xanthan gum solutions with (b) 1 mM NaCl, (c) 10 mM NaCl, (d) 50 mM NaCl, and (e) 600 mM NaCl. (f) Zero shear viscosity of xanthan gum solutions at different polymer and salt concentrations measured in this work (filled symbols) and reported in the literature~\cite{Liberatore2009} (hollow symbols).}
  \label{fgr_S2}
\end{figure}
}
\clearpage

\hypertarget{supp:fig_3}{
\begin{figure}[htbp]
\centering
  \includegraphics[width=0.99\textwidth]{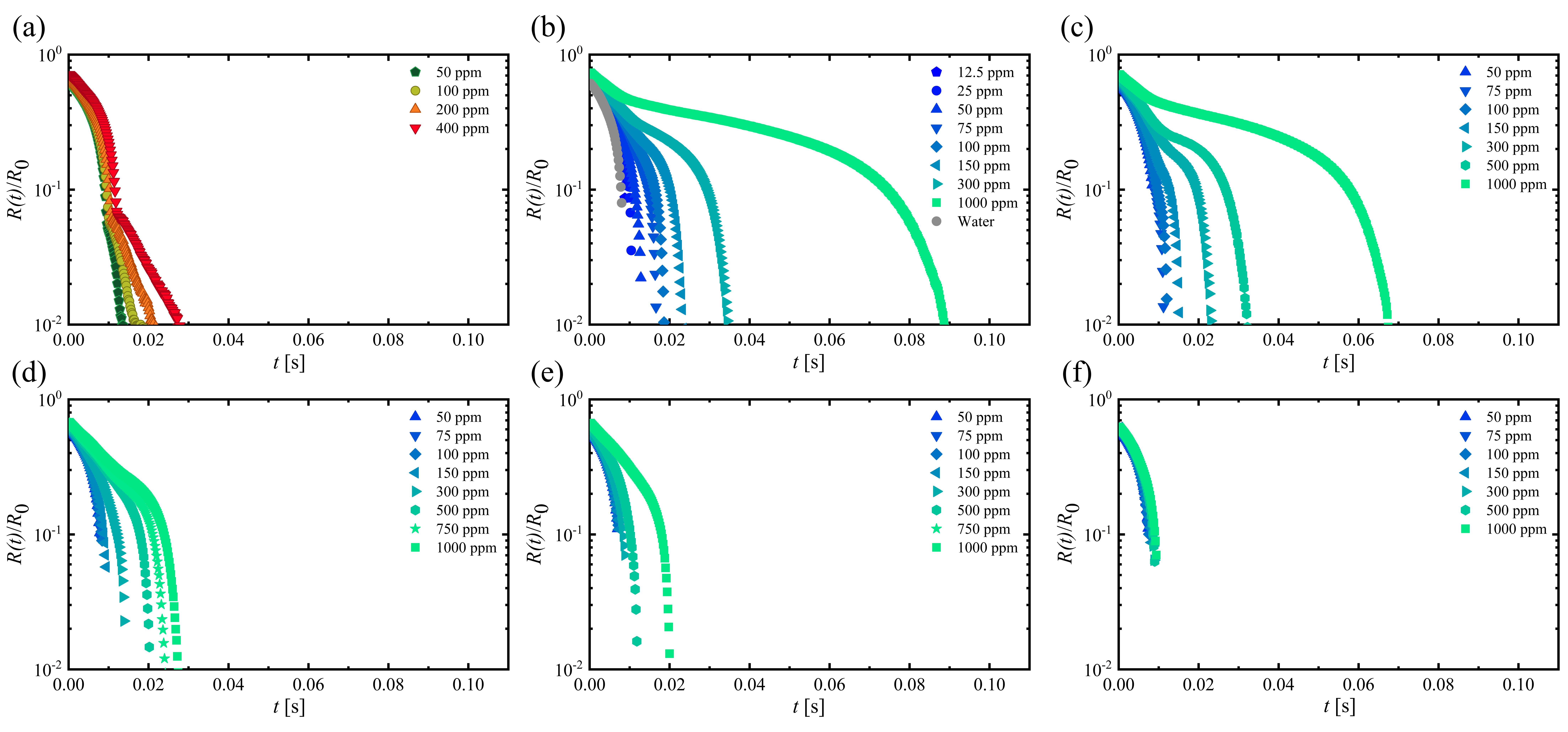}
  \caption{Temporal evolution of minimum neck radius in Dripping-onto-substrate (DoS) experiments for (a) different concentrations of PEO in DI water and different concentrations of xanthan gum in (b) DI water as solvent and solutions of (c) 1 mM NaCl, (d) 10 mM NaCl, (e) 50 mM NaCl, and (f) 600 mM NaCl in water as solvent. For clear comparison between the results for different solutions, the same maximum value of time (0.11 s) is used on the abscissa of all plots.}
  \label{fgr_S3}
\end{figure}
}

\hypertarget{supp:fig_4}{
\begin{figure*}[htbp]
\centering
  \includegraphics[width=0.99\textwidth]{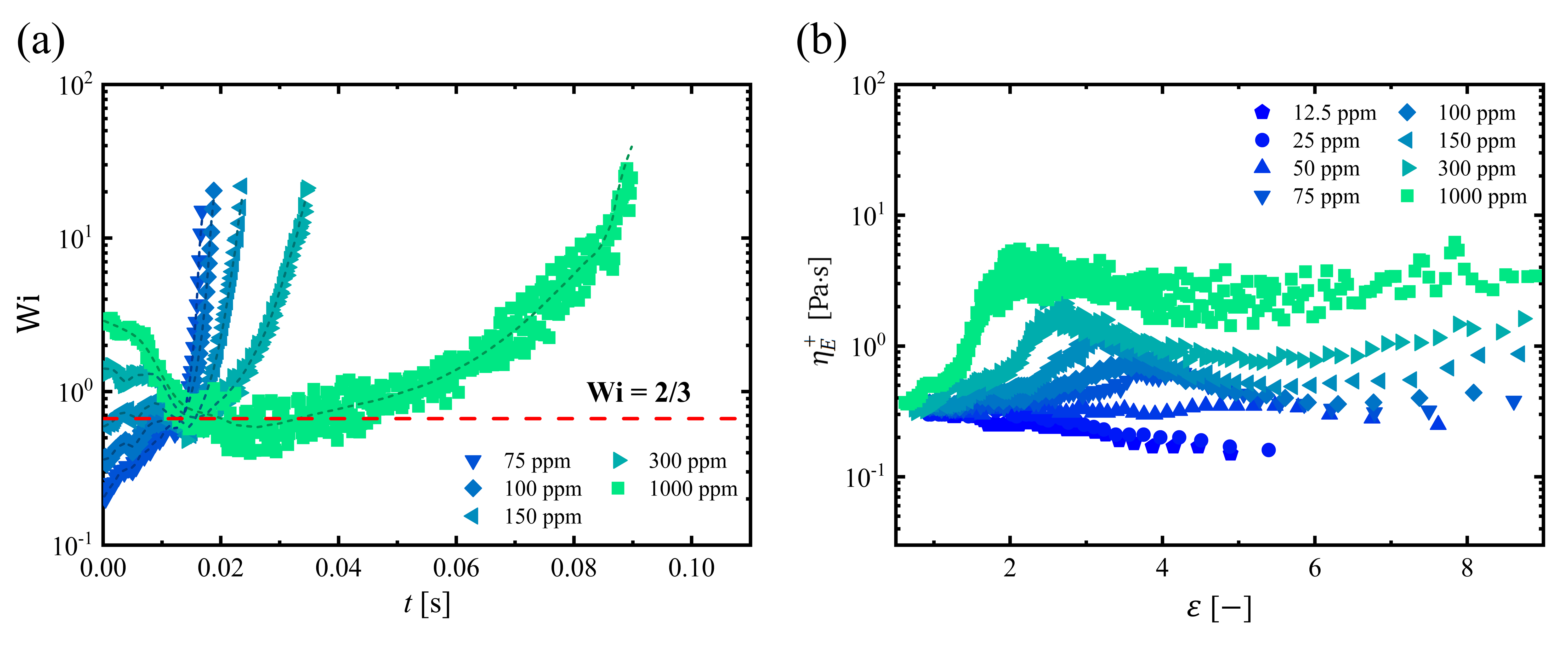}
  \caption{(a) Temporal evolution of the instantaneous Weissenberg number, $\textrm{Wi}(t) = \dot{\varepsilon}(t)\lambda_E$, for salt-free xanthan gum solutions that show transition to an elastocapillary response (exponential thinning in $R(t)$) during DoS experiments. (b) The transient extensional viscosity $\eta_E^{+}(t;c_{\textrm{x}})$ of salt-free xanthan gum solutions.}
  \label{fgr_S4}
\end{figure*}
}
\clearpage

\hypertarget{supp:fig_5}{
\begin{figure*}[htbp]
\centering
  \includegraphics[width=0.99\textwidth]{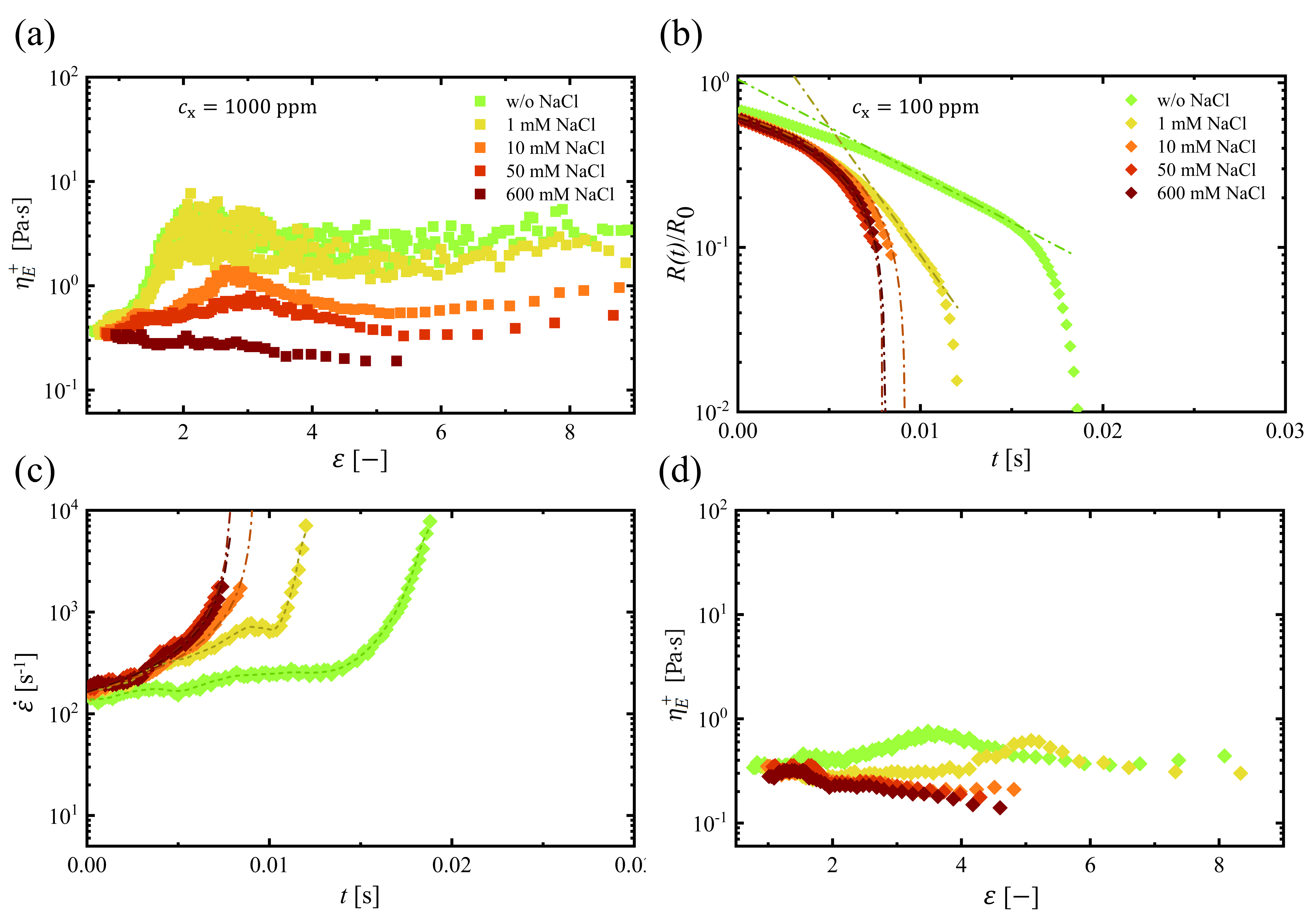}
  \caption{(a) Transient extensional viscosity $\eta_E^{+}(t;c_{\textrm{x}})$ of xanthan gum solutions with $c_{\textrm{x}}=1000$ ppm and different salt concentrations. Temporal evolution of (b) filament radius, (c) strain rate, and (d) extensional viscosity of xanthan gum solutions with $c_{\textrm{x}}=100$ ppm and different salt concentrations. The values of rheological parameters of xanthan gum solution with $c_{\textrm{x}}=100$ ppm in steady shear and transient extensional flows are summarized in Table\,\ref{tbl_s5}.}
  \label{fgr_S5}
\end{figure*}
}

\clearpage
\hypertarget{supp:tbl_1}{
\begin{table}[htbp]
\small
  \caption{Number of salt ions per unit xanthan gum molecule ($\times 10^3$)}
  \label{tbl_s1}
  \renewcommand{\arraystretch}{1.3}
  \begin{threeparttable}
  \begin{tabular*}{\textwidth}{@{\extracolsep{\fill}}lcrrrrrrrrr}
    \hline
    \hline
   \vcell{~~~\diagbox{$c_\textrm{NaCl} \textrm{ [mM]}$}{$c_\textrm{x}\textrm{ [ppm]}$}} & \multicolumn{1}{c}{\vcell{~~~~~~12.5}}   &\multicolumn{1}{c}{\vcell{25}}  & \multicolumn{1}{c}{\vcell{50}}     & \multicolumn{1}{c}{\vcell{75}}      & \multicolumn{1}{c}{\vcell{100}}         & \multicolumn{1}{c}{\vcell{150}}         & \multicolumn{1}{c}{\vcell{300}}         & \multicolumn{1}{c}{\vcell{500}}         & \multicolumn{1}{c}{\vcell{750}}        & \multicolumn{1}{c}{\vcell{1000}}     \\[-\rowheight]
    \printcelltop & \printcelltop & \printcelltop & \printcelltop & \printcelltop & \printcelltop & \printcelltop & \printcelltop & \printcelltop & \printcelltop & \printcelltop \\
\\[-0.7em]
    \hline
    ~~~~~~~~~~~~~1 & ~~~~~~~~160\tnote{\textrm{a}}   & 80.0    & 40.0    & 26.7    & 20.0    & 13.3   & 6.67    & 4.00    & 2.67    & 2.00   \\
    ~~~~~~~~~~~10& ~~~~~1600  & 800   & 400   & 267   & 200   & 133  & 66.7   & 40.0   & 26.7   & 20.0     \\
    ~~~~~~~~~~~50& ~~~~~8000  & 4000  & 2000  & 1333  & 1000  & 667  & 333  & 200  & 133  & 100   \\
    ~~~~~~~~~600& ~~~96000 & 48000 & 24000 & 16000 & 12000 & 8000 & 4000 & 2400 & 1600 & 1200 \\
    \hline
    \hline
    \end{tabular*}
    \begin{tablenotes}\footnotesize
        \item[a] Number of salt ions per unit xanthan gum molecule, $N_{\textrm{NaCl, XG}}$, calculated as, $N_{\textrm{NaCl, XG}}$ = $c_\textrm{NaCl}~[\textrm{mM}]\,/\,c_{\textrm{x}}~[\textrm{ppm}]= c_\textrm{NaCl}~[\textrm{mM}]\times M_{w}\,/\,c_{\textrm{x}}~[\textrm{mM}]$, where $M_{w} = 2 \times 10^6$ [g/mol] is the reported molecular weight of the xanthan gum sample used in the present work. 
    \end{tablenotes}
    \end{threeparttable}
\end{table}
}

\hypertarget{supp:tbl_2}{
\begin{table}[htbp]
\small
    \caption{Number of salt ions per repeat unit of xanthan gum$^a$}
    \label{tbl_s2}
    \renewcommand{\arraystretch}{1.3}
    \begin{threeparttable}
    \begin{tabular*}{\textwidth}{@{\extracolsep{\fill}}lcrrrrrrrrrr}
    \hline
    \hline
     \vcell{~~~\diagbox{$c_\textrm{NaCl} \textrm{ [mM]}$}{$c_\textrm{x}\textrm{ [ppm]}$}} & \multicolumn{1}{c}{\vcell{~~~~~12.5}}   &\multicolumn{1}{c}{\vcell{25}}  & \multicolumn{1}{c}{\vcell{50}}     & \multicolumn{1}{c}{\vcell{75}}      & \multicolumn{1}{c}{\vcell{100}}         & \multicolumn{1}{c}{\vcell{150}}         & \multicolumn{1}{c}{\vcell{300}}         & \multicolumn{1}{c}{\vcell{500}}         & \multicolumn{1}{c}{\vcell{750}}        & \multicolumn{1}{c}{\vcell{1000}}     \\[-\rowheight]
    \printcelltop & \printcelltop & \printcelltop & \printcelltop & \printcelltop & \printcelltop & \printcelltop & \printcelltop & \printcelltop & \printcelltop & \printcelltop \\
    \\[-0.7em]
    \hline
    ~~~~~~~~~~~~~1 & ~~~~~~~74.7\tnote{\textrm{b}}  & 37.4  & 18.7  & 12.5 & 9.34 & 6.23 & 3.11 & 1.87 & 1.25 & 0.93  \\
    ~~~~~~~~~~~10 & ~~~~~~747   & 374   & 187   & 125  & 93.4 & 62.3 & 31.1 & 18.7 & 12.5 & 9.34  \\
    ~~~~~~~~~~~50 & ~~~~3740  & 1870  & 934   & 623  & 467  & 311  & 156  & 93.4 & 62.3 & 46.7  \\
    ~~~~~~~~~600 & ~~44800 & 22400 & 11200 & 7470 & 5610 & 3740 & 1870 & 1120 & 747  & 560   \\
    \hline
    \hline
    \end{tabular*}
    \begin{tablenotes}\footnotesize
        \item[a] Molar mass of xanthan gum repeat unit, $M_n = 934$ g/mol
        \item[b] Number of salt ions per repeat unit of xanthan gum calculated as, $N_{\textrm{NaCl, XG}}\,/\,N_{m}$, where $N_{m}=2141$ is the number of repeat units in one xanthan gum molecule. 
    \end{tablenotes}
    \end{threeparttable}
\end{table}
}

\hypertarget{supp:tbl_3}{
\begin{table}[htbp]
\small
  \caption{Number of salt molecules per unit persistence length of xanthan gum}
  \label{tbl_s3}
  \renewcommand{\arraystretch}{1.3}
  \begin{threeparttable}
  \begin{tabular*}{\textwidth}{@{\extracolsep{\fill}}lcrrrrrrrrr}
    \hline
    \hline
    \vcell{~~~\diagbox{$c_\textrm{NaCl} \textrm{ [mM]}$}{$c_\textrm{x}\textrm{ [ppm]}$}} & \multicolumn{1}{c}{\vcell{~~~~~~~12.5}}   &\multicolumn{1}{c}{\vcell{25}}  & \multicolumn{1}{c}{\vcell{50}}     & \multicolumn{1}{c}{\vcell{75}}      & \multicolumn{1}{c}{\vcell{100}}         & \multicolumn{1}{c}{\vcell{150}}         & \multicolumn{1}{c}{\vcell{300}}         & \multicolumn{1}{c}{\vcell{500}}         & \multicolumn{1}{c}{\vcell{750}}        & \multicolumn{1}{c}{\vcell{1000}}     \\[-\rowheight]
    \printcelltop & \printcelltop & \printcelltop & \printcelltop & \printcelltop & \printcelltop & \printcelltop & \printcelltop & \printcelltop & \printcelltop & \printcelltop \\
    \\[-0.7em]
    \hline
    ~~~~~~~~~~~~~1 & ~~~~~~~~3890\tnote{\textrm{a}}    & 1940   & 972    & 648    & 486    & 324    & 162   & 97.2    & 64.8    & 48.6     \\
    ~~~~~~~~~~~10 & ~~~~~37400   & 18700  & 9340   & 6230   & 4670   & 3110   & 1560  & 934   & 623   & 467    \\
    ~~~~~~~~~~~50 & ~~~182000  & 91000  & 45500  & 30300  & 22700  & 15200  & 7580  & 4550  & 3030  & 2280   \\
    ~~~~~~~~~600 & ~1840000 & 919000 & 460000 & 306000 & 230000 & 153000 & 76600 & 46000 & 30600 & 23000  \\
    \hline
    \hline
    \end{tabular*}
    \begin{tablenotes}\footnotesize
        \item[a] Number of salt ions per unit persistence length of xanthan gum calculated as, $N_{\textrm{NaCl, XG}}\,/\,2N_{k}$, where $N_{k}$ is the number of Kuhn segments in a polymer chain. The values of $N_{k}$ are reported in Table\,\ref{tbl_s6} for different values of NaCl concentration.
    \end{tablenotes}
    \end{threeparttable}
\end{table}
}

\hypertarget{supp:tbl_4}{
\begin{table}[htbp]
\small
    \caption{Rheological parameters of 1000 ppm xanthan gum solutions at different salt concentrations in shear and extensional flows}
    \label{tbl_s4}
    \renewcommand{\arraystretch}{1.3}
    \begin{threeparttable}
    \begin{tabular*}{\textwidth}{@{\extracolsep{\fill}}cccc|cccccc}
        \hline
        \hline
        & \multicolumn{3}{c|}{Shear flow}& \multicolumn{5}{c}{Extensional flow}\\
        \hline
        $c_{\textrm{NaCl}}$~[mM] & $\eta_0$~[Pa$\cdot$s] & $\lambda_S$~[s] & $n$ & $Y$ & $x$ & $\lambda_E$~[ms] & $t_f$~[ms] & Oh\tnote{\textrm{a}}\\
        \hline
        ~~~~0 & 0.701 & 5.901 & 0.405 & - & - & 23.7 ± 1.82 & 91.3 ± 1.19 & 1.849 \\
        ~~~~1 & 0.353 & 10.611~~ & 0.507 & - & - & 19.3 ± 1.75 & 67.5 ± 4.43 & 0.931\\
        ~~10 & 0.255 & 9.552 & 0.523 & - & - & 6.02 ± 0.47 & 28.4 ± 1.60 & 0.673\\
        ~~50 & 0.103 & 2.101 & 0.544 & - & - & 3.60 ± 0.04 & 22.6 ± 3.30 & 0.271\\
        600 & 0.006 & 0.245 & 0.711 & 0.658 ± 0.002 & 0.764 ± 0.023 & - & 9.70 ± 0.20 & 0.016\\
        \hline
        \hline
    \end{tabular*}
    \begin{tablenotes}\footnotesize
        \item[a] The Ohnesorge number is calculated using the expression $\textrm{Oh} = \eta_0/(\rho\sigma R_0)^{1/2}$, where $R_0 = 2$ mm. 
    \end{tablenotes}
    \end{threeparttable}
\end{table}
}

\hypertarget{supp:tbl_5}{
\begin{table}[htbp]
\small
  \caption{Rheological parameters of 100 ppm xanthan gum solutions at different salt concentrations in shear and extensional flows}
  \label{tbl_s5}
  \renewcommand{\arraystretch}{1.3}
  \begin{tabular*}{\textwidth}{@{\extracolsep{\fill}}cccc|ccccc}
    \hline
    \hline
     & \multicolumn{3}{c|}{Shear flow}& \multicolumn{5}{c}{Extensional flow}\\
    \hline
    $c_{\textrm{NaCl}}$~[mM] & $\eta_0$~[Pa$\cdot$s] & $\lambda_S$~[s] & $n$ & $Y$ & $x$ & $\lambda_E$~[ms] & $t_f$~[ms] & Oh\\
    \hline
    ~~~~0 & 0.060 & 5.437 & 0.449 & - & - & 2.62 ± 0.23 &  18.9 ± 0.39 & 0.158\\
    ~~~~1 & 0.0029 & 0.180 & 0.681 & - & - & 0.82 ± 0.13 &  12.1 ± 0.01 & 0.008\\
    ~~10 & 0.0018 & 0.087 & 0.699 & 0.698 ± 0.002 & 0.759 ± 0.022 & - &  8.90 ± 0.12 & 0.005 \\
    ~~50 & 0.0016 & 0.129 & 0.713 & 0.724 ± 0.005 & 0.661 ± 0.005 & - & 7.88 ± 0.38 & 0.004\\
    600 & 0.0013 & 0.187 & 0.884 & 0.731 ± 0.005 & 0.665 ± 0.017 & - &  7.83 ± 0.17 & 0.003\\
     \hline
     \hline
  \end{tabular*}
\end{table}
}

\hypertarget{supp:tbl_6}{
\begin{table}[htbp]
\small
  \caption{Conformational parameters and finite extensibility of the flexible (PEO) and semi-flexible (xanthan gum) polymers used in this study}
  \label{tbl_s6}
  \renewcommand{\arraystretch}{1.3}
  \begin{threeparttable}
  \begin{tabular*}{\textwidth}{@{\extracolsep{\fill}}ccccccrrrrrr}
    \hline
    \hline
    \\[-1em]
    ~~Polymer 
    & \begin{tabular}[c]{@{}c@{}}$c_{\textrm{NaCl}}$\\{[}mM]\end{tabular} 
    & \begin{tabular}[c]{@{}c@{}}$M_w$\\{[}g/mol]\end{tabular}
    & \begin{tabular}[c]{@{}c@{}}$M_n$\\{[}g/mol]\end{tabular}
    & \begin{tabular}[c]{@{}c@{}}$N_m$\tnote{$\textrm{a}$} \\{}\end{tabular} 
    & \begin{tabular}[c]{@{}c@{}}$b_n$\\{[}nm]\end{tabular} 
    & \begin{tabular}[c]{@{}c@{}}$r_{\textrm{max}}$\\{[}nm]\end{tabular}
    & \begin{tabular}[c]{@{}c@{}}$\ell_p$\\{[}nm]\end{tabular}
    & \begin{tabular}[c]{@{}c@{}}$b_k$\tnote{$\textrm{b}$} \\{[}nm]\end{tabular}
    & \begin{tabular}[c]{@{}c@{}}$N_k$\tnote{$\textrm{c}$} \\{[}nm]\end{tabular}
    & \begin{tabular}[c]{@{}c@{}}$\left<r_{0}^{2} \right>^{1/2}$\\{[}nm]\end{tabular} 
    & \begin{tabular}[c]{@{}c@{}}$L~~~$\\{}\end{tabular}\\ 
    \hline
    ~~PEO & ~~0 & $2\times10^6$ & ~~44\tnote{$\textrm{d}$} & 45455 & 0.29\tnote{$\textrm{e}$} & 13182 & 0.48~\cite{Shell2020} & 0.96 & 13731.3& 112 & ~117 \\
    \hline
          & ~~0 &  &  &  &  &  & 210\tnote{$\textrm{h}$} & 420 & 5.1 & 901 & 2.38\\
          & ~~1 &  &  &  &  &  & 52 & 104 & 20.6 & 466 & 4.59\\
    ~~Xanthan gum & ~~10 & $2\times10^6$ & 934\tnote{$\textrm{f}$} & ~~2141 & 1.00\tnote{$\textrm{g}$} & 2141 & 50 & 100 & 21.4 & 457 & 4.68\\
          & 50  &  &  &  &  &  & 47.5 & 95 & 22.5 & 446 & 4.80\\
          & 600  &  &  &  &  &  & 41 & 82 & 26.1 & 415 & 5.16\\
    \hline
    \hline
  \end{tabular*}
\begin{tablenotes}\footnotesize
    \item[a] Number of monomers in a polymer chain, $N_m=M_w/M_n$. 
    \item[b] Kuhn length of a polymer chain, $b_k=2\ell_p$
    \item[c] Number of Kuhn segments in a polymer chain, $N_k=r_{max}/b_k$
    \item[d] Molar mass of PEO monomer, \ch{C2H4O} \cite{bailey2012poly}.
    \item[e] PEO monomer length derived from the combined bond lengths (\ch{C-C} and \ch{C-O}) and bond angle \cite{brandrup1999polymer}
    \item[f] Molar mass of xanthan gum monomer, \ch{C35H49O29} \cite{bhat2022advances}.
    \item[g] Xanthan gum repeat unit length derived from the length of two glucose units in xanthan gum monomer~\cite{rojas2016cellulose}.
    \item[h] The persistence lengths for 0, 1, 10 mM NaCl concentrations are reported in~\cite{Chauveteau1986}. The persistence lengths for 50 and 600 mM NaCl concentrations are interpolated based on the best power-law fit to the measurements reported in~\cite{Chauveteau1986}, as shown in Fig\hyperlink{supp:fig_5}{S5}.
\end{tablenotes}
\end{threeparttable}
\end{table}
}
\clearpage
\bibliography{reference.bib}

\begin{thebibliography}{77}%
\makeatletter
\providecommand \@ifxundefined [1]{%
 \@ifx{#1\undefined}
}%
\providecommand \@ifnum [1]{%
 \ifnum #1\expandafter \@firstoftwo
 \else \expandafter \@secondoftwo
 \fi
}%
\providecommand \@ifx [1]{%
 \ifx #1\expandafter \@firstoftwo
 \else \expandafter \@secondoftwo
 \fi
}%
\providecommand \natexlab [1]{#1}%
\providecommand \enquote  [1]{``#1''}%
\providecommand \bibnamefont  [1]{#1}%
\providecommand \bibfnamefont [1]{#1}%
\providecommand \citenamefont [1]{#1}%
\providecommand \href@noop [0]{\@secondoftwo}%
\providecommand \href [0]{\begingroup \@sanitize@url \@href}%
\providecommand \@href[1]{\@@startlink{#1}\@@href}%
\providecommand \@@href[1]{\endgroup#1\@@endlink}%
\providecommand \@sanitize@url [0]{\catcode `\\12\catcode `\$12\catcode `\&12\catcode `\#12\catcode `\^12\catcode `\_12\catcode `\%12\relax}%
\providecommand \@@startlink[1]{}%
\providecommand \@@endlink[0]{}%
\providecommand \url  [0]{\begingroup\@sanitize@url \@url }%
\providecommand \@url [1]{\endgroup\@href {#1}{\urlprefix }}%
\providecommand \urlprefix  [0]{URL }%
\providecommand \Eprint [0]{\href }%
\providecommand \doibase [0]{http://dx.doi.org/}%
\providecommand \selectlanguage [0]{\@gobble}%
\providecommand \bibinfo  [0]{\@secondoftwo}%
\providecommand \bibfield  [0]{\@secondoftwo}%
\providecommand \translation [1]{[#1]}%
\providecommand \BibitemOpen [0]{}%
\providecommand \bibitemStop [0]{}%
\providecommand \bibitemNoStop [0]{.\EOS\space}%
\providecommand \EOS [0]{\spacefactor3000\relax}%
\providecommand \BibitemShut  [1]{\csname bibitem#1\endcsname}%
\let\auto@bib@innerbib\@empty
\bibitem [{\citenamefont {Lapasin}(2012)}]{Lapasin2012}%
  \BibitemOpen
  \bibfield  {author} {\bibinfo {author} {\bibfnamefont {Romano}\ \bibnamefont {Lapasin}},\ }\href {https://doi.org/10.1007/978-1-4615-2185-3} {\emph {\bibinfo {title} {Rheology of industrial polysaccharides: theory and applications}}}\ (\bibinfo  {publisher} {Springer Science \& Business Media},\ \bibinfo {year} {2012})\BibitemShut {NoStop}%
\bibitem [{\citenamefont {Nsengiyumva}\ and\ \citenamefont {Alexandridis}(2022)}]{Alexandridis2022}%
  \BibitemOpen
  \bibfield  {author} {\bibinfo {author} {\bibfnamefont {Emmanuel~M}\ \bibnamefont {Nsengiyumva}}\ and\ \bibinfo {author} {\bibfnamefont {Paschalis}\ \bibnamefont {Alexandridis}},\ }\bibfield  {title} {\enquote {\bibinfo {title} {Xanthan gum in aqueous solutions: Fundamentals and applications},}\ }\href {https://doi.org/10.1016/j.ijbiomac.2022.06.189} {\bibfield  {journal} {\bibinfo  {journal} {Int. J. Biol. Macromol.}\ }\textbf {\bibinfo {volume} {216}},\ \bibinfo {pages} {583--604} (\bibinfo {year} {2022})}\BibitemShut {NoStop}%
\bibitem [{\citenamefont {Meka}\ \emph {et~al.}(2017)\citenamefont {Meka}, \citenamefont {Sing}, \citenamefont {Pichika}, \citenamefont {Nali}, \citenamefont {Kolapalli},\ and\ \citenamefont {Kesharwani}}]{Kesharwani2017}%
  \BibitemOpen
  \bibfield  {author} {\bibinfo {author} {\bibfnamefont {Venkata~S}\ \bibnamefont {Meka}}, \bibinfo {author} {\bibfnamefont {Manprit~KG}\ \bibnamefont {Sing}}, \bibinfo {author} {\bibfnamefont {Mallikarjuna~R}\ \bibnamefont {Pichika}}, \bibinfo {author} {\bibfnamefont {Srinivasa~R}\ \bibnamefont {Nali}}, \bibinfo {author} {\bibfnamefont {Venkata~RM}\ \bibnamefont {Kolapalli}}, \ and\ \bibinfo {author} {\bibfnamefont {Prashant}\ \bibnamefont {Kesharwani}},\ }\bibfield  {title} {\enquote {\bibinfo {title} {A comprehensive review on polyelectrolyte complexes},}\ }\href {https://doi.org/10.1016/j.drudis.2017.06.008} {\bibfield  {journal} {\bibinfo  {journal} {Drug Discov. Today}\ }\textbf {\bibinfo {volume} {22}},\ \bibinfo {pages} {1697--1706} (\bibinfo {year} {2017})}\BibitemShut {NoStop}%
\bibitem [{\citenamefont {Pilevaran}\ \emph {et~al.}(2021)\citenamefont {Pilevaran}, \citenamefont {Tavakolipour}, \citenamefont {Naji-Tabasi},\ and\ \citenamefont {Elhamirad}}]{Elhamirad2021}%
  \BibitemOpen
  \bibfield  {author} {\bibinfo {author} {\bibfnamefont {Majid}\ \bibnamefont {Pilevaran}}, \bibinfo {author} {\bibfnamefont {Hamid}\ \bibnamefont {Tavakolipour}}, \bibinfo {author} {\bibfnamefont {Sara}\ \bibnamefont {Naji-Tabasi}}, \ and\ \bibinfo {author} {\bibfnamefont {Amir~Hossein}\ \bibnamefont {Elhamirad}},\ }\bibfield  {title} {\enquote {\bibinfo {title} {Development of mechanical and thermal properties of whey protein–xanthan gum hydrogel by incorporation of basil seed gum nanoparticles, salt, and acidic ph},}\ }\href {https://doi.org/10.1007/s10971-021-05508-y} {\bibfield  {journal} {\bibinfo  {journal} {J. Solgel. Sci. Technol.}\ }\textbf {\bibinfo {volume} {98}},\ \bibinfo {pages} {76--83} (\bibinfo {year} {2021})}\BibitemShut {NoStop}%
\bibitem [{\citenamefont {Coria-Hernández}\ \emph {et~al.}(2021)\citenamefont {Coria-Hernández}, \citenamefont {Meléndez-Pérez}, \citenamefont {Méndez-Albores},\ and\ \citenamefont {Arjona-Román}}]{ArjonaRoman2021}%
  \BibitemOpen
  \bibfield  {author} {\bibinfo {author} {\bibfnamefont {Jonathan}\ \bibnamefont {Coria-Hernández}}, \bibinfo {author} {\bibfnamefont {Rosalía}\ \bibnamefont {Meléndez-Pérez}}, \bibinfo {author} {\bibfnamefont {Abraham}\ \bibnamefont {Méndez-Albores}}, \ and\ \bibinfo {author} {\bibfnamefont {José~Luis}\ \bibnamefont {Arjona-Román}},\ }\bibfield  {title} {\enquote {\bibinfo {title} {Effect of cryostructuring treatment on some properties of xanthan and karaya cryogels for food applications},}\ }\href {https://doi.org/10.3390/molecules26092788} {\bibfield  {journal} {\bibinfo  {journal} {Molecules}\ }\textbf {\bibinfo {volume} {26}},\ \bibinfo {pages} {2788} (\bibinfo {year} {2021})}\BibitemShut {NoStop}%
\bibitem [{\citenamefont {Parente}\ \emph {et~al.}(2015)\citenamefont {Parente}, \citenamefont {Ochoa~Andrade}, \citenamefont {Ares}, \citenamefont {Russo},\ and\ \citenamefont {Jim\'{e}nez‐Kairuz}}]{JimenezKairuz2015}%
  \BibitemOpen
  \bibfield  {author} {\bibinfo {author} {\bibfnamefont {ME}~\bibnamefont {Parente}}, \bibinfo {author} {\bibfnamefont {A}~\bibnamefont {Ochoa~Andrade}}, \bibinfo {author} {\bibfnamefont {G}~\bibnamefont {Ares}}, \bibinfo {author} {\bibfnamefont {F}~\bibnamefont {Russo}}, \ and\ \bibinfo {author} {\bibfnamefont {\'{A}}\ \bibnamefont {Jim\'{e}nez‐Kairuz}},\ }\bibfield  {title} {\enquote {\bibinfo {title} {Bioadhesive hydrogels for cosmetic applications},}\ }\href {https://doi.org/10.1111/ics.12227} {\bibfield  {journal} {\bibinfo  {journal} {Int. J. Cosmet. Sci.}\ }\textbf {\bibinfo {volume} {37}},\ \bibinfo {pages} {511--518} (\bibinfo {year} {2015})}\BibitemShut {NoStop}%
\bibitem [{\citenamefont {Rajappan}\ and\ \citenamefont {McKinley}(2019)}]{Mckinley2019}%
  \BibitemOpen
  \bibfield  {author} {\bibinfo {author} {\bibfnamefont {Anoop}\ \bibnamefont {Rajappan}}\ and\ \bibinfo {author} {\bibfnamefont {Gareth~H}\ \bibnamefont {McKinley}},\ }\bibfield  {title} {\enquote {\bibinfo {title} {Epidermal biopolysaccharides from plant seeds enable biodegradable turbulent drag reduction},}\ }\href {https://doi.org/10.1038/s41598-019-54521-3} {\bibfield  {journal} {\bibinfo  {journal} {Sci. Rep.}\ }\textbf {\bibinfo {volume} {9}},\ \bibinfo {pages} {18263} (\bibinfo {year} {2019})}\BibitemShut {NoStop}%
\bibitem [{\citenamefont {Pereira}\ \emph {et~al.}(2013)\citenamefont {Pereira}, \citenamefont {Andrade},\ and\ \citenamefont {Soares}}]{Soares2013}%
  \BibitemOpen
  \bibfield  {author} {\bibinfo {author} {\bibfnamefont {Anselmo~S}\ \bibnamefont {Pereira}}, \bibinfo {author} {\bibfnamefont {Rafhael~M}\ \bibnamefont {Andrade}}, \ and\ \bibinfo {author} {\bibfnamefont {Edson~J}\ \bibnamefont {Soares}},\ }\bibfield  {title} {\enquote {\bibinfo {title} {Drag reduction induced by flexible and rigid molecules in a turbulent flow into a rotating cylindrical double gap device: Comparison between poly (ethylene oxide), polyacrylamide, and xanthan gum},}\ }\href {https://doi.org/10.1016/j.jnnfm.2013.09.008} {\bibfield  {journal} {\bibinfo  {journal} {J. Nonnewton. Fluid Mech.}\ }\textbf {\bibinfo {volume} {202}},\ \bibinfo {pages} {72--87} (\bibinfo {year} {2013})}\BibitemShut {NoStop}%
\bibitem [{\citenamefont {Hong}\ \emph {et~al.}(2015)\citenamefont {Hong}, \citenamefont {Choi}, \citenamefont {Zhang}, \citenamefont {Renou},\ and\ \citenamefont {Grisel}}]{Grisel2015}%
  \BibitemOpen
  \bibfield  {author} {\bibinfo {author} {\bibfnamefont {Cheng~Hai}\ \bibnamefont {Hong}}, \bibinfo {author} {\bibfnamefont {Hyoung~Jin}\ \bibnamefont {Choi}}, \bibinfo {author} {\bibfnamefont {Ke}~\bibnamefont {Zhang}}, \bibinfo {author} {\bibfnamefont {Frederic}\ \bibnamefont {Renou}}, \ and\ \bibinfo {author} {\bibfnamefont {Michel}\ \bibnamefont {Grisel}},\ }\bibfield  {title} {\enquote {\bibinfo {title} {Effect of salt on turbulent drag reduction of xanthan gum},}\ }\href {https://doi.org/10.1016/j.carbpol.2014.12.015} {\bibfield  {journal} {\bibinfo  {journal} {Carbohydr. Polym.}\ }\textbf {\bibinfo {volume} {121}},\ \bibinfo {pages} {342--347} (\bibinfo {year} {2015})}\BibitemShut {NoStop}%
\bibitem [{\citenamefont {Tu}\ \emph {et~al.}(2022)\citenamefont {Tu}, \citenamefont {Shi}, \citenamefont {Li}, \citenamefont {Wang}, \citenamefont {Qiao}, \citenamefont {Jiang},\ and\ \citenamefont {Zhang}}]{Zhang2022}%
  \BibitemOpen
  \bibfield  {author} {\bibinfo {author} {\bibfnamefont {Wenyao}\ \bibnamefont {Tu}}, \bibinfo {author} {\bibfnamefont {Wenjuan}\ \bibnamefont {Shi}}, \bibinfo {author} {\bibfnamefont {Hao}\ \bibnamefont {Li}}, \bibinfo {author} {\bibfnamefont {Yixin}\ \bibnamefont {Wang}}, \bibinfo {author} {\bibfnamefont {Dongling}\ \bibnamefont {Qiao}}, \bibinfo {author} {\bibfnamefont {Fatang}\ \bibnamefont {Jiang}}, \ and\ \bibinfo {author} {\bibfnamefont {Binjia}\ \bibnamefont {Zhang}},\ }\bibfield  {title} {\enquote {\bibinfo {title} {Xanthan gum inclusion optimizes the sol-gel and mechanical properties of agar/konjac glucomannan system for designing core-shell structural capsules},}\ }\href {https://doi.org/10.1016/j.foodhyd.2021.107101} {\bibfield  {journal} {\bibinfo  {journal} {Food Hydrocoll.}\ }\textbf {\bibinfo {volume} {122}},\ \bibinfo {pages} {107101} (\bibinfo {year} {2022})}\BibitemShut {NoStop}%
\bibitem [{\citenamefont {Sugiura}\ \emph {et~al.}(2020)\citenamefont {Sugiura}, \citenamefont {Onuki}, \citenamefont {Fujita}, \citenamefont {Nakamura},\ and\ \citenamefont {Harada}}]{Harada2020}%
  \BibitemOpen
  \bibfield  {author} {\bibinfo {author} {\bibfnamefont {Daisuke}\ \bibnamefont {Sugiura}}, \bibinfo {author} {\bibfnamefont {Yoshinori}\ \bibnamefont {Onuki}}, \bibinfo {author} {\bibfnamefont {Yoshiaki}\ \bibnamefont {Fujita}}, \bibinfo {author} {\bibfnamefont {Akihiro}\ \bibnamefont {Nakamura}}, \ and\ \bibinfo {author} {\bibfnamefont {Tsutomu}\ \bibnamefont {Harada}},\ }\bibfield  {title} {\enquote {\bibinfo {title} {Effect of disintegrants on prolongation of tablet disintegration induced by immersion in xanthan gum-containing thickening solution: contribution of disintegrant interactions with disintegration fluids},}\ }\href {https://doi.org/10.1248/cpb.c20-00480} {\bibfield  {journal} {\bibinfo  {journal} {Chem. Pharm. Bull.}\ }\textbf {\bibinfo {volume} {68}},\ \bibinfo {pages} {1055--1060} (\bibinfo {year} {2020})}\BibitemShut {NoStop}%
\bibitem [{\citenamefont {Fu}\ \emph {et~al.}(2021)\citenamefont {Fu}, \citenamefont {Qin}, \citenamefont {Liu},\ and\ \citenamefont {Zhang}}]{Zhang2021}%
  \BibitemOpen
  \bibfield  {author} {\bibinfo {author} {\bibfnamefont {Xiaosheng}\ \bibnamefont {Fu}}, \bibinfo {author} {\bibfnamefont {Feifei}\ \bibnamefont {Qin}}, \bibinfo {author} {\bibfnamefont {Tao}\ \bibnamefont {Liu}}, \ and\ \bibinfo {author} {\bibfnamefont {Xiuxia}\ \bibnamefont {Zhang}},\ }\bibfield  {title} {\enquote {\bibinfo {title} {Enhanced oil recovery performance and solution properties of hydrophobic associative xanthan gum},}\ }\href {https://doi.org/10.1021/acs.energyfuels.1c02941} {\bibfield  {journal} {\bibinfo  {journal} {Energy Fuels}\ }\textbf {\bibinfo {volume} {36}},\ \bibinfo {pages} {181--194} (\bibinfo {year} {2021})}\BibitemShut {NoStop}%
\bibitem [{\citenamefont {Dey}\ and\ \citenamefont {Chatterji}(2023)}]{Chatterji2023}%
  \BibitemOpen
  \bibfield  {author} {\bibinfo {author} {\bibfnamefont {Rahul}\ \bibnamefont {Dey}}\ and\ \bibinfo {author} {\bibfnamefont {Biswa~Prasun}\ \bibnamefont {Chatterji}},\ }\bibfield  {title} {\enquote {\bibinfo {title} {Sources and methods of manufacturing xanthan by fermentation of various carbon sources},}\ }\href {https://doi.org/10.1002/btpr.3379} {\bibfield  {journal} {\bibinfo  {journal} {Biotechnol. Prog.}\ }\textbf {\bibinfo {volume} {39}},\ \bibinfo {pages} {e3379} (\bibinfo {year} {2023})}\BibitemShut {NoStop}%
\bibitem [{\citenamefont {Wyatt}\ and\ \citenamefont {Liberatore}(2009)}]{Liberatore2009}%
  \BibitemOpen
  \bibfield  {author} {\bibinfo {author} {\bibfnamefont {Nicholas~B}\ \bibnamefont {Wyatt}}\ and\ \bibinfo {author} {\bibfnamefont {Matthew~W}\ \bibnamefont {Liberatore}},\ }\bibfield  {title} {\enquote {\bibinfo {title} {Rheology and viscosity scaling of the polyelectrolyte xanthan gum},}\ }\href {https://doi.org/10.1002/app.31093} {\bibfield  {journal} {\bibinfo  {journal} {J. Appl. Polym. Sci.}\ }\textbf {\bibinfo {volume} {114}},\ \bibinfo {pages} {4076--4084} (\bibinfo {year} {2009})}\BibitemShut {NoStop}%
\bibitem [{\citenamefont {Muller}\ \emph {et~al.}(1986)\citenamefont {Muller}, \citenamefont {Aurhourrache}, \citenamefont {Lecourtier},\ and\ \citenamefont {Chauveteau}}]{Chauveteau1986}%
  \BibitemOpen
  \bibfield  {author} {\bibinfo {author} {\bibfnamefont {G}~\bibnamefont {Muller}}, \bibinfo {author} {\bibfnamefont {M}~\bibnamefont {Aurhourrache}}, \bibinfo {author} {\bibfnamefont {J}~\bibnamefont {Lecourtier}}, \ and\ \bibinfo {author} {\bibfnamefont {G}~\bibnamefont {Chauveteau}},\ }\bibfield  {title} {\enquote {\bibinfo {title} {Salt dependence of the conformation of a single-stranded xanthan},}\ }\href {https://doi.org/10.1016/0141-8130(86)90021-8} {\bibfield  {journal} {\bibinfo  {journal} {Int. J. Biol. Macromol.}\ }\textbf {\bibinfo {volume} {8}},\ \bibinfo {pages} {167--172} (\bibinfo {year} {1986})}\BibitemShut {NoStop}%
\bibitem [{\citenamefont {Stokke}\ and\ \citenamefont {Brant}(1990)}]{Brant1990}%
  \BibitemOpen
  \bibfield  {author} {\bibinfo {author} {\bibfnamefont {B.~T.}\ \bibnamefont {Stokke}}\ and\ \bibinfo {author} {\bibfnamefont {D.~A.}\ \bibnamefont {Brant}},\ }\bibfield  {title} {\enquote {\bibinfo {title} {The reliability of wormlike polysaccharide chain dimensions estimated from electron-micrographs},}\ }\href {https://doi.org/10.1002/bip.360301303} {\bibfield  {journal} {\bibinfo  {journal} {Biopolymers}\ }\textbf {\bibinfo {volume} {30}},\ \bibinfo {pages} {1161--1181} (\bibinfo {year} {1990})}\BibitemShut {NoStop}%
\bibitem [{\citenamefont {Jeanes}\ \emph {et~al.}(1961)\citenamefont {Jeanes}, \citenamefont {Pittsley},\ and\ \citenamefont {Senti}}]{Jeanes1961}%
  \BibitemOpen
  \bibfield  {author} {\bibinfo {author} {\bibfnamefont {Allene}\ \bibnamefont {Jeanes}}, \bibinfo {author} {\bibfnamefont {JE}~\bibnamefont {Pittsley}}, \ and\ \bibinfo {author} {\bibfnamefont {FR}~\bibnamefont {Senti}},\ }\bibfield  {title} {\enquote {\bibinfo {title} {Polysaccharide b‐1459: a new hydrocolloid polyelectrolyte produced from glucose by bacterial fermentation},}\ }\href {https://doi.org/10.1002/app.1961.070051704} {\bibfield  {journal} {\bibinfo  {journal} {J. Appl. Polym. Sci.}\ }\textbf {\bibinfo {volume} {5}},\ \bibinfo {pages} {519--526} (\bibinfo {year} {1961})}\BibitemShut {NoStop}%
\bibitem [{\citenamefont {Whitcomb}\ and\ \citenamefont {Macosko}(1978)}]{Macosko1978}%
  \BibitemOpen
  \bibfield  {author} {\bibinfo {author} {\bibfnamefont {Patrick~James}\ \bibnamefont {Whitcomb}}\ and\ \bibinfo {author} {\bibfnamefont {CW}~\bibnamefont {Macosko}},\ }\bibfield  {title} {\enquote {\bibinfo {title} {Rheology of xanthan gum},}\ }\href {https://doi.org/10.1122/1.549485} {\bibfield  {journal} {\bibinfo  {journal} {J. Rheol.}\ }\textbf {\bibinfo {volume} {22}},\ \bibinfo {pages} {493--505} (\bibinfo {year} {1978})}\BibitemShut {NoStop}%
\bibitem [{\citenamefont {Wyatt}\ \emph {et~al.}(2011)\citenamefont {Wyatt}, \citenamefont {Gunther},\ and\ \citenamefont {Liberatore}}]{Liberatore2011}%
  \BibitemOpen
  \bibfield  {author} {\bibinfo {author} {\bibfnamefont {Nicholas~B}\ \bibnamefont {Wyatt}}, \bibinfo {author} {\bibfnamefont {Casey~M}\ \bibnamefont {Gunther}}, \ and\ \bibinfo {author} {\bibfnamefont {Matthew~W}\ \bibnamefont {Liberatore}},\ }\bibfield  {title} {\enquote {\bibinfo {title} {Increasing viscosity in entangled polyelectrolyte solutions by the addition of salt},}\ }\href {https://doi.org/10.1016/j.polymer.2011.03.053} {\bibfield  {journal} {\bibinfo  {journal} {Polymer}\ }\textbf {\bibinfo {volume} {52}},\ \bibinfo {pages} {2437--2444} (\bibinfo {year} {2011})}\BibitemShut {NoStop}%
\bibitem [{\citenamefont {Missi}\ \emph {et~al.}(2024)\citenamefont {Missi}, \citenamefont {Montillet}, \citenamefont {Capron}, \citenamefont {Bellettre},\ and\ \citenamefont {Burghelea}}]{missi2024thermo}%
  \BibitemOpen
  \bibfield  {author} {\bibinfo {author} {\bibfnamefont {Elia}\ \bibnamefont {Missi}}, \bibinfo {author} {\bibfnamefont {Agn{\`e}s}\ \bibnamefont {Montillet}}, \bibinfo {author} {\bibfnamefont {Isabelle}\ \bibnamefont {Capron}}, \bibinfo {author} {\bibfnamefont {J{\'e}r{\^o}me}\ \bibnamefont {Bellettre}}, \ and\ \bibinfo {author} {\bibfnamefont {Teodor}\ \bibnamefont {Burghelea}},\ }\bibfield  {title} {\enquote {\bibinfo {title} {Thermo-rheological properties of xanthan solutions: from shear thinning to elasto-viscoplastic behavior},}\ }\href {https://doi.org/10.1039/D4SM00714J} {\bibfield  {journal} {\bibinfo  {journal} {Soft Matter}\ } (\bibinfo {year} {2024})}\BibitemShut {NoStop}%
\bibitem [{\citenamefont {Garcıa-Ochoa}\ \emph {et~al.}(2000)\citenamefont {Garcıa-Ochoa}, \citenamefont {Santos}, \citenamefont {Casas},\ and\ \citenamefont {Gómez}}]{Gomez2000}%
  \BibitemOpen
  \bibfield  {author} {\bibinfo {author} {\bibfnamefont {F}~\bibnamefont {Garcıa-Ochoa}}, \bibinfo {author} {\bibfnamefont {VE}~\bibnamefont {Santos}}, \bibinfo {author} {\bibfnamefont {JA}~\bibnamefont {Casas}}, \ and\ \bibinfo {author} {\bibfnamefont {E}~\bibnamefont {Gómez}},\ }\bibfield  {title} {\enquote {\bibinfo {title} {Xanthan gum: production, recovery, and properties},}\ }\href {https://doi.org/10.1016/S0734-9750(00)00050-1} {\bibfield  {journal} {\bibinfo  {journal} {Biotechnol. Adv.}\ }\textbf {\bibinfo {volume} {18}},\ \bibinfo {pages} {549--579} (\bibinfo {year} {2000})}\BibitemShut {NoStop}%
\bibitem [{\citenamefont {Morris}(2019)}]{Morris2019}%
  \BibitemOpen
  \bibfield  {author} {\bibinfo {author} {\bibfnamefont {Edwin~R}\ \bibnamefont {Morris}},\ }\bibfield  {title} {\enquote {\bibinfo {title} {Ordered conformation of xanthan in solutions and “weak gels”: Single helix, double helix–or both?}}\ }\href {https://doi.org/10.1016/j.foodhyd.2017.11.036} {\bibfield  {journal} {\bibinfo  {journal} {Food Hydrocoll.}\ }\textbf {\bibinfo {volume} {86}},\ \bibinfo {pages} {18--25} (\bibinfo {year} {2019})}\BibitemShut {NoStop}%
\bibitem [{\citenamefont {Sato}\ \emph {et~al.}(1984{\natexlab{a}})\citenamefont {Sato}, \citenamefont {Norisuye},\ and\ \citenamefont {Fujita}}]{Fujita1984scattering}%
  \BibitemOpen
  \bibfield  {author} {\bibinfo {author} {\bibfnamefont {Takahiro}\ \bibnamefont {Sato}}, \bibinfo {author} {\bibfnamefont {Takashi}\ \bibnamefont {Norisuye}}, \ and\ \bibinfo {author} {\bibfnamefont {Hiroshi}\ \bibnamefont {Fujita}},\ }\bibfield  {title} {\enquote {\bibinfo {title} {Double-stranded helix of xanthan in dilute solution: Evidence from light scattering},}\ }\href {https://doi.org/10.1295/polymj.16.341} {\bibfield  {journal} {\bibinfo  {journal} {Polym. J.}\ }\textbf {\bibinfo {volume} {16}},\ \bibinfo {pages} {341--350} (\bibinfo {year} {1984}{\natexlab{a}})}\BibitemShut {NoStop}%
\bibitem [{\citenamefont {Sato}\ \emph {et~al.}(1984{\natexlab{b}})\citenamefont {Sato}, \citenamefont {Norisuye},\ and\ \citenamefont {Fujita}}]{Fujita1984}%
  \BibitemOpen
  \bibfield  {author} {\bibinfo {author} {\bibfnamefont {Takahiro}\ \bibnamefont {Sato}}, \bibinfo {author} {\bibfnamefont {Takashi}\ \bibnamefont {Norisuye}}, \ and\ \bibinfo {author} {\bibfnamefont {Hiroshi}\ \bibnamefont {Fujita}},\ }\bibfield  {title} {\enquote {\bibinfo {title} {Double-stranded helix of xanthan: dimensional and hydrodynamic properties in 0.1 m aqueous sodium chloride},}\ }\href {https://doi.org/10.1021/ma00142a043} {\bibfield  {journal} {\bibinfo  {journal} {Macromolecules}\ }\textbf {\bibinfo {volume} {17}},\ \bibinfo {pages} {2696--2700} (\bibinfo {year} {1984}{\natexlab{b}})}\BibitemShut {NoStop}%
\bibitem [{\citenamefont {Holzwarth}(1978)}]{Holzwarth1978}%
  \BibitemOpen
  \bibfield  {author} {\bibinfo {author} {\bibfnamefont {George}\ \bibnamefont {Holzwarth}},\ }\bibfield  {title} {\enquote {\bibinfo {title} {Molecular weight of xanthan polysaccharide},}\ }\href {https://doi.org/10.1016/S0008-6215(00)83250-4} {\bibfield  {journal} {\bibinfo  {journal} {Carbohydr. Res.}\ }\textbf {\bibinfo {volume} {66}},\ \bibinfo {pages} {173--186} (\bibinfo {year} {1978})}\BibitemShut {NoStop}%
\bibitem [{\citenamefont {Rodd}\ \emph {et~al.}(2000)\citenamefont {Rodd}, \citenamefont {Dunstan},\ and\ \citenamefont {Boger}}]{Boger2000}%
  \BibitemOpen
  \bibfield  {author} {\bibinfo {author} {\bibfnamefont {AB}~\bibnamefont {Rodd}}, \bibinfo {author} {\bibfnamefont {DE}~\bibnamefont {Dunstan}}, \ and\ \bibinfo {author} {\bibfnamefont {DV}~\bibnamefont {Boger}},\ }\bibfield  {title} {\enquote {\bibinfo {title} {Characterisation of xanthan gum solutions using dynamic light scattering and rheology},}\ }\href {https://doi.org/10.1016/S0144-8617(99)00156-3} {\bibfield  {journal} {\bibinfo  {journal} {Carbohydr. Polym.}\ }\textbf {\bibinfo {volume} {42}},\ \bibinfo {pages} {159--174} (\bibinfo {year} {2000})}\BibitemShut {NoStop}%
\bibitem [{\citenamefont {Tinland}\ and\ \citenamefont {Rinaudo}(1989)}]{Rinaudo1989}%
  \BibitemOpen
  \bibfield  {author} {\bibinfo {author} {\bibfnamefont {B}~\bibnamefont {Tinland}}\ and\ \bibinfo {author} {\bibfnamefont {M}~\bibnamefont {Rinaudo}},\ }\bibfield  {title} {\enquote {\bibinfo {title} {Dependence of the stiffness of the xanthan chain on the external salt concentration},}\ }\href {https://doi.org/10.1021/ma00194a058} {\bibfield  {journal} {\bibinfo  {journal} {Macromolecules}\ }\textbf {\bibinfo {volume} {22}},\ \bibinfo {pages} {1863--1865} (\bibinfo {year} {1989})}\BibitemShut {NoStop}%
\bibitem [{\citenamefont {Higiro}\ \emph {et~al.}(2007)\citenamefont {Higiro}, \citenamefont {Herald}, \citenamefont {Alavi},\ and\ \citenamefont {Bean}}]{Bean2007}%
  \BibitemOpen
  \bibfield  {author} {\bibinfo {author} {\bibfnamefont {J}~\bibnamefont {Higiro}}, \bibinfo {author} {\bibfnamefont {TJ}~\bibnamefont {Herald}}, \bibinfo {author} {\bibfnamefont {S}~\bibnamefont {Alavi}}, \ and\ \bibinfo {author} {\bibfnamefont {Scott}\ \bibnamefont {Bean}},\ }\bibfield  {title} {\enquote {\bibinfo {title} {Rheological study of xanthan and locust bean gum interaction in dilute solution: Effect of salt},}\ }\href {https://doi.org/10.1016/j.foodres.2006.02.002} {\bibfield  {journal} {\bibinfo  {journal} {Biotechnol. Adv.}\ }\textbf {\bibinfo {volume} {40}},\ \bibinfo {pages} {435--447} (\bibinfo {year} {2007})}\BibitemShut {NoStop}%
\bibitem [{\citenamefont {Brunchi}\ \emph {et~al.}(2014)\citenamefont {Brunchi}, \citenamefont {Morariu},\ and\ \citenamefont {Bercea}}]{Bercea2014}%
  \BibitemOpen
  \bibfield  {author} {\bibinfo {author} {\bibfnamefont {Cristina-Eliza}\ \bibnamefont {Brunchi}}, \bibinfo {author} {\bibfnamefont {Simona}\ \bibnamefont {Morariu}}, \ and\ \bibinfo {author} {\bibfnamefont {Maria}\ \bibnamefont {Bercea}},\ }\bibfield  {title} {\enquote {\bibinfo {title} {Intrinsic viscosity and conformational parameters of xanthan in aqueous solutions: Salt addition effect},}\ }\href {https://doi.org/10.1016/j.colsurfb.2014.07.023} {\bibfield  {journal} {\bibinfo  {journal} {Colloids Surf. B Biointerfaces}\ }\textbf {\bibinfo {volume} {122}},\ \bibinfo {pages} {512--519} (\bibinfo {year} {2014})}\BibitemShut {NoStop}%
\bibitem [{\citenamefont {Jimenez}\ \emph {et~al.}(2020)\citenamefont {Jimenez}, \citenamefont {Martínez~Narváez},\ and\ \citenamefont {Sharma}}]{Sharma2020pof}%
  \BibitemOpen
  \bibfield  {author} {\bibinfo {author} {\bibfnamefont {Leidy~Nallely}\ \bibnamefont {Jimenez}}, \bibinfo {author} {\bibfnamefont {Carina~DV}\ \bibnamefont {Martínez~Narváez}}, \ and\ \bibinfo {author} {\bibfnamefont {Vivek}\ \bibnamefont {Sharma}},\ }\bibfield  {title} {\enquote {\bibinfo {title} {Capillary breakup and extensional rheology response of food thickener cellulose gum (nacmc) in salt-free and excess salt solutions},}\ }\href {https://doi.org/10.1063/1.5128254} {\bibfield  {journal} {\bibinfo  {journal} {Phys. Fluids}\ }\textbf {\bibinfo {volume} {32}},\ \bibinfo {pages} {012113} (\bibinfo {year} {2020})}\BibitemShut {NoStop}%
\bibitem [{\citenamefont {Haward}\ \emph {et~al.}(2012)\citenamefont {Haward}, \citenamefont {Sharma}, \citenamefont {Butts}, \citenamefont {McKinley},\ and\ \citenamefont {Rahatekar}}]{Rahatekar2012}%
  \BibitemOpen
  \bibfield  {author} {\bibinfo {author} {\bibfnamefont {Simon~J}\ \bibnamefont {Haward}}, \bibinfo {author} {\bibfnamefont {Vivek}\ \bibnamefont {Sharma}}, \bibinfo {author} {\bibfnamefont {Craig~P}\ \bibnamefont {Butts}}, \bibinfo {author} {\bibfnamefont {Gareth~H}\ \bibnamefont {McKinley}}, \ and\ \bibinfo {author} {\bibfnamefont {Sameer~S}\ \bibnamefont {Rahatekar}},\ }\bibfield  {title} {\enquote {\bibinfo {title} {Shear and extensional rheology of cellulose/ionic liquid solutions},}\ }\href {https://doi.org/10.1021/bm300407q} {\bibfield  {journal} {\bibinfo  {journal} {Biomacromolecules}\ }\textbf {\bibinfo {volume} {13}},\ \bibinfo {pages} {1688--1699} (\bibinfo {year} {2012})}\BibitemShut {NoStop}%
\bibitem [{\citenamefont {Cussuol}\ \emph {et~al.}(2023)\citenamefont {Cussuol}, \citenamefont {Soares}, \citenamefont {Siqueira}, \citenamefont {Moreira},\ and\ \citenamefont {Dalmaschio}}]{cussuol2023polymer}%
  \BibitemOpen
  \bibfield  {author} {\bibinfo {author} {\bibfnamefont {Jordan~D}\ \bibnamefont {Cussuol}}, \bibinfo {author} {\bibfnamefont {Edson~J}\ \bibnamefont {Soares}}, \bibinfo {author} {\bibfnamefont {Renato~N}\ \bibnamefont {Siqueira}}, \bibinfo {author} {\bibfnamefont {Kelly~CCSR}\ \bibnamefont {Moreira}}, \ and\ \bibinfo {author} {\bibfnamefont {Cleocir~J}\ \bibnamefont {Dalmaschio}},\ }\bibfield  {title} {\enquote {\bibinfo {title} {Polymer drag reduction regeneration},}\ }\href {https://doi.org/10.1016/j.jnnfm.2023.105126} {\bibfield  {journal} {\bibinfo  {journal} {J. Nonnewton. Fluid Mech.}\ }\textbf {\bibinfo {volume} {321}},\ \bibinfo {pages} {105126} (\bibinfo {year} {2023})}\BibitemShut {NoStop}%
\bibitem [{\citenamefont {Larson}\ and\ \citenamefont {Desai}(2015)}]{larson2015modeling}%
  \BibitemOpen
  \bibfield  {author} {\bibinfo {author} {\bibfnamefont {Ronald~G}\ \bibnamefont {Larson}}\ and\ \bibinfo {author} {\bibfnamefont {Priyanka~S}\ \bibnamefont {Desai}},\ }\bibfield  {title} {\enquote {\bibinfo {title} {Modeling the rheology of polymer melts and solutions},}\ }\href {https://doi.org/10.1146/annurev-fluid-010814-014612} {\bibfield  {journal} {\bibinfo  {journal} {Annu. Rev. Fluid Mech.}\ }\textbf {\bibinfo {volume} {47}},\ \bibinfo {pages} {47--65} (\bibinfo {year} {2015})}\BibitemShut {NoStop}%
\bibitem [{\citenamefont {Millero}\ \emph {et~al.}(2008)\citenamefont {Millero}, \citenamefont {Feistel}, \citenamefont {Wright},\ and\ \citenamefont {McDougall}}]{McDougall2008}%
  \BibitemOpen
  \bibfield  {author} {\bibinfo {author} {\bibfnamefont {Frank~J}\ \bibnamefont {Millero}}, \bibinfo {author} {\bibfnamefont {Rainer}\ \bibnamefont {Feistel}}, \bibinfo {author} {\bibfnamefont {Daniel~G}\ \bibnamefont {Wright}}, \ and\ \bibinfo {author} {\bibfnamefont {Trevor~J}\ \bibnamefont {McDougall}},\ }\bibfield  {title} {\enquote {\bibinfo {title} {The composition of standard seawater and the definition of the reference-composition salinity scale},}\ }\href {https://doi.org/10.1016/j.dsr.2007.10.001} {\bibfield  {journal} {\bibinfo  {journal} {Deep Sea Res. 1 Oceanogr. Res. Pap.}\ }\textbf {\bibinfo {volume} {55}},\ \bibinfo {pages} {50--72} (\bibinfo {year} {2008})}\BibitemShut {NoStop}%
\bibitem [{\citenamefont {Rodd}\ \emph {et~al.}(2005)\citenamefont {Rodd}, \citenamefont {Scott}, \citenamefont {Cooper-White},\ and\ \citenamefont {McKinley}}]{McKinley2005}%
  \BibitemOpen
  \bibfield  {author} {\bibinfo {author} {\bibfnamefont {Lucy~E}\ \bibnamefont {Rodd}}, \bibinfo {author} {\bibfnamefont {Timothy~P}\ \bibnamefont {Scott}}, \bibinfo {author} {\bibfnamefont {Justin~J}\ \bibnamefont {Cooper-White}}, \ and\ \bibinfo {author} {\bibfnamefont {Gareth~H}\ \bibnamefont {McKinley}},\ }\bibfield  {title} {\enquote {\bibinfo {title} {Capillary break-up rheometry of low-viscosity elastic fluids},}\ }\href {https://doi.org/10.1515/arh-2005-0001} {\bibfield  {journal} {\bibinfo  {journal} {Appl. Rheol.}\ }\textbf {\bibinfo {volume} {15}},\ \bibinfo {pages} {12--27} (\bibinfo {year} {2005})}\BibitemShut {NoStop}%
\bibitem [{\citenamefont {Clasen}\ \emph {et~al.}(2006)\citenamefont {Clasen}, \citenamefont {Plog}, \citenamefont {Kulicke}, \citenamefont {Owens}, \citenamefont {Macosko}, \citenamefont {Scriven}, \citenamefont {Verani},\ and\ \citenamefont {McKinley}}]{clasen2006dilute}%
  \BibitemOpen
  \bibfield  {author} {\bibinfo {author} {\bibfnamefont {Christian}\ \bibnamefont {Clasen}}, \bibinfo {author} {\bibfnamefont {JP}~\bibnamefont {Plog}}, \bibinfo {author} {\bibfnamefont {W-M}\ \bibnamefont {Kulicke}}, \bibinfo {author} {\bibfnamefont {M}~\bibnamefont {Owens}}, \bibinfo {author} {\bibfnamefont {Chris}\ \bibnamefont {Macosko}}, \bibinfo {author} {\bibfnamefont {LE}~\bibnamefont {Scriven}}, \bibinfo {author} {\bibfnamefont {M}~\bibnamefont {Verani}}, \ and\ \bibinfo {author} {\bibfnamefont {Gareth~H}\ \bibnamefont {McKinley}},\ }\bibfield  {title} {\enquote {\bibinfo {title} {How dilute are dilute solutions in extensional flows?}}\ }\href {https://doi.org/10.1122/1.2357595} {\bibfield  {journal} {\bibinfo  {journal} {J. Rheol.}\ }\textbf {\bibinfo {volume} {50}},\ \bibinfo {pages} {849--881} (\bibinfo {year} {2006})}\BibitemShut {NoStop}%
\bibitem [{\citenamefont {Clasen}\ \emph {et~al.}(2012)\citenamefont {Clasen}, \citenamefont {Phillips}, \citenamefont {Palangetic},\ and\ \citenamefont {Vermant}}]{clasen2012dispensing}%
  \BibitemOpen
  \bibfield  {author} {\bibinfo {author} {\bibfnamefont {Christian}\ \bibnamefont {Clasen}}, \bibinfo {author} {\bibfnamefont {Paul~M}\ \bibnamefont {Phillips}}, \bibinfo {author} {\bibfnamefont {Ljiljana}\ \bibnamefont {Palangetic}}, \ and\ \bibinfo {author} {\bibfnamefont {Jan}\ \bibnamefont {Vermant}},\ }\bibfield  {title} {\enquote {\bibinfo {title} {Dispensing of rheologically complex fluids: The map of misery},}\ }\href {https://doi.org/10.1002/aic.13704} {\bibfield  {journal} {\bibinfo  {journal} {AIChE J.}\ }\textbf {\bibinfo {volume} {58}},\ \bibinfo {pages} {3242--3255} (\bibinfo {year} {2012})}\BibitemShut {NoStop}%
\bibitem [{\citenamefont {Dobrynin}\ \emph {et~al.}(1995)\citenamefont {Dobrynin}, \citenamefont {Colby},\ and\ \citenamefont {Rubinstein}}]{Rubinstein1995}%
  \BibitemOpen
  \bibfield  {author} {\bibinfo {author} {\bibfnamefont {Andrey~V}\ \bibnamefont {Dobrynin}}, \bibinfo {author} {\bibfnamefont {Ralph~H}\ \bibnamefont {Colby}}, \ and\ \bibinfo {author} {\bibfnamefont {Michael}\ \bibnamefont {Rubinstein}},\ }\bibfield  {title} {\enquote {\bibinfo {title} {Scaling theory of polyelectrolyte solutions},}\ }\href {https://doi.org/10.1021/ma00110a021} {\bibfield  {journal} {\bibinfo  {journal} {Macromolecules}\ }\textbf {\bibinfo {volume} {28}},\ \bibinfo {pages} {1859--1871} (\bibinfo {year} {1995})}\BibitemShut {NoStop}%
\bibitem [{\citenamefont {De~Gennes}(1979)}]{DeGennes1979}%
  \BibitemOpen
  \bibfield  {author} {\bibinfo {author} {\bibfnamefont {Pierre-Gilles}\ \bibnamefont {De~Gennes}},\ }\href@noop {} {\emph {\bibinfo {title} {Scaling concepts in polymer physics}}}\ (\bibinfo  {publisher} {Cornell University Press},\ \bibinfo {year} {1979})\BibitemShut {NoStop}%
\bibitem [{\citenamefont {Ying}\ and\ \citenamefont {Chu}(1987)}]{Chu1987}%
  \BibitemOpen
  \bibfield  {author} {\bibinfo {author} {\bibfnamefont {Qicong}\ \bibnamefont {Ying}}\ and\ \bibinfo {author} {\bibfnamefont {Benjamin}\ \bibnamefont {Chu}},\ }\bibfield  {title} {\enquote {\bibinfo {title} {Overlap concentration of macromolecules in solution},}\ }\href {https://doi.org/10.1021/ma00168a023} {\bibfield  {journal} {\bibinfo  {journal} {Macromolecules}\ }\textbf {\bibinfo {volume} {20}},\ \bibinfo {pages} {362--366} (\bibinfo {year} {1987})}\BibitemShut {NoStop}%
\bibitem [{\citenamefont {Doi}\ \emph {et~al.}(1988)\citenamefont {Doi}, \citenamefont {Edwards},\ and\ \citenamefont {Edwards}}]{Edwards1988}%
  \BibitemOpen
  \bibfield  {author} {\bibinfo {author} {\bibfnamefont {Masao}\ \bibnamefont {Doi}}, \bibinfo {author} {\bibfnamefont {Sam~F}\ \bibnamefont {Edwards}}, \ and\ \bibinfo {author} {\bibfnamefont {Samuel~Frederick}\ \bibnamefont {Edwards}},\ }\href@noop {} {\emph {\bibinfo {title} {The theory of polymer dynamics}}},\ Vol.~\bibinfo {volume} {73}\ (\bibinfo  {publisher} {Oxford University Press},\ \bibinfo {year} {1988})\BibitemShut {NoStop}%
\bibitem [{\citenamefont {Dobrynin}\ and\ \citenamefont {Rubinstein}(2005)}]{Rubinstein2005}%
  \BibitemOpen
  \bibfield  {author} {\bibinfo {author} {\bibfnamefont {Andrey~V}\ \bibnamefont {Dobrynin}}\ and\ \bibinfo {author} {\bibfnamefont {Michael}\ \bibnamefont {Rubinstein}},\ }\bibfield  {title} {\enquote {\bibinfo {title} {Theory of polyelectrolytes in solutions and at surfaces},}\ }\href {https://doi.org/10.1016/j.progpolymsci.2005.07.006} {\bibfield  {journal} {\bibinfo  {journal} {Prog. Polym. Sci.}\ }\textbf {\bibinfo {volume} {30}},\ \bibinfo {pages} {1049--1118} (\bibinfo {year} {2005})}\BibitemShut {NoStop}%
\bibitem [{\citenamefont {Coviello}\ \emph {et~al.}(1986)\citenamefont {Coviello}, \citenamefont {Kajiwara}, \citenamefont {Burchard}, \citenamefont {Dentini},\ and\ \citenamefont {Crescenzi}}]{coviello1986solution}%
  \BibitemOpen
  \bibfield  {author} {\bibinfo {author} {\bibfnamefont {Tommasina}\ \bibnamefont {Coviello}}, \bibinfo {author} {\bibfnamefont {K}~\bibnamefont {Kajiwara}}, \bibinfo {author} {\bibfnamefont {W}~\bibnamefont {Burchard}}, \bibinfo {author} {\bibfnamefont {Mariella}\ \bibnamefont {Dentini}}, \ and\ \bibinfo {author} {\bibfnamefont {Vittorio}\ \bibnamefont {Crescenzi}},\ }\bibfield  {title} {\enquote {\bibinfo {title} {Solution properties of xanthan. 1. dynamic and static light scattering from native and modified xanthans in dilute solutions},}\ }\href {https://doi.org/10.1021/ma00165a027} {\bibfield  {journal} {\bibinfo  {journal} {Macromolecules}\ }\textbf {\bibinfo {volume} {19}},\ \bibinfo {pages} {2826--2831} (\bibinfo {year} {1986})}\BibitemShut {NoStop}%
\bibitem [{\citenamefont {Camesano}\ and\ \citenamefont {Wilkinson}(2001)}]{camesano2001single}%
  \BibitemOpen
  \bibfield  {author} {\bibinfo {author} {\bibfnamefont {Terri~A}\ \bibnamefont {Camesano}}\ and\ \bibinfo {author} {\bibfnamefont {Kevin~J}\ \bibnamefont {Wilkinson}},\ }\bibfield  {title} {\enquote {\bibinfo {title} {Single molecule study of xanthan conformation using atomic force microscopy},}\ }\href {https://doi.org/10.1021/bm015555g} {\bibfield  {journal} {\bibinfo  {journal} {Biomacromolecules}\ }\textbf {\bibinfo {volume} {2}},\ \bibinfo {pages} {1184--1191} (\bibinfo {year} {2001})}\BibitemShut {NoStop}%
\bibitem [{\citenamefont {Macosko}(1994)}]{Macosko1994}%
  \BibitemOpen
  \bibfield  {author} {\bibinfo {author} {\bibfnamefont {Christopher~W}\ \bibnamefont {Macosko}},\ }\href@noop {} {\emph {\bibinfo {title} {Rheology: Principles, Measurements, and Applications}}}\ (\bibinfo  {publisher} {Wiley},\ \bibinfo {year} {1994})\BibitemShut {NoStop}%
\bibitem [{\citenamefont {Ewoldt}\ \emph {et~al.}(2015)\citenamefont {Ewoldt}, \citenamefont {Johnston},\ and\ \citenamefont {Caretta}}]{Caretta2015}%
  \BibitemOpen
  \bibfield  {author} {\bibinfo {author} {\bibfnamefont {Randy~H}\ \bibnamefont {Ewoldt}}, \bibinfo {author} {\bibfnamefont {Michael~T}\ \bibnamefont {Johnston}}, \ and\ \bibinfo {author} {\bibfnamefont {Lucas~M}\ \bibnamefont {Caretta}},\ }\enquote {\bibinfo {title} {Complex fluids in biological systems: Experiment, theory, and computation},}\ \ (\bibinfo  {publisher} {Springer},\ \bibinfo {year} {2015})\ Chap.\ \bibinfo {chapter} {Experimental challenges of shear rheology: How to avoid bad data}, pp.\ \bibinfo {pages} {207--241}\BibitemShut {NoStop}%
\bibitem [{\citenamefont {Dinic}\ \emph {et~al.}(2015)\citenamefont {Dinic}, \citenamefont {Zhang}, \citenamefont {Jimenez},\ and\ \citenamefont {Sharma}}]{Sharma2015}%
  \BibitemOpen
  \bibfield  {author} {\bibinfo {author} {\bibfnamefont {Jelena}\ \bibnamefont {Dinic}}, \bibinfo {author} {\bibfnamefont {Yiran}\ \bibnamefont {Zhang}}, \bibinfo {author} {\bibfnamefont {Leidy~Nallely}\ \bibnamefont {Jimenez}}, \ and\ \bibinfo {author} {\bibfnamefont {Vivek}\ \bibnamefont {Sharma}},\ }\bibfield  {title} {\enquote {\bibinfo {title} {Extensional relaxation times of dilute, aqueous polymer solutions},}\ }\href {https://doi.org/10.1021/acsmacrolett.5b00393} {\bibfield  {journal} {\bibinfo  {journal} {ACS Macro Lett.}\ }\textbf {\bibinfo {volume} {4}},\ \bibinfo {pages} {804--808} (\bibinfo {year} {2015})}\BibitemShut {NoStop}%
\bibitem [{\citenamefont {Dinic}\ \emph {et~al.}(2017)\citenamefont {Dinic}, \citenamefont {Jimenez},\ and\ \citenamefont {Sharma}}]{Sharma2017}%
  \BibitemOpen
  \bibfield  {author} {\bibinfo {author} {\bibfnamefont {Jelena}\ \bibnamefont {Dinic}}, \bibinfo {author} {\bibfnamefont {Leidy~Nallely}\ \bibnamefont {Jimenez}}, \ and\ \bibinfo {author} {\bibfnamefont {Vivek}\ \bibnamefont {Sharma}},\ }\bibfield  {title} {\enquote {\bibinfo {title} {Pinch-off dynamics and dripping-onto-substrate (dos) rheometry of complex fluids},}\ }\href {https://doi.org/10.1039/C6LC01155A} {\bibfield  {journal} {\bibinfo  {journal} {Lab Chip}\ }\textbf {\bibinfo {volume} {17}},\ \bibinfo {pages} {460--473} (\bibinfo {year} {2017})}\BibitemShut {NoStop}%
\bibitem [{\citenamefont {Doshi}\ \emph {et~al.}(2003)\citenamefont {Doshi}, \citenamefont {Suryo}, \citenamefont {Yildirim}, \citenamefont {McKinley},\ and\ \citenamefont {Basaran}}]{McKinley2003}%
  \BibitemOpen
  \bibfield  {author} {\bibinfo {author} {\bibfnamefont {Pankaj}\ \bibnamefont {Doshi}}, \bibinfo {author} {\bibfnamefont {Ronald}\ \bibnamefont {Suryo}}, \bibinfo {author} {\bibfnamefont {Ozgur~E}\ \bibnamefont {Yildirim}}, \bibinfo {author} {\bibfnamefont {Gareth~H}\ \bibnamefont {McKinley}}, \ and\ \bibinfo {author} {\bibfnamefont {Osman~A}\ \bibnamefont {Basaran}},\ }\bibfield  {title} {\enquote {\bibinfo {title} {Scaling in pinch-off of generalized newtonian fluids},}\ }\href {https://doi.org/10.1016/S0377-0257(03)00081-8} {\bibfield  {journal} {\bibinfo  {journal} {J. Nonnewton. Fluid Mech.}\ }\textbf {\bibinfo {volume} {113}},\ \bibinfo {pages} {1--27} (\bibinfo {year} {2003})}\BibitemShut {NoStop}%
\bibitem [{\citenamefont {Zhang}\ and\ \citenamefont {Calabrese}(2022)}]{Calabrese2022}%
  \BibitemOpen
  \bibfield  {author} {\bibinfo {author} {\bibfnamefont {Diana~Y}\ \bibnamefont {Zhang}}\ and\ \bibinfo {author} {\bibfnamefont {Michelle~A}\ \bibnamefont {Calabrese}},\ }\bibfield  {title} {\enquote {\bibinfo {title} {Temperature-controlled dripping-onto-substrate (dos) extensional rheometry of polymer micelle solutions},}\ }\href {https://doi.org/10.1039/D2SM00377E} {\bibfield  {journal} {\bibinfo  {journal} {Soft Matter}\ }\textbf {\bibinfo {volume} {18}},\ \bibinfo {pages} {3993--4008} (\bibinfo {year} {2022})}\BibitemShut {NoStop}%
\bibitem [{\citenamefont {Mart\'{i}nez~Narv\'{a}ez}\ \emph {et~al.}(2021)\citenamefont {Mart\'{i}nez~Narv\'{a}ez}, \citenamefont {Dinic}, \citenamefont {Lu}, \citenamefont {Wang}, \citenamefont {Rock}, \citenamefont {Sun},\ and\ \citenamefont {Sharma}}]{Sharma2021}%
  \BibitemOpen
  \bibfield  {author} {\bibinfo {author} {\bibfnamefont {Carina~DV}\ \bibnamefont {Mart\'{i}nez~Narv\'{a}ez}}, \bibinfo {author} {\bibfnamefont {Jelena}\ \bibnamefont {Dinic}}, \bibinfo {author} {\bibfnamefont {Xinyu}\ \bibnamefont {Lu}}, \bibinfo {author} {\bibfnamefont {Chao}\ \bibnamefont {Wang}}, \bibinfo {author} {\bibfnamefont {Reza}\ \bibnamefont {Rock}}, \bibinfo {author} {\bibfnamefont {Hao}\ \bibnamefont {Sun}}, \ and\ \bibinfo {author} {\bibfnamefont {Vivek}\ \bibnamefont {Sharma}},\ }\bibfield  {title} {\enquote {\bibinfo {title} {Rheology and pinching dynamics of associative polysaccharide solutions},}\ }\href {https://doi.org/10.1021/acs.macromol.0c02751} {\bibfield  {journal} {\bibinfo  {journal} {Macromolecules}\ }\textbf {\bibinfo {volume} {54}},\ \bibinfo {pages} {6372--6388} (\bibinfo {year} {2021})}\BibitemShut {NoStop}%
\bibitem [{\citenamefont {Day}\ \emph {et~al.}(1998)\citenamefont {Day}, \citenamefont {Hinch},\ and\ \citenamefont {Lister}}]{day1998self}%
  \BibitemOpen
  \bibfield  {author} {\bibinfo {author} {\bibfnamefont {Richard~F}\ \bibnamefont {Day}}, \bibinfo {author} {\bibfnamefont {E~John}\ \bibnamefont {Hinch}}, \ and\ \bibinfo {author} {\bibfnamefont {John~R}\ \bibnamefont {Lister}},\ }\bibfield  {title} {\enquote {\bibinfo {title} {Self-similar capillary pinchoff of an inviscid fluid},}\ }\href {https://doi.org/10.1103/PhysRevLett.80.704} {\bibfield  {journal} {\bibinfo  {journal} {Phys. Rev. Lett.}\ }\textbf {\bibinfo {volume} {80}},\ \bibinfo {pages} {704} (\bibinfo {year} {1998})}\BibitemShut {NoStop}%
\bibitem [{\citenamefont {Dinic}\ and\ \citenamefont {Sharma}(2019)}]{Sharma2019}%
  \BibitemOpen
  \bibfield  {author} {\bibinfo {author} {\bibfnamefont {Jelena}\ \bibnamefont {Dinic}}\ and\ \bibinfo {author} {\bibfnamefont {Vivek}\ \bibnamefont {Sharma}},\ }\bibfield  {title} {\enquote {\bibinfo {title} {Macromolecular relaxation, strain, and extensibility determine elastocapillary thinning and extensional viscosity of polymer solutions},}\ }\href {https://doi.org/10.1073/pnas.1820277116} {\bibfield  {journal} {\bibinfo  {journal} {Proc. Natl. Acad. Sci. U. S. A.}\ }\textbf {\bibinfo {volume} {116}},\ \bibinfo {pages} {8766--8774} (\bibinfo {year} {2019})}\BibitemShut {NoStop}%
\bibitem [{\citenamefont {Zinelis}\ \emph {et~al.}(2024)\citenamefont {Zinelis}, \citenamefont {Abadie}, \citenamefont {McKinley},\ and\ \citenamefont {Matar}}]{McKinley2024}%
  \BibitemOpen
  \bibfield  {author} {\bibinfo {author} {\bibfnamefont {Konstantinos}\ \bibnamefont {Zinelis}}, \bibinfo {author} {\bibfnamefont {Thomas}\ \bibnamefont {Abadie}}, \bibinfo {author} {\bibfnamefont {Gareth~H}\ \bibnamefont {McKinley}}, \ and\ \bibinfo {author} {\bibfnamefont {Omar}\ \bibnamefont {Matar}},\ }\bibfield  {title} {\enquote {\bibinfo {title} {The fluid dynamics of a viscoelastic fluid dripping onto a substrate},}\ }\href {https://doi.org/10.1039/D4SM00406J} {\bibfield  {journal} {\bibinfo  {journal} {Soft Matter}\ } (\bibinfo {year} {2024})}\BibitemShut {NoStop}%
\bibitem [{\citenamefont {Deblais}\ \emph {et~al.}(2018)\citenamefont {Deblais}, \citenamefont {Herrada}, \citenamefont {Hauner}, \citenamefont {Velikov}, \citenamefont {Van~Roon}, \citenamefont {Kellay}, \citenamefont {Eggers},\ and\ \citenamefont {Bonn}}]{Bonn2018}%
  \BibitemOpen
  \bibfield  {author} {\bibinfo {author} {\bibfnamefont {Antoine}\ \bibnamefont {Deblais}}, \bibinfo {author} {\bibfnamefont {MA}~\bibnamefont {Herrada}}, \bibinfo {author} {\bibfnamefont {Ines}\ \bibnamefont {Hauner}}, \bibinfo {author} {\bibfnamefont {Krassimir~P}\ \bibnamefont {Velikov}}, \bibinfo {author} {\bibfnamefont {T}~\bibnamefont {Van~Roon}}, \bibinfo {author} {\bibfnamefont {Hamid}\ \bibnamefont {Kellay}}, \bibinfo {author} {\bibfnamefont {Jens}\ \bibnamefont {Eggers}}, \ and\ \bibinfo {author} {\bibfnamefont {Daniel}\ \bibnamefont {Bonn}},\ }\bibfield  {title} {\enquote {\bibinfo {title} {Viscous effects on inertial drop formation},}\ }\href {https://doi.org/10.1103/PhysRevLett.121.254501} {\bibfield  {journal} {\bibinfo  {journal} {Phys. Rev. Lett.}\ }\textbf {\bibinfo {volume} {121}},\ \bibinfo {pages} {254501} (\bibinfo {year} {2018})}\BibitemShut {NoStop}%
\bibitem [{\citenamefont {Norisuye}(1993)}]{norisuye1993semiflexible}%
  \BibitemOpen
  \bibfield  {author} {\bibinfo {author} {\bibfnamefont {Takashi}\ \bibnamefont {Norisuye}},\ }\bibfield  {title} {\enquote {\bibinfo {title} {Semiflexible polymers in dilute solution},}\ }\href {https://doi.org/10.1016/0079-6700(93)90017-7} {\bibfield  {journal} {\bibinfo  {journal} {Prog. Polym. Sci.}\ }\textbf {\bibinfo {volume} {18}},\ \bibinfo {pages} {543--584} (\bibinfo {year} {1993})}\BibitemShut {NoStop}%
\bibitem [{\citenamefont {Dinic}\ and\ \citenamefont {Sharma}(2020)}]{Sharma2020}%
  \BibitemOpen
  \bibfield  {author} {\bibinfo {author} {\bibfnamefont {Jelena}\ \bibnamefont {Dinic}}\ and\ \bibinfo {author} {\bibfnamefont {Vivek}\ \bibnamefont {Sharma}},\ }\bibfield  {title} {\enquote {\bibinfo {title} {Flexibility, extensibility, and ratio of kuhn length to packing length govern the pinching dynamics, coil-stretch transition, and rheology of polymer solutions},}\ }\href {https://doi.org/10.1021/acs.macromol.0c00076} {\bibfield  {journal} {\bibinfo  {journal} {Macromolecules}\ }\textbf {\bibinfo {volume} {53}},\ \bibinfo {pages} {4821--4835} (\bibinfo {year} {2020})}\BibitemShut {NoStop}%
\bibitem [{\citenamefont {Fox}\ and\ \citenamefont {Weisberg}(2018)}]{fox2018r}%
  \BibitemOpen
  \bibfield  {author} {\bibinfo {author} {\bibfnamefont {John}\ \bibnamefont {Fox}}\ and\ \bibinfo {author} {\bibfnamefont {Sanford}\ \bibnamefont {Weisberg}},\ }\href@noop {} {\emph {\bibinfo {title} {An R companion to applied regression}}}\ (\bibinfo  {publisher} {Sage publications},\ \bibinfo {year} {2018})\BibitemShut {NoStop}%
\bibitem [{\citenamefont {Lee}\ \emph {et~al.}(2008)\citenamefont {Lee}, \citenamefont {Venable}, \citenamefont {MacKerell},\ and\ \citenamefont {Pastor}}]{Pastor2008}%
  \BibitemOpen
  \bibfield  {author} {\bibinfo {author} {\bibfnamefont {Hwankyu}\ \bibnamefont {Lee}}, \bibinfo {author} {\bibfnamefont {Richard~M}\ \bibnamefont {Venable}}, \bibinfo {author} {\bibfnamefont {Alexander~D}\ \bibnamefont {MacKerell}}, \ and\ \bibinfo {author} {\bibfnamefont {Richard~W}\ \bibnamefont {Pastor}},\ }\bibfield  {title} {\enquote {\bibinfo {title} {Molecular dynamics studies of polyethylene oxide and polyethylene glycol: hydrodynamic radius and shape anisotropy},}\ }\href {https://doi.org/10.1529/biophysj.108.133025} {\bibfield  {journal} {\bibinfo  {journal} {Biophys. J.}\ }\textbf {\bibinfo {volume} {95}},\ \bibinfo {pages} {1590--1599} (\bibinfo {year} {2008})}\BibitemShut {NoStop}%
\bibitem [{\citenamefont {Sherck}\ \emph {et~al.}(2020)\citenamefont {Sherck}, \citenamefont {Webber}, \citenamefont {Brown}, \citenamefont {Keller}, \citenamefont {Barry}, \citenamefont {DeStefano}, \citenamefont {Jiao}, \citenamefont {Segalman}, \citenamefont {Fredrickson},\ and\ \citenamefont {Shell}}]{Shell2020}%
  \BibitemOpen
  \bibfield  {author} {\bibinfo {author} {\bibfnamefont {Nicholas}\ \bibnamefont {Sherck}}, \bibinfo {author} {\bibfnamefont {Thomas}\ \bibnamefont {Webber}}, \bibinfo {author} {\bibfnamefont {Dennis~Robinson}\ \bibnamefont {Brown}}, \bibinfo {author} {\bibfnamefont {Timothy}\ \bibnamefont {Keller}}, \bibinfo {author} {\bibfnamefont {Mikayla}\ \bibnamefont {Barry}}, \bibinfo {author} {\bibfnamefont {Audra}\ \bibnamefont {DeStefano}}, \bibinfo {author} {\bibfnamefont {Sally}\ \bibnamefont {Jiao}}, \bibinfo {author} {\bibfnamefont {Rachel~A}\ \bibnamefont {Segalman}}, \bibinfo {author} {\bibfnamefont {Glenn~H}\ \bibnamefont {Fredrickson}}, \ and\ \bibinfo {author} {\bibfnamefont {M~Scott}\ \bibnamefont {Shell}},\ }\bibfield  {title} {\enquote {\bibinfo {title} {End-to-end distance probability distributions of dilute poly (ethylene oxide) in aqueous solution},}\ }\href {https://doi.org/10.1021/jacs.0c08709} {\bibfield  {journal} {\bibinfo  {journal} {J. Am. Chem. Soc.}\ }\textbf {\bibinfo {volume} {142}},\
  \bibinfo {pages} {19631--19641} (\bibinfo {year} {2020})}\BibitemShut {NoStop}%
\bibitem [{\citenamefont {De~Gennes}(1974)}]{de1974coil}%
  \BibitemOpen
  \bibfield  {author} {\bibinfo {author} {\bibfnamefont {P~G}\ \bibnamefont {De~Gennes}},\ }\bibfield  {title} {\enquote {\bibinfo {title} {Coil-stretch transition of dilute flexible polymers under ultrahigh velocity gradients},}\ }\href {https://doi.org/10.1063/1.1681018} {\bibfield  {journal} {\bibinfo  {journal} {J.\ Chem.\ Phys.}\ }\textbf {\bibinfo {volume} {60}},\ \bibinfo {pages} {5030--5042} (\bibinfo {year} {1974})}\BibitemShut {NoStop}%
\bibitem [{\citenamefont {McKinley}(2005)}]{mckinley2005visco}%
  \BibitemOpen
  \bibfield  {author} {\bibinfo {author} {\bibfnamefont {Gareth~H}\ \bibnamefont {McKinley}},\ }\bibfield  {title} {\enquote {\bibinfo {title} {Visco-elasto-capillary thinning and break-up of complex fluids},}\ }\href {http://hdl.handle.net/1721.1/18085} {\bibfield  {journal} {\bibinfo  {journal} {Annu. Rheol. Rev.}\ ,\ \bibinfo {pages} {1--48}} (\bibinfo {year} {2005})}\BibitemShut {NoStop}%
\bibitem [{\citenamefont {Entov}\ and\ \citenamefont {Hinch}(1997)}]{entov1997effect}%
  \BibitemOpen
  \bibfield  {author} {\bibinfo {author} {\bibfnamefont {VM}~\bibnamefont {Entov}}\ and\ \bibinfo {author} {\bibfnamefont {EJ}~\bibnamefont {Hinch}},\ }\bibfield  {title} {\enquote {\bibinfo {title} {Effect of a spectrum of relaxation times on the capillary thinning of a filament of elastic liquid},}\ }\href {https://doi.org/10.1016/S0377-0257(97)00022-0} {\bibfield  {journal} {\bibinfo  {journal} {J. Nonnewton. Fluid Mech.}\ }\textbf {\bibinfo {volume} {72}},\ \bibinfo {pages} {31--53} (\bibinfo {year} {1997})}\BibitemShut {NoStop}%
\bibitem [{\citenamefont {Walters}\ \emph {et~al.}(1990)\citenamefont {Walters}, \citenamefont {Bhatti},\ and\ \citenamefont {Mori}}]{walters1990influence}%
  \BibitemOpen
  \bibfield  {author} {\bibinfo {author} {\bibfnamefont {K}~\bibnamefont {Walters}}, \bibinfo {author} {\bibfnamefont {AQ}~\bibnamefont {Bhatti}}, \ and\ \bibinfo {author} {\bibfnamefont {N}~\bibnamefont {Mori}},\ }\enquote {\bibinfo {title} {Recent developments in structured continua},}\ \ (\bibinfo  {publisher} {Pitman},\ \bibinfo {year} {1990})\ Chap.\ \bibinfo {chapter} {The influence of polymer conformation on the rheological properties of aqueous polymer solutions}, pp.\ \bibinfo {pages} {3242--3255}\BibitemShut {NoStop}%
\bibitem [{\citenamefont {Larson}(1999)}]{Larson1999}%
  \BibitemOpen
  \bibfield  {author} {\bibinfo {author} {\bibfnamefont {Ronald~G}\ \bibnamefont {Larson}},\ }\href@noop {} {\emph {\bibinfo {title} {The structure and rheology of complex fluids}}},\ Vol.\ \bibinfo {volume} {150}\ (\bibinfo  {publisher} {Oxford University Press},\ \bibinfo {address} {New York},\ \bibinfo {year} {1999})\BibitemShut {NoStop}%
\bibitem [{\citenamefont {Green}\ and\ \citenamefont {Tobolsky}(1946)}]{green1946new}%
  \BibitemOpen
  \bibfield  {author} {\bibinfo {author} {\bibfnamefont {Melville~S}\ \bibnamefont {Green}}\ and\ \bibinfo {author} {\bibfnamefont {Arthur~V}\ \bibnamefont {Tobolsky}},\ }\bibfield  {title} {\enquote {\bibinfo {title} {A new approach to the theory of relaxing polymeric media},}\ }\href {https://doi.org/10.1063/1.1724109} {\bibfield  {journal} {\bibinfo  {journal} {J. Chem. Phys.}\ }\textbf {\bibinfo {volume} {14}},\ \bibinfo {pages} {80--92} (\bibinfo {year} {1946})}\BibitemShut {NoStop}%
\bibitem [{\citenamefont {Tanaka}\ and\ \citenamefont {Edwards}(1992)}]{tanaka1992viscoelastic}%
  \BibitemOpen
  \bibfield  {author} {\bibinfo {author} {\bibfnamefont {Fumihide}\ \bibnamefont {Tanaka}}\ and\ \bibinfo {author} {\bibfnamefont {Samuel~Frederick}\ \bibnamefont {Edwards}},\ }\bibfield  {title} {\enquote {\bibinfo {title} {Viscoelastic properties of physically crosslinked networks: Part 1. non-linear stationary viscoelasticity},}\ }\href {https://doi.org/10.1016/0377-0257(92)80027-U} {\bibfield  {journal} {\bibinfo  {journal} {J. Nonnewton. Fluid Mech.}\ }\textbf {\bibinfo {volume} {43}},\ \bibinfo {pages} {247--271} (\bibinfo {year} {1992})}\BibitemShut {NoStop}%
\bibitem [{\citenamefont {Tripathi}\ \emph {et~al.}(2006)\citenamefont {Tripathi}, \citenamefont {Tam},\ and\ \citenamefont {McKinley}}]{tripathi2006rheology}%
  \BibitemOpen
  \bibfield  {author} {\bibinfo {author} {\bibfnamefont {Anubhav}\ \bibnamefont {Tripathi}}, \bibinfo {author} {\bibfnamefont {Kam~C}\ \bibnamefont {Tam}}, \ and\ \bibinfo {author} {\bibfnamefont {Gareth~H}\ \bibnamefont {McKinley}},\ }\bibfield  {title} {\enquote {\bibinfo {title} {Rheology and dynamics of associative polymers in shear and extension: Theory and experiments},}\ }\href {https://doi.org/10.1021/ma051614x} {\bibfield  {journal} {\bibinfo  {journal} {Macromolecules}\ }\textbf {\bibinfo {volume} {39}},\ \bibinfo {pages} {1981--1999} (\bibinfo {year} {2006})}\BibitemShut {NoStop}%
\bibitem [{\citenamefont {Schafer}(2011)}]{schafer2011savitzky}%
  \BibitemOpen
  \bibfield  {author} {\bibinfo {author} {\bibfnamefont {Ronald~W}\ \bibnamefont {Schafer}},\ }\bibfield  {title} {\enquote {\bibinfo {title} {What is a savitzky-golay filter?[lecture notes]},}\ }\href {https://doi.org/10.1109/MSP.2011.941097} {\bibfield  {journal} {\bibinfo  {journal} {IEEE Signal Process. Mag.}\ }\textbf {\bibinfo {volume} {28}},\ \bibinfo {pages} {111--117} (\bibinfo {year} {2011})}\BibitemShut {NoStop}%
\bibitem [{\citenamefont {Landau}\ and\ \citenamefont {Lifshitz}(2013)}]{landau2013statistical}%
  \BibitemOpen
  \bibfield  {author} {\bibinfo {author} {\bibfnamefont {Lev~Davidovich}\ \bibnamefont {Landau}}\ and\ \bibinfo {author} {\bibfnamefont {Evgenii~Mikhailovich}\ \bibnamefont {Lifshitz}},\ }\href@noop {} {\emph {\bibinfo {title} {Statistical Physics: Volume 5}}},\ Vol.~\bibinfo {volume} {5}\ (\bibinfo  {publisher} {Elsevier},\ \bibinfo {year} {2013})\BibitemShut {NoStop}%
\bibitem [{\citenamefont {Pincus}\ \emph {et~al.}(2020)\citenamefont {Pincus}, \citenamefont {Rodger},\ and\ \citenamefont {Prakash}}]{pincus2020viscometric}%
  \BibitemOpen
  \bibfield  {author} {\bibinfo {author} {\bibfnamefont {I}~\bibnamefont {Pincus}}, \bibinfo {author} {\bibfnamefont {Alison}\ \bibnamefont {Rodger}}, \ and\ \bibinfo {author} {\bibfnamefont {J~Ravi}\ \bibnamefont {Prakash}},\ }\bibfield  {title} {\enquote {\bibinfo {title} {Viscometric functions and rheo-optical properties of dilute polymer solutions: Comparison of fene-fraenkel dumbbells with rodlike models},}\ }\href {https://doi.org/10.1016/j.jnnfm.2020.104395} {\bibfield  {journal} {\bibinfo  {journal} {J. Nonnewton. Fluid Mech.}\ }\textbf {\bibinfo {volume} {285}},\ \bibinfo {pages} {104395} (\bibinfo {year} {2020})}\BibitemShut {NoStop}%
\bibitem [{\citenamefont {Pincus}\ \emph {et~al.}(2023)\citenamefont {Pincus}, \citenamefont {Rodger},\ and\ \citenamefont {Ravi~Prakash}}]{pincus2023dilute}%
  \BibitemOpen
  \bibfield  {author} {\bibinfo {author} {\bibfnamefont {I}~\bibnamefont {Pincus}}, \bibinfo {author} {\bibfnamefont {A}~\bibnamefont {Rodger}}, \ and\ \bibinfo {author} {\bibfnamefont {J}~\bibnamefont {Ravi~Prakash}},\ }\bibfield  {title} {\enquote {\bibinfo {title} {Dilute polymer solutions under shear flow: Comprehensive qualitative analysis using a bead-spring chain model with a fene-fraenkel spring},}\ }\href {https://doi.org/10.1122/8.0000517} {\bibfield  {journal} {\bibinfo  {journal} {J. Rheol.}\ }\textbf {\bibinfo {volume} {67}},\ \bibinfo {pages} {373--402} (\bibinfo {year} {2023})}\BibitemShut {NoStop}%
\bibitem [{\citenamefont {Tirtaatmadja}\ \emph {et~al.}(2006)\citenamefont {Tirtaatmadja}, \citenamefont {McKinley},\ and\ \citenamefont {Cooper-White}}]{McKinley2006}%
  \BibitemOpen
  \bibfield  {author} {\bibinfo {author} {\bibfnamefont {Viyada}\ \bibnamefont {Tirtaatmadja}}, \bibinfo {author} {\bibfnamefont {Gareth~H}\ \bibnamefont {McKinley}}, \ and\ \bibinfo {author} {\bibfnamefont {Justin~J}\ \bibnamefont {Cooper-White}},\ }\bibfield  {title} {\enquote {\bibinfo {title} {Drop formation and breakup of low viscosity elastic fluids: Effects of molecular weight and concentration},}\ }\href {https://doi.org/10.1063/1.2190469} {\bibfield  {journal} {\bibinfo  {journal} {Phys. Fluids}\ }\textbf {\bibinfo {volume} {18}},\ \bibinfo {pages} {043101} (\bibinfo {year} {2006})}\BibitemShut {NoStop}%
\bibitem [{\citenamefont {Bailey}(2012)}]{bailey2012poly}%
  \BibitemOpen
  \bibfield  {author} {\bibinfo {author} {\bibfnamefont {FE~Jr}\ \bibnamefont {Bailey}},\ }\href@noop {} {\emph {\bibinfo {title} {Poly (ethylene oxide)}}}\ (\bibinfo  {publisher} {Elsevier},\ \bibinfo {year} {2012})\BibitemShut {NoStop}%
\bibitem [{\citenamefont {Brandrup}\ \emph {et~al.}(1999)\citenamefont {Brandrup}, \citenamefont {Immergut}, \citenamefont {Grulke}, \citenamefont {Abe},\ and\ \citenamefont {Bloch}}]{brandrup1999polymer}%
  \BibitemOpen
  \bibfield  {author} {\bibinfo {author} {\bibfnamefont {Johannes}\ \bibnamefont {Brandrup}}, \bibinfo {author} {\bibfnamefont {Edmund~H}\ \bibnamefont {Immergut}}, \bibinfo {author} {\bibfnamefont {Eric~A}\ \bibnamefont {Grulke}}, \bibinfo {author} {\bibfnamefont {Akihiro}\ \bibnamefont {Abe}}, \ and\ \bibinfo {author} {\bibfnamefont {Daniel~R}\ \bibnamefont {Bloch}},\ }\href@noop {} {\emph {\bibinfo {title} {Polymer Handbook}}},\ Vol.~\bibinfo {volume} {89}\ (\bibinfo  {publisher} {Wiley New York},\ \bibinfo {year} {1999})\BibitemShut {NoStop}%
\bibitem [{\citenamefont {Bhat}\ \emph {et~al.}(2022)\citenamefont {Bhat}, \citenamefont {Wani}, \citenamefont {Mir},\ and\ \citenamefont {Masoodi}}]{bhat2022advances}%
  \BibitemOpen
  \bibfield  {author} {\bibinfo {author} {\bibfnamefont {Iqra~Mohiuddin}\ \bibnamefont {Bhat}}, \bibinfo {author} {\bibfnamefont {Shoib~Mohmad}\ \bibnamefont {Wani}}, \bibinfo {author} {\bibfnamefont {Sajad~Ahmad}\ \bibnamefont {Mir}}, \ and\ \bibinfo {author} {\bibfnamefont {FA}~\bibnamefont {Masoodi}},\ }\bibfield  {title} {\enquote {\bibinfo {title} {Advances in xanthan gum production, modifications and its applications},}\ }\href {https://doi.org/10.1016/j.bcab.2022.102328} {\bibfield  {journal} {\bibinfo  {journal} {Biocatal. Agric. Biotechnol.}\ }\textbf {\bibinfo {volume} {42}},\ \bibinfo {pages} {102328} (\bibinfo {year} {2022})}\BibitemShut {NoStop}%
\bibitem [{\citenamefont {Rojas}(2016)}]{rojas2016cellulose}%
  \BibitemOpen
  \bibfield  {author} {\bibinfo {author} {\bibfnamefont {Orlando~J}\ \bibnamefont {Rojas}},\ }\href@noop {} {\emph {\bibinfo {title} {Cellulose chemistry and properties: fibers, nanocelluloses and advanced materials}}},\ Vol.\ \bibinfo {volume} {271}\ (\bibinfo  {publisher} {Springer},\ \bibinfo {year} {2016})\BibitemShut {NoStop}%
\end{thebibliography}%

\end{document}